\documentclass[aps,pra,floatfix,showpacs,twocolumn,preprintnumbers,tightenlines,superscriptaddress]{revtex4-1}

\usepackage{papersstyle}
\usepackage[nice]{nicefrac}
\usepackage{lipsum}
\usepackage{makecell}
\usepackage{booktabs}

\usepackage{xcolor}

\usepackage{amsmath}

\usepackage[capitalise]{cleveref}

\newcommand{\Imag}{\mathrm{Im}}         

\newcommand{\br}[1]{\left(#1\right)}
\newcommand{\ket}[1]{|#1\rangle}
\newcommand{\bra}[1]{\langle#1|}
\newcommand{\tr}{\mathrm{Tr}}
\newcommand{\Tr}[1]{\tr\br{#1}}

\newcommand{\us}{\ensuremath{\mu\mathrm{s}}}
\newcommand{\ns}{\ensuremath{\mathrm{ns}}}
\newcommand{\kHz}{\ensuremath{\mathrm{kHz}}}
\newcommand{\MHz}{\ensuremath{\mathrm{MHz}}}
\newcommand{\GHz}{\ensuremath{\mathrm{GHz}}}
\newcommand{\mK}{\ensuremath{\mathrm{mK}}}
\newcommand{\dB}{\ensuremath{\mathrm{dB}}}

\newcommand{\EC}{\ensuremath{E_{\rm C}}}

\newcommand{\Techo}{T_{2,\mathrm{echo}}}
\newcommand{\Tone}{T_{1}}
\newcommand{\Ttwostar}{T_{2}^{\ast}}

\begin{document}

\title{Chip-to-chip entanglement of transmon qubits using engineered measurement fields}

\author{C.~Dickel}
\author{J.~J.~Wesdorp}
\affiliation{QuTech, Delft University of Technology, Delft, The Netherlands}
\affiliation{Kavli Institute of Nanoscience, Delft University of Technology, Lorentzweg 1, 2628 CJ Delft, The Netherlands}
\author{N.~K.~Langford}
\affiliation{QuTech, Delft University of Technology, Delft, The Netherlands}
\affiliation{Kavli Institute of Nanoscience, Delft University of Technology, Lorentzweg 1, 2628 CJ Delft, The Netherlands}
\affiliation{School of Mathematical and Physical Sciences, University of Technology Sydney, Ultimo, New South Wales 2007, Australia}
\author{S.~Peiter}
\author{R.~Sagastizabal}
\author{A.~Bruno}
\affiliation{QuTech, Delft University of Technology, Delft, The Netherlands}
\affiliation{Kavli Institute of Nanoscience, Delft University of Technology, Lorentzweg 1, 2628 CJ Delft, The Netherlands}
\author{B.~Criger}
\affiliation{Computer Engineering, Delft University of Technology, Mekelweg 4, 2628 CD Delft, The Netherlands}
\affiliation{Kavli Institute of Nanoscience, Delft University of Technology, Lorentzweg 1, 2628 CJ Delft, The Netherlands}
\author{F.~Motzoi}
\affiliation{Department of Physics and Astronomy, Aarhus University, DK-8000 Aarhus C, Denmark}
\author{L.~DiCarlo}
\email{l.dicarlo@tudelft.nl}
\affiliation{QuTech, Delft University of Technology, Delft, The Netherlands}
\affiliation{Kavli Institute of Nanoscience, Delft University of Technology, Lorentzweg 1, 2628 CJ Delft, The Netherlands}

\begin{abstract}
While the on-chip processing power in circuit QED devices is growing rapidly, an open challenge is to establish high-fidelity quantum links between qubits on different chips.
Here, we show entanglement between transmon qubits on different cQED chips with $49\%$ concurrence and $73\%$ Bell-state fidelity.
We engineer a half-parity measurement by successively reflecting a coherent microwave field off two nearly-identical transmon-resonator systems.
By ensuring the measured output field does not distinguish $\vert 01 \rangle$ from $\vert 10 \rangle$, unentangled superposition states are probabilistically projected onto entangled states in the odd-parity subspace.
We use in-situ tunability and an additional weakly coupled driving field on the second resonator to overcome imperfect matching due to fabrication variations.
To demonstrate the flexibility of this approach, we also produce an even-parity entangled state of similar quality, by engineering the matching of outputs for the $\vert 00 \rangle$ and $\vert 11 \rangle$ states.
The protocol is characterized over a range of measurement strengths using quantum state tomography showing good agreement with a comprehensive theoretical model.
\end{abstract}

\maketitle

\section{Introduction}

The quest for large-scale quantum information processors is inspiring a multitude of architectures over a range of different qubit platforms that can be divided into two broad categories: monolithic~\cite{Divincenzo09,Helmer09cavity,Versluis17,Hill15,Li17} and modular~\cite{Monroe13,Nickerson13,Nemoto14,Brecht16}.
Monolithic architectures, in particular 2D lattices of qubits, are suitable for implementing the surface code~\cite{Bravyi98,Fowler12}, but designers face challenges with fabrication yield, connectivity and cross-talk on large-scale devices.
In contrast, modular architectures promise switchboard-like all-to-all connectivity, reduce design complexity and even correlated noise to the module scale, but face the challenge of distributing entanglement between nodes.
While local entangling operations inevitably outperform their remote counterparts, the challenges of scaling up suggest that a future quantum computer will require a hybrid architecture, which balances local speed and fidelity with the benefits of modularity.

In circuit quantum electrodynamics (cQED)~\cite{Blais04}, entanglement distribution schemes have mainly relied on two mechanisms:
Firstly, entanglement by measurement~\cite{Kerckhoff09,Barrett05} with either coherent~\cite{Roch14,Chantasri16,Roy15} or Fock-states~\cite{Narla16,Ohm17}, where a nonlocal entangling measurement is implemented by measuring photonic modes that have interacted with the qubits.
Secondly, pitch-and-catch schemes~\cite{Wenner14, Pfaff17}, where qubit-qubit entanglement is created by photons traveling from one qubit to another.
Since these protocols rely on photonic quantum information carriers, photon loss can limit either the achievable entanglement or the success rate.
Modest entanglement can be bolstered by entanglement distillation to produce high-fidelity quantum links.
Ultimately, the important figures of merit defining the performance of entanglement distribution protocols are entanglement generation rate and entanglement fidelity.
Experiments have primarily focused on qubits embedded in separate 3D superconducting cavities~\cite{Roch14,Chantasri16,Narla16}, which allows separate fabrication and selection of qubits and cavities, and tuning of the cavity coupling to input ports.
The effort to locally scale up to many-qubit experiments on the other hand has largely happened ``on chip'' \cite{Kelly15,Riste15,Song17,Takita17}, where both qubits and resonators are patterned in superconducting thin films and where fast, high-fidelity multiqubit gates have been demonstrated~\cite{Barends14}.
In these 2D cQED devices, fabrication variability impedes the precise parameter matching required for many entanglement protocols, but these devices are arguably better suited for integration and scale-up.
Therefore, generating rapid, high-fidelity entanglement chip-to-chip enables the exploration of interesting modular architectures in cQED.

Here, we entangle two transmon qubits on separate 2D-cQED chips by engineering a half-parity measurement using the bounce-bounce entanglement-by-measurement protocol~\cite{Kerckhoff09, Roch14, Chantasri16}.
A perfect odd half-parity measurement probabilistically projects a maximum superposition state into the $\ket{00}$, $\ket{11}$ or an entangled superposition of $\ket{01}$ and $\ket{10}$.
Distinguishability between $\ket{01}$ and $\ket{10}$, caused by differences between the two chips, leads to dephasing of the resulting entangled state and therefore degrades the entanglement.
Two innovations make the protocol robust to fabrication variations.
Firstly, adding resonator tuning qubits for frequency matching overcomes imperfect resonator frequency targeting.
Secondly, we use an additional weakly-coupled port of the second resonator to apply a compensation pulse and reduce any distinguishability in the output fields for $\ket{01}$ and $\ket{10}$.
We demonstrate the versatility of this technique by also matching the outputs for $\ket{00}$ and $\ket{11}$ to create an even-parity Bell state with similar performance.
We characterize the performance of our protocol in aggregate by comparing the output states at different measurement strengths against a comprehensive model of the experiment.

\section{Experiment Overview and Extended Bounce-Bounce Protocol}
\label{sec:bounce_bounce}

\begin{figure}
  \begin{center}
  \includegraphics[width=\columnwidth]{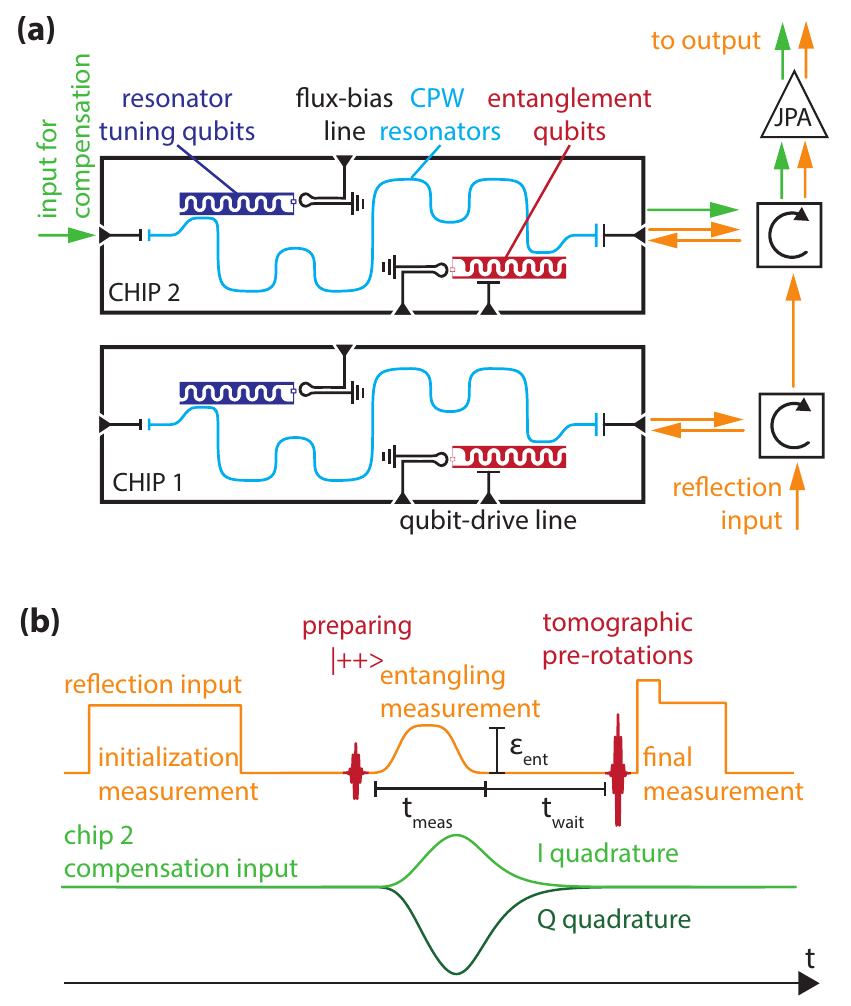}
  \end{center}
  \caption{
    \textbf{(a)} Bounce-bounce entanglement scheme.
    A microwave field (orange arrows) from the reflection input successively reflects on two CPW resonators (light blue) on separate chips via two circulators and is then amplified using a JPA.
    Each resonator is dispersively coupled to a transmon qubit (red).
    Additional tuning qubits (dark blue) are used to match the resonator frequencies via their dispersive shifts.
    A weakly-coupled input port to the resonator on the second chip is used to inject a compensation field (green arrow) to reduce distinguishability caused by a mismatch between parameters of different qubit-resonator systems.
    \textbf{(b)} Pulse scheme of the experiment.
    An initial measurement is used to condition on qubits in the ground state.
    Then, the entanglement qubits are prepared in the $\left \vert ++ \right \rangle$ state.
    Entangling measurement pulses are applied through the reflection and compensation input.
    After waiting for the photons to leak out of the resonator, quantum state tomography is performed by applying an over-complete set of pre-rotations and a final measurement.
}
  \label{fig:fig1}
\end{figure}

The bounce-bounce approach to entanglement was proposed as a continuous two-qubit parity measurement in cavity quantum electrodynamics \cite{Kerckhoff09}.
The qubit parity is mapped on a coherent state that successively reflects from two cavities and is then read out with a continuous homodyne measurement, leaving the two qubits entangled.
Our setup [\cref{fig:fig1}\textbf{(a)}] consists of two nominally identical chips each containing two transmon qubits both coupled to a coplanar waveguide (CPW) resonator.
The two qubits in red will be entangled in the protocol.
The $\nicefrac{\lambda}{2}$ CPW resonator is strongly coupled to a feed line on one side and weakly on the other.
This asymmetric coupling directs most of the photons on a single path that leads through the circulators to a Josephson parametric amplifier (JPA)~\cite{Castellanos-Beltran08}, realizing a high-fidelity measurement of the output field.
Details on the experimental setup and device fabrication can be found in Appendices \ref{sec:setup} and \ref{sec:parameters}.

In order to understand the measurement central to this experiment, it is useful to first consider the standard cQED measurement for a single qubit-resonator system in the dispersive limit~\cite{Gambetta08}.
In this limit, the qubit-resonator interaction simplifies to a qubit-dependent, dispersive shift $\chi$ of the resonator frequency.
Under a coherent drive, the resonator therefore follows qubit-dependent coherent-state trajectories $\ket{\alpha_i(t)}$ with classical equations of motion for $\alpha_i(t)$ that depend on system parameters and the time-dependent drive.
This entangles the resonator and qubit, creating the state $a \ket{0}\ket{\alpha_{0}} + b \ket{1}\ket{\alpha_{1}}$ for a qubit initially in $a\ket{0}+b\ket{1}$.
As photons leak out of the resonator carrying qubit-state information, the qubit becomes more mixed, with coherence decaying according to the measurement-induced dephasing rate $\Gamma_{\mathrm{m}}$:
\begin{equation}\label{eq:measurement_induced_dephasing}
\Gamma_{\mathrm{m}} = 2 \chi \int{ \Imag \left [ \alpha_{0}(t) \alpha^{*}_{1}(t) \right ] \mathrm{d}t}.
\end{equation}
The dephasing seen by the qubit can be controlled by the coherent cavity drive, but the ability to infer the qubit state from the measured time-varying output signal, or transient, also depends on the noise added by the detection chain.
Importantly, because the output field is directly related to the intracavity field, if the cavity starts and ends in the vacuum state, the dephasing can also be related to the measured average transient difference~\cite{Bultink17}.

In a multi-qubit context, these concepts were generalized to realize entangling measurements~\cite{Kerckhoff09,Tornberg10,Motzoi15}.
For a joint measurement, selectively tuning the distinguishability between different state-dependent output transients can give dramatically different dephasing rates for different two-qubit coherence terms.
For example, minimizing the dephasing between $\ket{01}$ and $\ket{10}$ creates a half-parity measurement that selectively preserves superpositions in the odd subspace, while giving distinct outcomes for $\ket{00}$ and $\ket{11}$.
Thus, this measurement projects a separable maximum superposition state to an entangled odd-parity Bell state with 50\% probability, with the corresponding measured outcome heralding successful entanglement generation.

In the bounce-bounce scheme, a perfect half-parity measurement requires no intra-cavity loss $\eta_l$ and identical qubit-cavity pairs.
In 2D-cQED devices, however, fabrication variability makes precise parameter matching infeasible, and a more sophisticated approach is required.
In our experiment, we introduced two techniques to mitigate these effects.
Firstly, the variable dispersive shifts from two additional tuning qubits [dark blue in \cref{fig:fig1}\textbf{(a)}] are used to match the fundamental frequencies of the two resonators [see \cref{sec:tuning_qubits} for details].
Secondly, to minimize any remaining transient distinguishability due to different resonator linewidths or dispersive shifts, we apply a compensation pulse to an additional, weakly coupled input port at the back of the second resonator [denoted by green arrows in \cref{fig:fig1}\textbf{(a)}].
Effectively, interacting with only one resonator, this compensation pulse adds coherently to the reflected field from the bounce-bounce path [orange arrows in \cref{fig:fig1}] and can be shaped to conditionally displace the target trajectories to remove residual transient distinguishability.

For a given input pulse and system parameters, the optimal compensation pulse shape can be solved directly from the classical field equations in the Fourier domain [see \cref{sec:modeling} for detailed derivation].
In this approach, the qubit state dependent output field $y^{ij}\left(\omega\right)$ is a linear function of the reflection input field  $\epsilon^{\mathrm{s}}(\omega)$ and the transmission compensation field $\epsilon^{\mathrm{w}}(\omega)$ via
\begin{equation}
\label{eq:conceptual_output_solution}
y^{ij}(\omega) = H_{\mathrm{refl}}^{ij}(\omega, \vec{p}) \epsilon^{\mathrm{s}}(\omega) + H_{\mathrm{trans}}^{j}(\omega, \vec{p}) \epsilon^{\mathrm{w}}(\omega),
\end{equation}
where $i, j \in \{ 0, 1\}$ denote the state of the first and second qubit, and where $H_{\mathrm{refl}}^{ij}(\omega, \vec{p})$ and $H_{\mathrm{trans}}^j(\omega, \vec{p})$ are complex valued transfer functions that denote the individual system response to each input.
The system parameter vector $\vec{p}$ consists of, for each chip, the resonator linewidth $\bar{\kappa}$ =  $\kappa^s$ + $\kappa^w$ + $\kappa^I$, with terms for the weakly and strongly coupled ports and the intrinsic losses, the dispersive shift $\chi$, and the resonator-drive detuning $\Delta$, as well as $\eta_l$ and $\phi$, the interchip loss and acquired phase (see \cref{tab:Device_Params} for the measured values).
This approach was tested by comparing predicted and measured output fields for various input fields [see \cref{fig:SOM_transients}].
To ensure a measurement does not distinguish in the odd [even] subspace, we require $y^{01}(t) = y^{10}(t)$ [$y^{00}(t) = y^{11}(t)$] at all times.
This gives a linear equation $\epsilon^{\mathrm{w}}\left(\omega\right) = H^\mathrm{comp}(\omega, \vec{p}) \epsilon^{\mathrm{s}}\left(\omega\right)$ where $H^\mathrm{comp}(\omega,\vec{p})$ relates the transmission input to the reflection input.
The classical solutions were then used to implement master equation (ME) and stochastic master equation (SME) models in the polaron frame incorporating the effect of qubit decoherence and post-selection on the measurement result, respectively~\cite{Roch14,Motzoi15} [see \cref{sec:modeling} for details].

\section{Experimental pulse sequence and compensation pulse tune-up}
\label{sec:compensation_field}

\begin{figure*}[]
  \begin{center}
  \includegraphics[width=\textwidth]{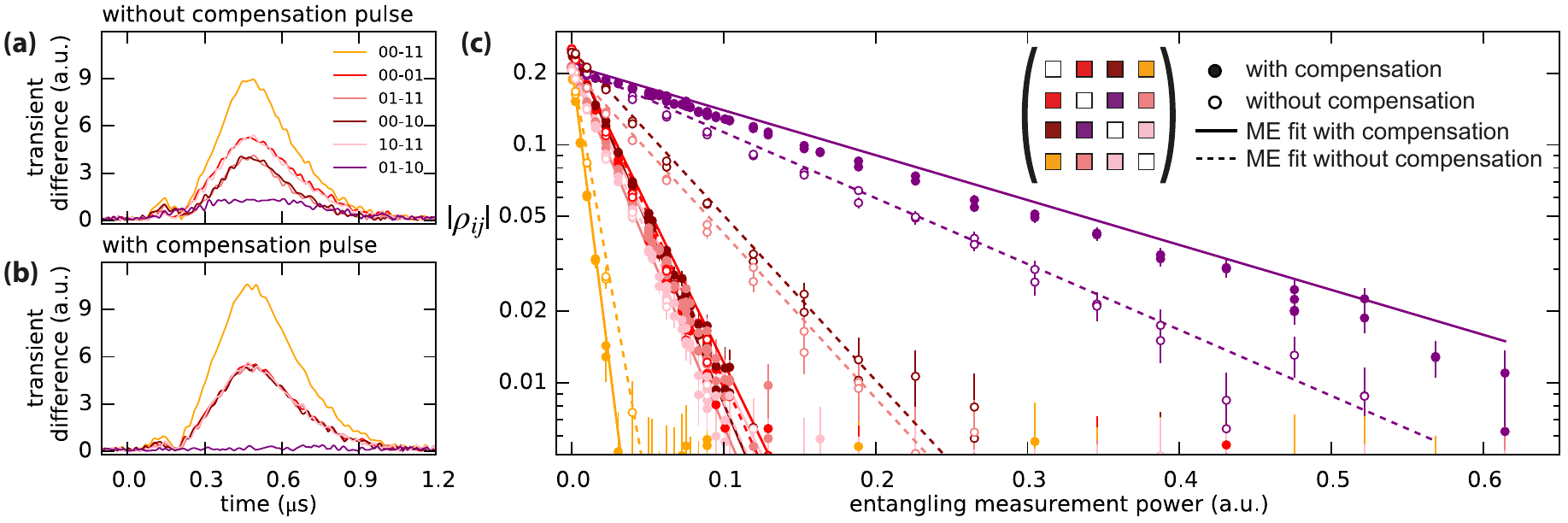}
  \end{center}
  \caption{
      \textbf{(a)} and \textbf{(b)} Average output transient differences $\vert y_{ij}-y_{kl}\vert$ for different pairs of initial states with and without the compensation pulse.
      For the ideal half-parity measurement, the difference between the $\vert 01 \rangle$ and  $\vert 10 \rangle$ outputs is zero at all times, which we realize with the compensation pulse.
      Additionally, the output difference for the other states is increased.
      \textbf{(c)} Measurement-induced dephasing giving a decay of the coherence elements of the unconditioned density matrix as a function of measurement power.
      Dispersive readout gives exponential decay as a function of power as suggested by \cref{eq:measurement_induced_dephasing}.
      A master-equation model is fitted to data with inter-chip loss $\eta_{\mathrm{l}}$ and amplitude scaling factor as only free parameters.
      Residual dephasing is largely explained by $\eta_{\mathrm{l}}=11.8\%$ (obtained from fit).
  }
  \label{fig:fig2}
\end{figure*}

The experimental pulse sequence [\cref{fig:fig1}\textbf{(b)}] is designed to faithfully characterize the entangling measurement using quantum state tomography (QST) with a joint readout \cite{Chow10,Filipp09}.
We first apply a projective measurement to be able to filter out residual qubit excitations.
While conditioning on the initial measurement reduces residual excitation, any remaining residual excitation can lead to an overestimate of the achieved entanglement by QST, an effect which we correct for (see \cref{sec:tomography}).
Next, we prepare the two qubits in the maximum superposition state $\ket{++}$, a tensor product with both qubits in the state $\ket{+} = (\ket{0} + \ket{1})/\sqrt{2}$.
Qubit gates are applied to the entanglement qubits via a capacitively coupled drive line (see \cref{sec:tune-up} for qubit tune-up and performance).
Then, we apply the entangling measurement, to probabilistically project the maximum superposition state to an entangled state.
To verify the entanglement, we perform QST by applying an overcomplete set of different pre-rotations on the qubits followed by a final measurement.
All measurements consist of coherent microwave drives that populate the resonators with photons.
The initial and final measurements are tuned for high single-shot fidelity (and avoiding measurement-induced excitations in case of the initial measurement).

The entanglement measurement strength can be varied either by changing the measurement amplitude or the duration, as marked in \cref{fig:fig1}\textbf{(b)}.
However, the simple equations of motion for the resonator state are only valid in the absence of qubit relaxation, thus the measurement time should be much shorter than the qubit lifetime $\Tone$.
As the shorter $\Tone = 9~\us$, we use a $300~\ns$ measurement pulse with a smoothed square envelope to ease bandwidth requirements on the compensation pulse.
The pulse is too short for the resonators to reach steady state.
While the resonator-qubit system is in an entangled state with non-vacuum coherent states in the resonator, reliable gates on the qubits are not possible.
Accordingly, the entangling protocol is only completed once the photons have left the resonators.
We wait $700~\ns$ for the resonators to empty before doing tomography, fixing the duration of the entangling protocol to $1~\us$.
Thus, measurement-independent qubit decoherence is fixed and the tomography as a function of measurement amplitude reveals the action of the measurement.

To realize the optimum compensation pulse $\epsilon^{\mathrm{w}}(\omega)$ for an input  $\epsilon^{\mathrm{s}}(\omega)$, the parameters $\vec{p}$ need to be determined.
Precise measurements are not straightforward for several parameters, such as $\kappa_\mathrm{W}$ and $\kappa_\mathrm{I}$, the power difference of the two drives (due to small unknown differences in line attenuation and in the two mixers), $\eta_\mathrm{l}$ and the phase shift that the signal acquires between the two chips.
To tune up the optimal compensation pulse, we minimize the transient difference between the odd (even) subspace $\ket{01}$ ($\ket{00}$) and $\ket{10}$ ($\ket{11}$) normalized by the sum of the other transient differences to keep the impact on readout fidelity to a minimum.
This is optimized by iteratively varying $\vec{p}$ using a combination of hands-on and hands-off optimization~\cite{Nelder65}.

We can look at the transient differences for all state pairs with and without compensation pulse [see \cref{fig:fig2}\textbf{(a,b)}], to determine how good the compensation works.
Just using the reflection input, there is still a mismatch between the output fields for $\vert 01 \rangle$ and  $\vert 10 \rangle$ [difference $\vert y_{ij}-y_{kl}\vert$].
The improvement of the compensation pulse is two-fold, increasing the transient difference for the states we want to dephase and minimizing it for the subspace we want to preserve.
The optimization is performed close to the optimal amplitude for entanglement after initial experiments.
Since the transient-difference signal is noisy and affected by qubit relaxation, in future experiments it may prove more efficient to optimize on a qubit-based signal such as the dephasing itself or the acquired phase shift between the target states.

\section{Experimental Results}
\label{sec:results}

\begin{figure*}[ht]
  \begin{center}
  \includegraphics[width=\textwidth]{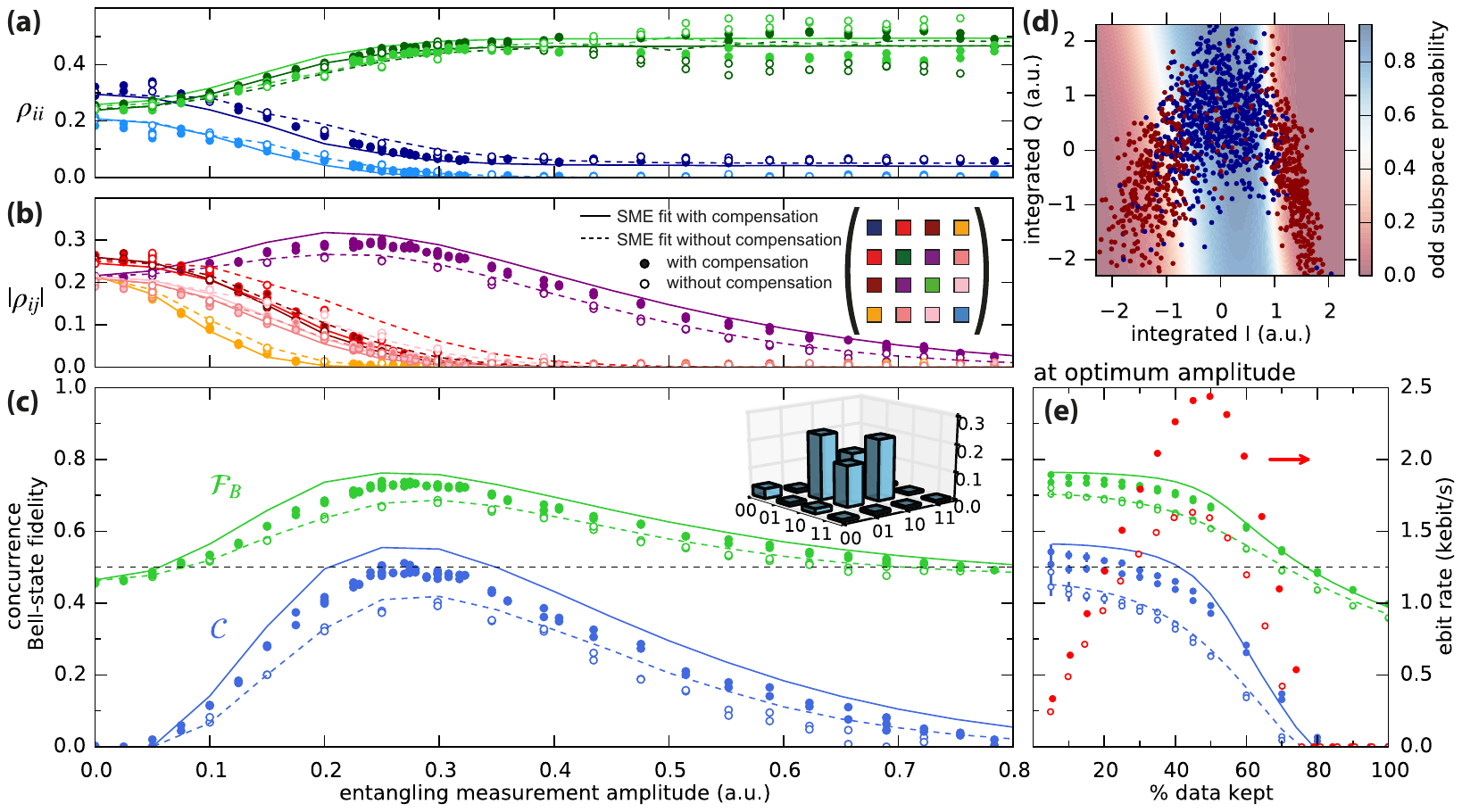}
  \end{center}
  \caption{
      \textbf{(a, b)} Evolution of the conditional density matrix $\rho$ as a function of measurement amplitude with and without the compensation pulse.
      We keep $25\%$ of the data based on the measurement outcome.
      SME simulation using the ME parameters with $\eta_\mathrm{m} = 50\%$ shows good agreement with the data.
      \textbf{(c)} Concurrence $\mathcal{C}$ and Bell-state fidelity $\mathcal{F}_\mathrm{B}$ as a function of measurement amplitude comparing both cases.
      Inset shows $|\rho|$ with compensation at optimum amplitude.
      \textbf{(d)} Thresholding the data using machine learning.
      Example data of integrated calibration-point output at optimum amplitude with even (red points) and odd (blue points) subspace data points.
      A neural network classifier is trained on calibration points giving a learned odd-subspace probability landscape (color scale).
      $90\%$ of the calibration data is used to train, $10\%$ is used to estimate the assignment fidelity, here giving $85\%$.
      The classifier is used to select the fraction of data with the highest odd-subspace probability.
      \textbf{(e)} $\mathcal{C}$ and $\mathcal{F}_\mathrm{B}$ at the optimum amplitude as a function of data kept with and without the compensation pulse.
      We also compute the ebit rate (red) as described in the main text.
  }
  \label{fig:fig3}
\end{figure*}

We now confirm the effect of the transient matching on the qubits, using QST to reconstruct the density matrix after measurement.
Measurement-induced dephasing leads to an exponential decay in the coherence elements of the density matrix as a function of measurement power as shown in \cref{fig:fig2}(\textbf{c}).
The better transient matching with the compensation pulse results in reduced dephasing in the wanted subspace while enhancing the dephasing of the unwanted coherence elements over the entire amplitude range.
Measurement power was rescaled for the independently measured mixer nonlinearity.
We plot the ME simulation results (see \cref{sec:modeling}) for both the compensation and no-compensation case, showing good agreement with the data.
When fitting the ME we fixed the estimates for all parameters in $\vec{p}$ from independent measurements except for $\eta_\mathrm{l}$ and a scaling factor between the input power on the arbitrary waveform generator (AWG) that time-shapes a microwave carrier and the power that arrives at the experiment.
We performed a single fit of the measurement induced dephasing with the ME simulation to all 6 independent complex off-diagonal density matrix elements as a function of amplitude (populations remain constant).
The full density matrix data and fits can be found in \cref{fig:SOM_full_density_matrix_fits} (\cref{sec:modeling}).

Due to the finite $\eta_\mathrm{l}$, there is dephasing even for perfectly matched transients.
We find that the datasets with and without compensation pulse are well described for  $\eta_\mathrm{l}=11.8\%$ in the model.
This power loss is partially explained by the circulators, which are specified to give 3-4\% , with the connectors to the printed circuit board (PCB) also likely to contribute significantly.

We now shift from looking only at the selective dephasing in the unconditional density matrix evolution to looking at the density matrix conditioned on the measurement outcome.
The new variable to consider in this context is $\eta_\mathrm{m}$, which determines the signal-to-noise ratio (SNR) for state determination based on the measurement outcome.
We use the following entanglement measures as figures of merit: concurrence $\mathcal{C}$~\cite{Wootters98}, Bell-state fidelity $\mathcal{F}_\mathrm{B}$ and the ebit rate, discussed below.

For good qubit readout at low photon numbers we require a low-noise amplifier.
The amplifier is a JPA that we operate in phase-sensitive mode with a single strong pump tone (see \cref{sec:tune-up} for tune-up procedure).
This results approximately in a homodyne measurement that is effectively only sensitive along one quadrature, due to the squeezing of the amplifier.
The single-quadrature sensitivity puts an interesting constraint on the output fields: it penalizes having a signal that oscillates between quadratures.
For this reason, it is beneficial to place the measurement tone at the symmetry point between the ground and excited state frequencies of the resonators.
To simultaneously reach this condition for both resonators, they need to be lined up using the tuning qubits.
This is irrelevant for the measurement-induced dephasing, as resonator frequency differences can be taken into account in the compensation.

In addition to optimally employing the JPA and achieving the symmetric readout condition, digital processing of the output traces with integration weights is used to further increase the SNR.
For a binary readout problem, the weight function for optimally distinguishing the states is the average transient difference in each quadrature (in the absence of qubit decay)~\cite{Ryan15}.
In this case (\cref{fig:fig2}), the shape of the transient difference for different pairs of states is similar, such that we can economize.
We used the mean of the transient difference for $\ket{01}$-$\ket{00}$, $\ket{10}$-$\ket{00}$, $\ket{01}$-$\ket{11}$, and $\ket{10}$-$\ket{11}$ as integration weights for the I and Q quadratures separately, giving a complex data point for each run of the experiment.

The binary decision whether a measurement result corresponds to the odd-subspace is a textbook classification problem.
We relied on a machine-learning based approach~\cite{Magesan15}, training a neural network classifier~\cite{scikit-learn} on calibration points [\cref{fig:fig3}\textbf{(d)}], which proved more robust than an approach based on Gaussian fits and linear boundaries in phase space.
The calibration points for even (odd) parity are the red (blue) points.
The color scale indicates the odd-subspace probability landscape learned by the neural network.
For a given wanted percentage of data kept, the experimental runs with the highest odd-subspace probability were kept.
Note that the classifier could also be trained on full single-shot traces, in which case the integration weights would be unnecessary.

The conditional density matrix evolution keeping $25\%$ of the data is shown in \cref{fig:fig3}\textbf{(a,b)} as a function of the measurement amplitude.
As the measurement becomes stronger, the ability to threshold out the even subspace increases as shown by the reduction in even population and increase in odd populations.
The wanted odd subspace coherence element first increases due to the selection, and is eventually limited by the measurement induced dephasing.
Qubit relaxation during measurement leads to a residual population in the $\ket{00}$ state. Note that early relaxation events will lead to $\ket{00}$ outcomes and will be filtered out.
The data shows good agreement with an SME simulation with the same parameters as the ME simulation.
The SME assumes a perfect signal-quadrature measurement with no squeezing as described in~\cite{Roch14, Motzoi15}.
We find that $\eta_\mathrm{m}=50\%$ gives good agreement with the experiment for the no-compensation case.

We now extract different entanglement measures from the conditioned density matrix keeping $25\%$ of the data [\cref{fig:fig3}\textbf{(c)}].
While $\mathcal{C}$ can be directly computed, $\mathcal{F}_\mathrm{B}$ requires finding the odd (or even) Bell state with the highest overlap.
A non-zero $\mathcal{C}$ signals entanglement, as does a $\mathcal{F}_\mathrm{B}$ larger than 0.5.
Both $\mathcal{F}_\mathrm{B}$ and $\mathcal{C}$ peak at a common amplitude, which is characterized by a balance between good SNR and low measurement-induced dephasing in the odd-parity subspace.
Improvements in $\eta_\mathrm{m}$ would shift the optimum to lower amplitudes and improve the result.
The compensation pulse dataset clearly outperforms the no-compensation case but falls slightly below the theory which assumes a perfect compensation pulse.
It is possible that the JPA tune-up gave a slightly lower $\eta_\mathrm{m}$ but the ME simulation in \cref{fig:fig2}\textbf{(c)} already shows signs of the sub-optimal compensation and the maximum $\mathcal{C}$ coincides for both cases.
At high amplitudes, $\eta_\mathrm{m}$ likely starts to suffer from from the onset of compression in the JPA.
We reach an optimum $\mathcal{C}=0.49\pm0.01$ and $\mathcal{F}_\mathrm{B}=0.731\pm0.003$ with the compensation pulse, and $\mathcal{C}=0.40\pm0.01$ and $\mathcal{F}_\mathrm{B}=0.683\pm0.003$ without.
The error bars are derived from Monte Carlo simulations based on a coin-toss model of multinomial sampling statistics.
Point by point fluctuations seem to exceed the statistical errors, possibly due to JPA related fluctuations in quantum efficiency, drift in qubit coherence time and thermal excitations.

It is also interesting to look at the entanglement measures at the optimum amplitude as a function of the data kept when selecting on the entangling measurement [\cref{fig:fig3}\textbf{(e)}].
In addition to $\mathcal{C}$ and $\mathcal{F}_\mathrm{B}$, we also compute the ebit rate, which is the product of the logarithmic negativity~\cite{Horodecki09}, the fraction of the data we keep and the experimental repetition rate of $10~\kHz$.
The logarithmic negativity gives an upper bound for the distillable entanglement of a state~\cite{Vidal02}.
The ebit rate is a relatively conservative estimate of an actually achievable entanglement rate, as the entire experimental sequence takes less than $5~\us$ including an initialization measurement, which in principle could be combined with active feedback for faster qubit initialization~\cite{Riste12}, and the QST.
Thus, repetition rates on the order of  $200~\kHz$ should be achievable, corresponding to ebit rates around $40~\kHz$, comparable to qubit coherence times.
The ebit rate peaks when keeping 50$\%$ of the data, which corresponds to $\mathcal{C}=0.38\pm0.01$ and $\mathcal{F}_\mathrm{B}=0.668\pm0.003$.

\begin{figure}
  \begin{center}
  \includegraphics[width=\columnwidth]{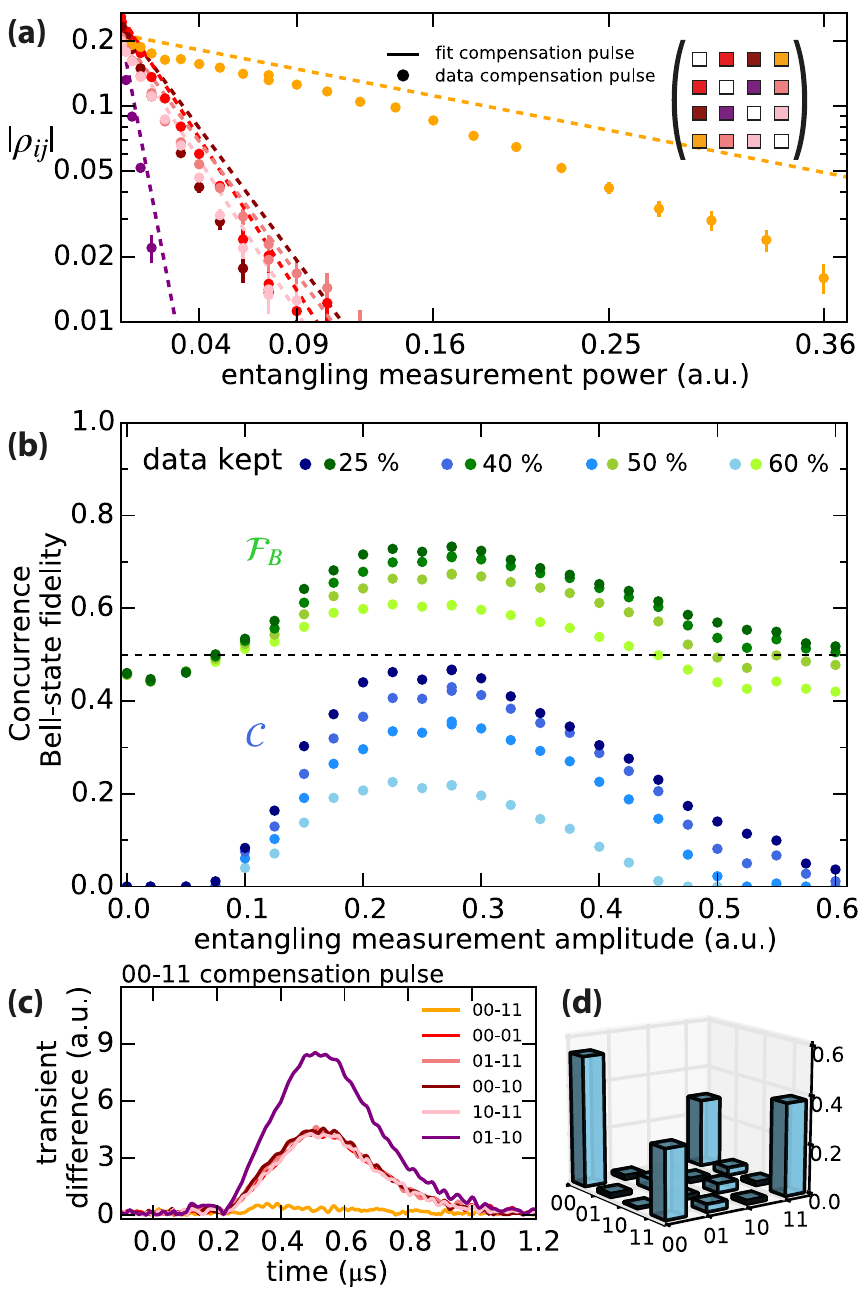}
  \end{center}
  \caption{
    \textbf{(a)} Measurement-induced dephasing for the even-subspace compensation pulse.
    \textbf{(b)} Concurrence (blue points) and even-Bell-state fidelity (green points) as a function of amplitude for different amounts of data kept.
    \textbf{(c)} Transient matching for the even subspace at optimum concurrence.
    \textbf{(d)} Best even-Bell-state density matrix keeping 25\% of the data.
  }
  \label{fig:fig4}
\end{figure}

As mentioned in \cref{sec:compensation_field}, by simply changing the form of the compensation pulse, it can be used to minimize the measurement-induced dephasing for any pair of states.
To demonstrate this, we also implemented the compensation pulse that produces identical output for $\ket{00}$ and $\ket{11}$.
The results are summarized in \cref{fig:fig4}.
In this case, the compensation pulse has to be stronger, as we match the two states that are naturally most distinguishable.
While the transient matching in \cref{fig:fig4}\textbf{(c)} is comparable to the odd case, the measurement-induced dephasing shows a stronger deviation from the model.
This is most likely due to mixer imperfections, such as skewness and non-linearity, which were not independently calibrated for both mixers.
These effects were likely more detrimental with higher mixer voltages for the even compensation pulse, but probably also contributed to not reaching the optimum in the odd case.
Nonetheless, we realize an even-parity entangled state almost matching the odd-parity performance and outperforming the no-compensation case reaching a $\mathcal{C}=0.47\pm0.01$ and $\mathcal{F}_\mathrm{B}=0.732\pm0.005$ when keeping $25\%$ of the data.
The model predicts identical performance for an optimally tuned compensation pulse.

\section{Conclusion}
\label{sec:Conclusion}

We have shown that the bounce-bounce scheme can be implemented in a 2D-circuit QED setup, achieving state-of-the-art remote entanglement for superconducting qubits.
The two chips in the experiment are not identical: tuning qubits are used to match the resonator frequencies.
The resonator linewidths are significantly different but the additional compensation pulse allows the matching of the transients to realize either an odd or an even half-parity measurement.
The experiment is not limited by the resonator linewidth as a steady state of the resonators is never reached.
Our current implementation leaves room for improvement in the limiting $\eta_\mathrm{l}$ and $\eta_\mathrm{m}$.

Managing the photon loss to improve the achieved entanglement is difficult, but there are several obvious improvements.
One circulator can be removed without compromising performance, as done in~\cite{Chantasri16}.
Developments of on-chip circulators~\cite{Chapman17} and better parametric amplifiers might lead to improvements in $\eta_\mathrm{l}$ and $\eta_\mathrm{m}$, respectively.
The loss could also be managed with quantum-error-correction-like protocols that make use of ancilla qubits~\cite{Roy16}.

A current maximum of $50\%$ success probability would either require several pairs of qubits where the protocol is performed in parallel or several entangling attempts.
The protocol can be sped up employing faster ramp-up and ramp-down pulses.
An entanglement generation time  $1~\us$ would be promising for quantum network operation given qubits with demonstrated $\sim50~\us$ coherence times.
With further improvements, a cQED realization of entanglement distillation~\cite{Deutsch96,Bennett96b} should come within reach.
Also, in this two-qubit/two-cavity bounce-bounce configuration, entanglement generation via bath engineering~\cite{Shankar13, Motzoi16,Kimchi-Schwartz16} and feedback-control schemes~\cite{Martin15} can be further explored to achieve steady-state entanglement.

\begin{acknowledgments}
We thank B. Tarasinski, N. Bultink, A. Rol, and F. Luthi for helpful discussions, and D. Rist\`{e} for help with the initial experimental plan and comments on the manuscript.
We thank D. Thoen and A. Endo for providing the NbTiN thin films, W. Kindel and K. W. Lehnert for the JPA, and A. Wallraff for a precision coil to bias it.
This research was supported by the Netherlands Organization for Scientific Research as part of the Frontiers of Nanoscience program (NWO/OCW), the ERC Synergy grant QC-lab and U.S. Army Research Laboratory (ARL) Center for Distributed Quantum Information (CDQI) program through cooperative Agreement No. W911NF-15-2-0061.
\end{acknowledgments}

\clearpage
\appendix

\section{Experimental setup}
\label{sec:setup}

\begin{figure}
  \centering
  \includegraphics[width=\columnwidth]{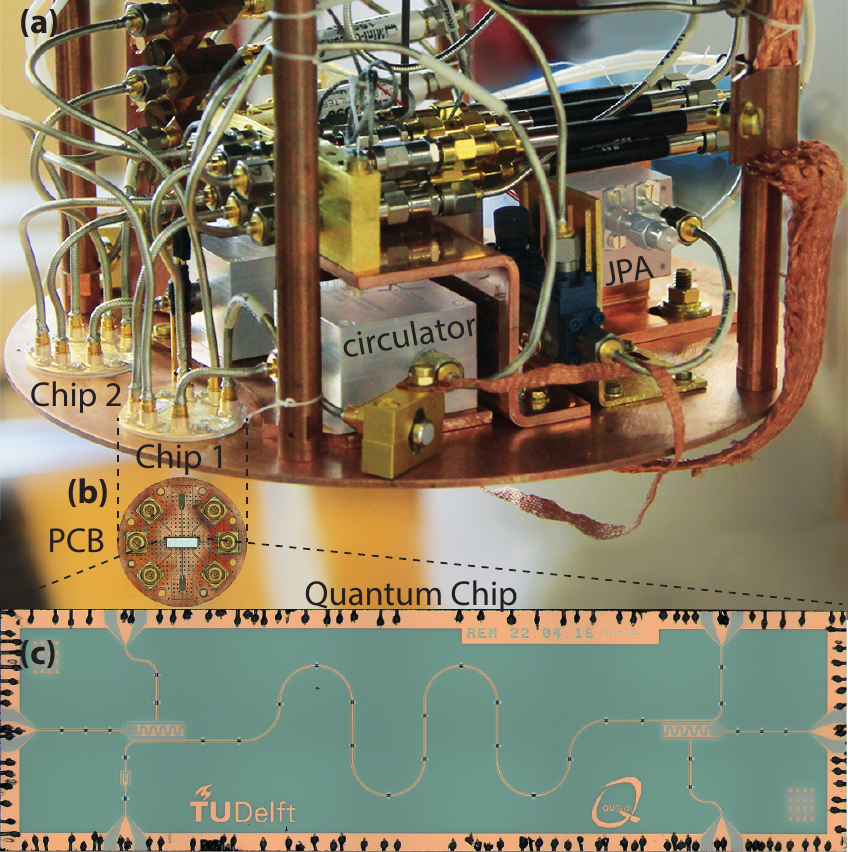}
  \caption{
    Photographs of the setup.
    \textbf{(a)} Cold finger of the dilution refrigerator with the 2 chips, circulators and JPA.
    \textbf{(b)} Bird's eye view of the PCB.
    \textbf{(c)} Microscope image of the 7~mm~$\times$~2~mm chip.
  }
  \label{fig:SOM_Setup_picture}
\end{figure}

\begin{figure}
  \centering
  \includegraphics[width=\columnwidth]{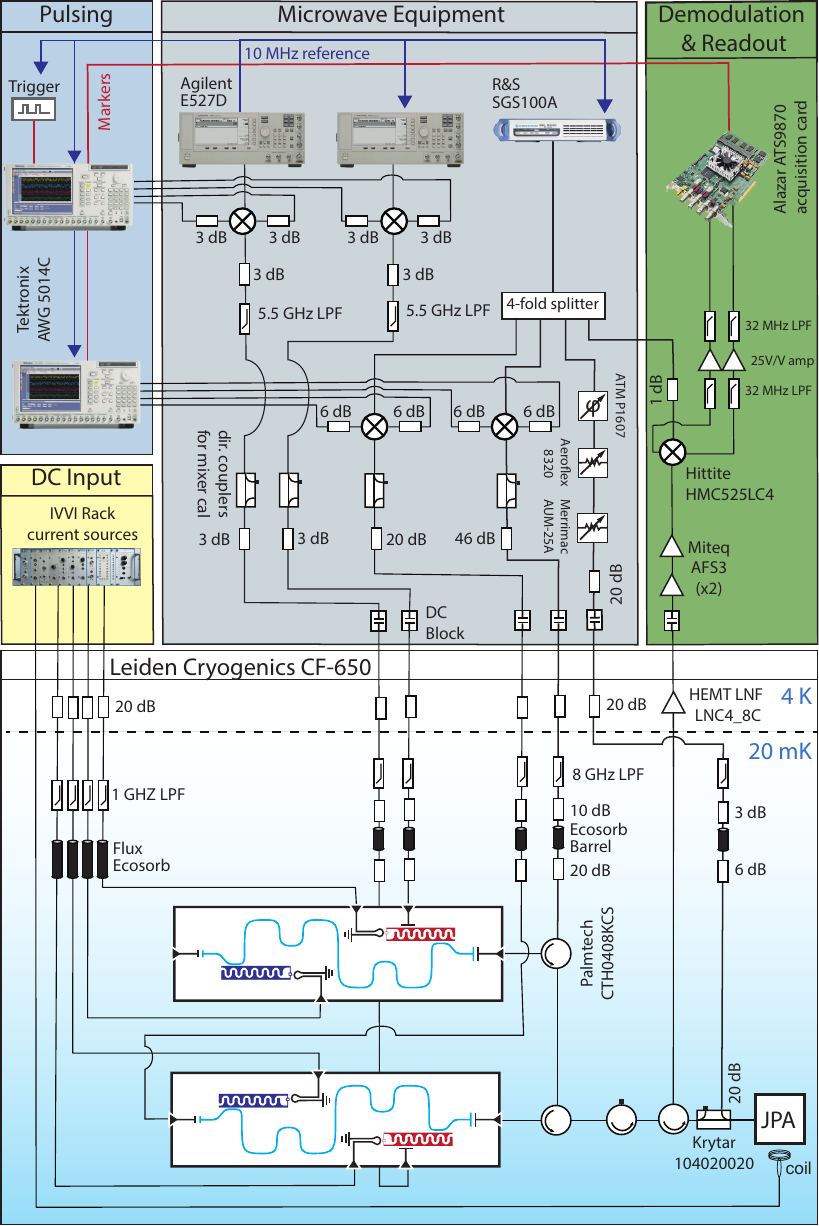}
  \caption{
    Detailed schematic of the experimental setup.
  }
  \label{fig:SOM_wirind_diagram}
\end{figure}

Both chips were attached to the cold finger of a Leiden Cryogenics CF-650 dilution refrigerator as seen in \cref{fig:SOM_Setup_picture}.
The temperature of the cold finger during the experiment was around $35~\mK$.
For radiation shielding, the entire setup is enclosed within a copper can coated with a mixture of Stycast 2850 and silicon carbide granules (15 to 1000 nm diameter) used for infrared absorption~\cite{Barends11}.
To shield against external magnetic fields, the can is enclosed by an aluminum can and two Cryophy cans.

A detailed wiring diagram of the experiment can be found in \cref{fig:SOM_wirind_diagram}.
Microwave lines are filtered using $\sim 60~\dB$ of attenuation using both commercial cryogenic attenuators and home-made Eccosorb filters for infrared absorption.
Flux-bias lines are also filtered using commercial low-pass filters and Eccosorb filters with a stronger absorption, in principle allowing for fast control of qubit frequencies, even though in this experiment only static biasing was used.
The JPA is mounted with an additional circulator to prevent leakage of the resonant pump tone back to the experiment.
This can be improved in future experiments, as double-pumping or pump-canceling schemes could have been used in place of the additional circulator, likely improving the quantum efficiency.

\section{Device Fabrication and parameters}
\label{sec:parameters}

The devices were fabricated with the same process as those in~\cite{Langford17}.
Device parameters can be found in \cref{tab:Device_Params}.
Bare resonator frequencies are close to the target frequency, resonator targeting is discussed in more detail in \cref{sec:tuning_qubits}.
The difference in $\kappa_{c}$ between the two chips with identical base-layer patterns that come from the same die is surprising and suggests that either wire-bonds or packaging play a role.
Likely this is also the cause of the $\kappa_c$ value being off target.

The qubit frequencies are well matched for this pair of devices.
Usually Josephson junction fabrication leads to an expected relative spread of several percent in qubit frequencies~\cite{Pop12}.
Pairs of matching qubits can be selected from the room-temperature resistances of the Josephson junctions~\cite{Ambegaokar63}, which in this case differed by 1\%.
However, the absolute frequencies were not on target, due to systematic shifts in the junction parameters between different fabrication runs.
Reducing the statistical spread and systematic variations between Josephson junction fabrication runs remains an outstanding challenge for future many-qubit devices.

\begin{table}
\centering
\resizebox{0.8\columnwidth}{!}{
\begin{tabular}{lrrr}
\hline
\text{Parameter} & \text{Target}       & \text{Chip 1}   & \text{Chip 2}    \\
\hline

$f_{\mathrm{r,bare}}$   &   6.27~GHz     & 6.344~GHz      & 6.339~GHz  \\

$\nicefrac{\kappa}{2\pi}$   &   2~MHz     & 3.01~MHz     & 4.53~MHz  \\

$f_{\mathrm{q, max}}$   &   5.57~GHz     & 5.23~GHz       & 5.24~GHz    \\

$\nicefrac{E_{\mathrm{c}}}{h}$   &   280~MHz     & 293~MHz       & 293~MHz   \\

$\nicefrac{\chi}{2\pi}$   &  -1~MHz     & -335~kHz      & -335~kHz     \\
\hline
\end{tabular}
}
\caption{
    Key device parameters and designed target values.
    Large difference in resonator $\kappa$ is either an effect of the wire-bonds or an effect of sample packaging.
    For both qubits, $\chi$ is given at the upper sweet spot, where they are operated throughout the experiment.
    }
\label{tab:Device_Params}
\end{table}

\section{Qubit Tuneup and performance}
\label{sec:tune-up}

\begin{figure}
  \centering
  \includegraphics[width=\columnwidth]{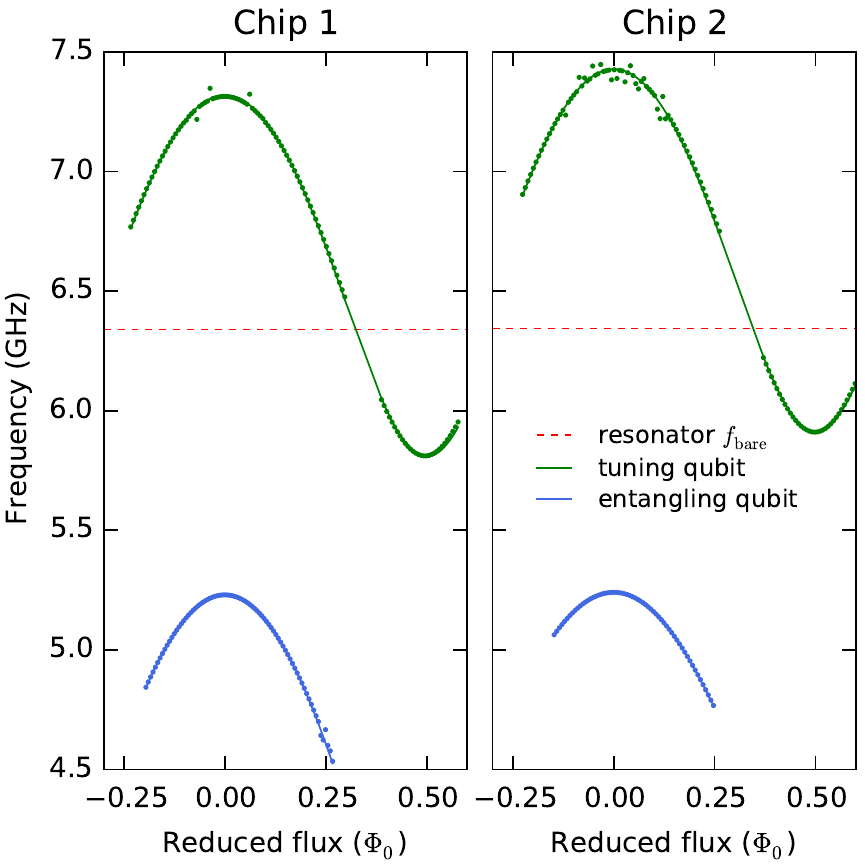}
  \caption{
    Flux-dependent frequency for the qubits in the experiment.
  }
  \label{fig:SOM_dac_arches}
\end{figure}

A plot of the qubit frequencies as a function of flux through their SQUID loops is shown in \cref{fig:SOM_dac_arches}.
For this dataset, two-tone spectroscopy was performed after decoupling the flux bias lines from an initial $2\%$ on-chip crosstalk to $< 0.2\%$ using a compensation matrix.
All qubits in the experiment have SQUID loops with asymmetric Josephson junctions, leading to a top and bottom sweet-spot and reducing the sensitivity to flux noise.
For the entanglement qubits, the bottom sweet-spot is estimated to be at $\sim 4~\GHz$.

Single-qubit rotations on the entanglement qubits were implemented using DRAG pulses~\cite{Motzoi09,Chow10b} using the first AWG.
A sideband modulation of $-100~\MHz$ was used to put the carrier leakage above the qubit frequency.
Gaussian pulses comprise $4\sigma$ with a total duration of $20~\ns$.
The AllXY sequence~\cite{thesisReed13} was used to tune up the DRAG parameter.
$\Tone$, $\Ttwostar$ and $\Techo$ measurements, as well as AllXY sequences and readout fidelity measurements were performed interleaved with the experimental runs in order to monitor performance.
The cross-driving isolation from chip 1 to chip 2 was estimated to be larger than 30~dB by trying to measure a Rabi oscillation on the chip 2 qubit through the drive-line of the chip 1 qubit.
During this procedure, the chip 1 qubit frequency was detuned.
The isolation from chip 2 to chip 1 should be $\sim$40 dB larger due to the directionality of the circulators, but this was not confirmed by measurement.

\begin{figure}
  \centering
  \includegraphics[width=\columnwidth]{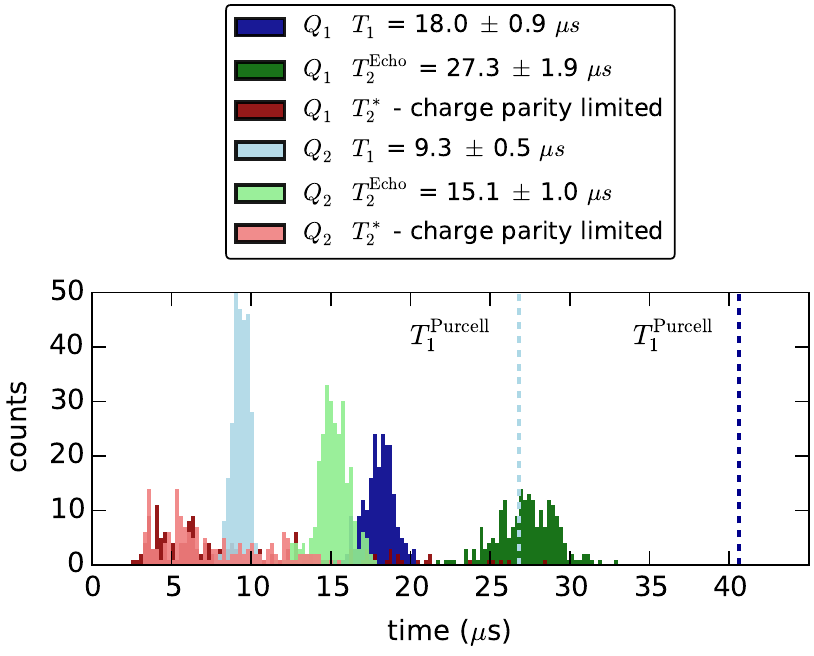}
  \caption{
    Coherence-time histograms for the entanglement qubits at operating point.
    Data was taken intermittently over a $24~\mathrm{h}$ interval with almost 300 data points per quantity.
  }
  \label{fig:SOM_coherence_times}
\end{figure}

The entanglement qubits were operated at their flux sweet-spots which maximized coherence and dispersive shift.
Coherence times can be found in \cref{fig:SOM_coherence_times}.
$\Tone$ was a factor $\sim 2$ below the Purcell limit $\Tone^{\mathrm{Purcell}}=\nicefrac{\Delta^2}{g^2 \kappa}$ for both qubits, with dielectric loss likely to be the other limiting factor.
The charging energy of the transmon $\nicefrac{\EC}{h} = 293~\MHz$ was higher than the design value.
The resulting maximum charge-parity splitting was measured to be $\approx 66~\kHz$ from the beating pattern measured in Ramsey experiments.
However, this frequency uncertainty does not become a limiting factor on the timescale of the experiment.

One microwave source was split four ways to generate the carriers for the bounce-bounce and compensation input, the JPA pump tone and the local oscillator for demodulation.
Readout pulses were defined using the second AWG.
The qubit readout using the JPA was optimized for separation between all four computational states.
A sequence preparing all four states with subsequent readout was used and single shots were collected.
We then optimized JPA pump power, flux-bias setting and pump phase, minimizing the overlap between the resulting probability distributions~\cite{Bhattacharyya46}.
Single-shot fidelities for the final readout for each individual qubit were generally on the order of 95-99\%.
A quantum efficiency $\eta_\mathrm{m}=50\%$ gives good agreement between the SME simulation and the experimental data in \cref{fig:fig3}.
This is below the limit expected from photon loss according to component specifications but consistent with reported values in other experiments~\cite{Roch14,deLange14}.
In principle, $\eta_\mathrm{m}=100\%$ can be achieved, but finite JPA gain and bandwidth as well as photon loss on the way to the JPA (for this setup $~25\%$ is expected from component specifications) limit the quantum efficiency.
Using a phase-insensitive, higher-bandwidth amplifier such as the traveling wave parametric amplifier~\cite{Macklin15} would result in imposing a 50\% upper-limit on the achievable quantum efficiency by that definition making the protocol more sensitive to loss between the chips - the residual source of measurement-induced dephasing.

In future experiments, it would be helpful to fully calibrate the mixer non-linearity and skewness across the experimental range or use step attenuators in order to realize more linear sweeps of the readout power.
Mixer imperfections impact the experiment in several ways but can in principle be completely corrected.
We did cancel the carrier leakage of the mixers in the experiments using fixed DC voltages, as it would lead to additional photon shot noise.
In addition, there is non-linearity in the output power, which manifests in our mixers as reduced output at high voltages.
We only corrected this effect in post-processing when we realized the severity but accordingly the compensation pulses, where two mixers with different amplitudes were involved, got affected.
Another effect that would start playing a role in the compensation cases is mixer skewness, which we did not account for.
In future experiments all these things can be measured and fixed by adjusting the AWG pulses.

\section{Comprehensive modeling of the experiment}
\label{sec:modeling}

We now describe the modeling for this experiment both for the output fields and the density matrix evolution.
It is natural to begin with the classical equations of motion in~\cref{sec:classical_equations_of_motion}, since the full two qubit two cavity ME can be reduced to a qubit only ME using the resonator field solutions and a polaron transformation~\cite{Motzoi15}.
The classical equations of motion are also used to derive the compensation pulse in \cref{sec_modeling_compensation_pulse}.
We then describe the ME in~\cref{sec:master_equation} and finally add a stochastic term to model post-selection of the measurement results in~\cref{sec:stochastic_master_equation}.

\subsection{Classical equations of motion}\label{sec:classical_equations_of_motion}
In the dispersive regime and in the absence of qubit relaxation, the resonator field modeling reduces to qubit state dependent harmonic oscillators.
We generally work in a rotating frame of the coherent measurement drive.
Making use of the cascaded nature of our system we can derive the Heisenberg equation of motion for the system using input output theory~~\cite{walls2007quantum}. Taking the expectation value immediately we end up with the following set of classical qubit-state-dependent coupled linear differential equations
\begin{equation}\label{eq:cascaded_classical_system_of_equations}
\begin{aligned}
    \dot\alpha^\pm\left(t\right) &= \left(-i\left(\Delta_{\mathrm{1}}\pm\chi_{\mathrm{1}}\right)-\frac{1}{2}\overline{\kappa}_{\mathrm{1}}\right) \alpha^\pm\left(t\right)+\sqrt{\kappa^{\mathrm{s}}_{\mathrm{1}}}\epsilon^{\mathrm{s}}\left(t\right)\\
    z^\pm\left(t\right) &= \sqrt{\kappa_{\mathrm{1}}^{\mathrm{s}}}\alpha^\pm\left(t\right)-\epsilon^{\mathrm{s}}\left(t\right)\\
    \dot\beta^{\pm\pm}\left(t\right) &= \left(-i\left(\Delta_{\mathrm{2}}\pm\chi_{\mathrm{2}}\right)-\frac{1}{2}\overline{\kappa}_{\mathrm{2}}\right) \beta^{\pm\pm}\left(t\right)\\
    &+\sqrt{\kappa^{\mathrm{s}}_{\mathrm{2}}\eta_\mathrm{l}}e^{i\phi}z^\pm\left(t\right) + \sqrt{\kappa^{\mathrm{w}}_{\mathrm{2}}}  \epsilon^{\mathrm{w}}\left(t\right) \\
    y^{\pm\pm}\left(t\right) &= -\sqrt{\kappa^{\mathrm{s}}_{\mathrm{2}}\eta_\mathrm{l}}e^{i\phi}z^\pm\left(t\right) + \sqrt{\kappa^{\mathrm{s}}_{\mathrm{2}}}\beta^{\pm\pm}\left(t\right),
\end{aligned}
\end{equation}
where the qubit 0(1) state is denoted by $+(-)$, $\alpha^\pm, \beta^{\pm\pm}$ denote the two qubit-state-dependent coherent states inside resonator 1 and 2 respectively, $z\left(t\right)$ denotes the reflected output field of resonator 1, $y\left(t\right)$ denotes the monitored output field after reflection off both resonators.
Driving fields $\epsilon^{\mathrm{s}}\left(t\right)$, $\epsilon^{\mathrm{w}}\left(t\right)$ are the reflection input at the strongly-coupled resonator ports and the transmission input at the weakly-coupled port of the second resonator, respectively.
The system parameters are the resonator linewidths $\overline{\kappa}_i = \kappa^{\mathrm{s}}_i + \kappa^{\mathrm{w}}_i + \kappa^{\mathrm{I}}_i$ with contributions from the two ports and the intrinsic loss, the dispersive shifts $\chi_i$, the resonator detunings from the measurement tone $\Delta_i$.
Between the chips, the field undergoes a power loss $\eta_\mathrm{l}$ and a phase shift $\phi$.

The above set of equations describes a linear time invariant system, so it can be readily solved in the Fourier domain.
The solutions are written using transfer functions for the single-qubit-resonator systems as
\begin{equation}
\label{eq:theory_cascaded_system_of_equations_solution_fourier_domain}
\begin{aligned}
\alpha^\pm\left(\omega\right) &= H_{\mathrm{1}}^\pm\left(\omega\right)\epsilon^{\mathrm{s}}\left(\omega\right)\\
z^\pm\left(\omega\right) &= H_{1^{\mathrm{R}}}^\pm\left(\omega\right)\epsilon^{\mathrm{s}}\left(\omega\right) \\
\beta^{\pm\pm}\left(\omega\right) &= \sqrt{\eta_\mathrm{l}}e^{i\phi}H_{\mathrm{2}}^{\pm}\left(\omega\right)H_{1^{\mathrm{R}}}^\pm\left(\omega\right)\epsilon^{\mathrm{s}}\left(\omega\right)  + \sqrt{\frac{\kappa^{\mathrm{w}}_{\mathrm{2}}}{\kappa_{\mathrm{2}}^{\mathrm{s}}}} H_{\mathrm{2}}^\pm \epsilon^{\mathrm{w}}\left(\omega\right)\\
y^{\pm\pm}\left(\omega\right) &= \sqrt{\eta_\mathrm{l}}e^{i\phi}H_{1^{\mathrm{R}}}^{\pm}\left(\omega\right)H_{2^R}^{\pm}\epsilon^{\mathrm{s}}\left(\omega\right) + \sqrt{\kappa^{\mathrm{w}}_{\mathrm{2}}}H_{\mathrm{2}}^{\pm}\left(\omega\right)\epsilon^{\mathrm{w}}\left(\omega\right),
\end{aligned}
\end{equation}
where $H_j^\pm \left(\omega\right) = \frac{ \sqrt{\kappa_j^{\mathrm{s}}}} {i\omega + i\left(\Delta_j \pm \chi_j\right) + \frac{1}{2}\overline{\kappa_j}}$, $j\in\{1,2\}$ are the transfer functions into resonator 1 and 2 and~$H_{j^R}^\pm \left(\omega\right) = \sqrt{\kappa_j^{\mathrm{s}}}H_j^\pm \left(\omega\right) - 1$ is the transfer function after reflection from them.
This approach shows clearly that cascading systems entails a simple multiplication of their transfer functions.

\begin{figure}
  \centering
  \includegraphics[width=\columnwidth]{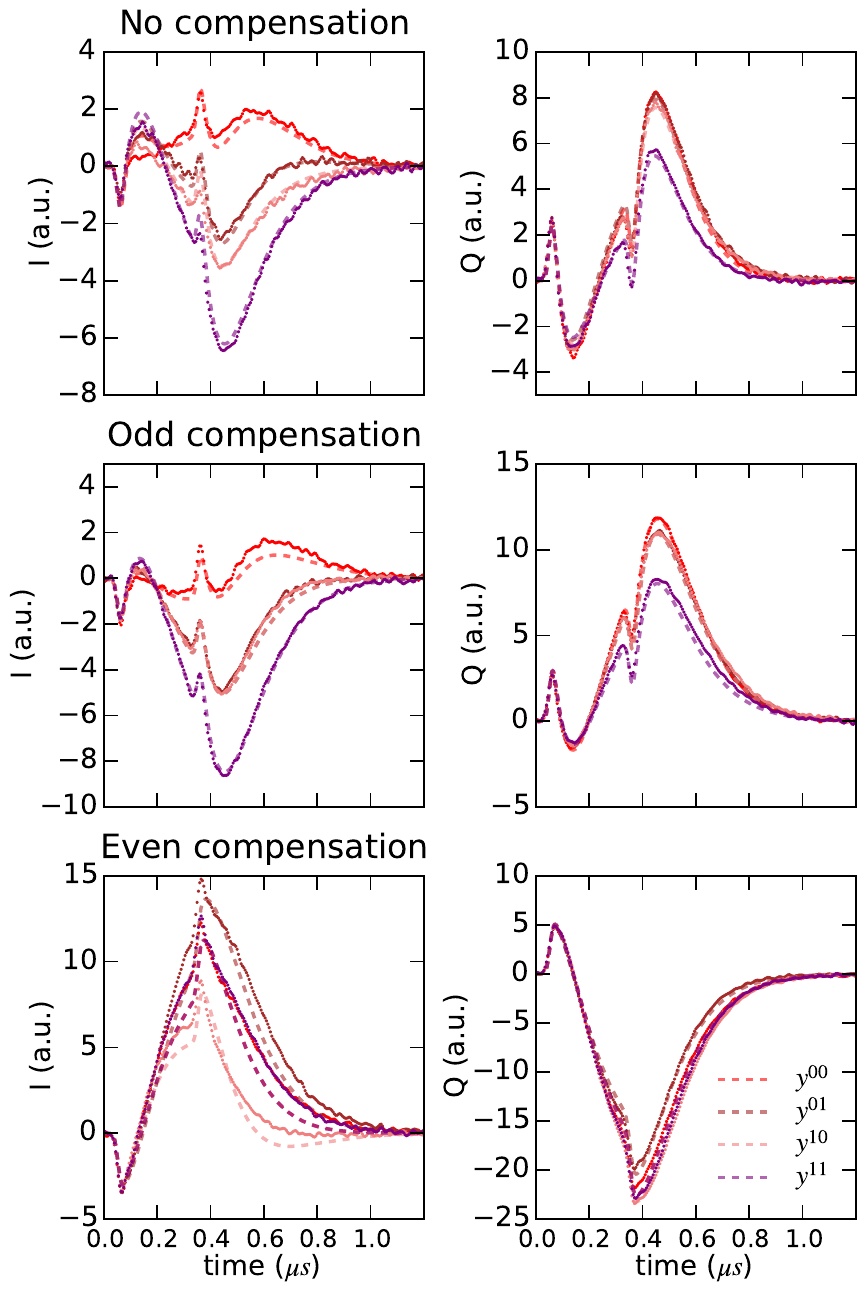}
  \caption{
    Output transients and model fits for the uncompensated, the odd-compensation and the even compensation case, each at maximum-concurrence amplitude.
    The signal was rotated to maximize the difference of the matched states in the I-quadrature for visual clarity.
  }
  \label{fig:SOM_transients}
\end{figure}

\begin{figure}
  \centering
  \includegraphics[width=\columnwidth]{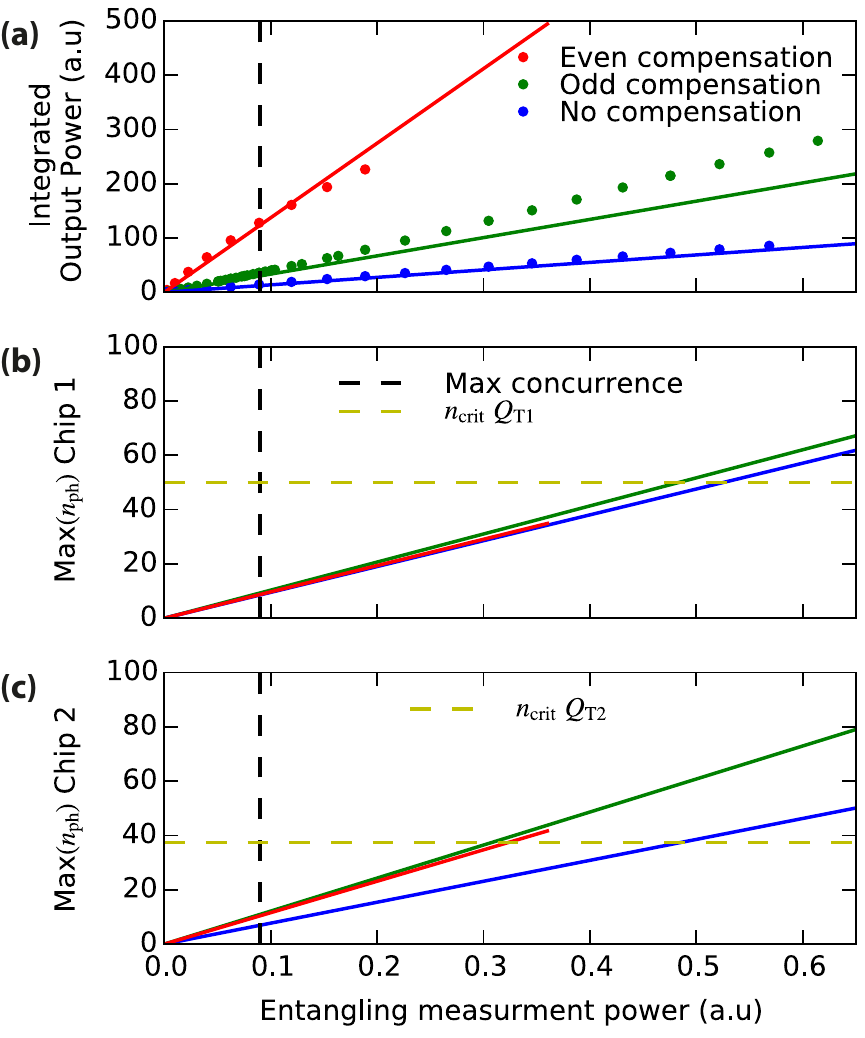}
  \caption{
    \textbf{(a)} Integrated output power in experiment (dots) and theory (lines) for the three cases to confirm the modeling of the output fields.
    We again find the best agreement for the no-compensation case.
    In the relevant regime up to the concurrence maximum we find good agreement.
    \textbf{(b)} and \textbf{(c)} maximum photon numbers in each resonators extracted from the model for all three cases.
    Critical photon number imposed by the tuning qubit $Q_{\mathrm{T}i}$ on chip $i$ is marked by the horizontal line for each resonator (see \cref{sec:tuning_qubits}).
  }
  \label{fig:SOM_integrated_output_nphoton}
\end{figure}

We use these equations to fit the measured output transients at the optimum entangling measurement amplitude.
The results for all three cases can be found in \cref{fig:SOM_transients}.
To further compare experiment and model, we can look at the integrated output power which is to a good approximation qubit state independent.
This is shown in \cref{fig:SOM_integrated_output_nphoton}\textbf{(a)}.
For low powers, particularly up to the point of maximum $\mathcal{C}$, we find good agreement with theory.
The no-compensation case shows the expected linear behavior.
For the odd-compensation case we find deviations at high powers while the even compensation case shows a general systematic deviation from linearity, likely due to mixer imperfections.
Using the fitted amplitude scaling factor from the ME (\cref{sec:master_equation}),  we can use the qubits as photon-meters and estimate the photon numbers in the resonators as a function of input power for both resonators, found in \cref{fig:SOM_integrated_output_nphoton}\textbf{(b)} and \textbf{(c)}.

For this work, we did not attempt active ramp-up and ring-down pulses for the resonators as done in \cite{McClure16,Bultink16}, but the transfer function mechanism provides a simple way to do this.
The transfer functions relate the drive to the resonator fields.
Any ansatz for the driving field at the strong port with enough free parameters can be used derive a pulse where the resonator photon numbers are ramped up and reset to zero faster than the resonators ring-up and ring-down time.
The number of parameters necessary is given by the number of different qubit states for which the ramp-up and ramp-down is supposed to work.

\subsection{Compensating pulse solution}\label{sec_modeling_compensation_pulse}

Using the compensation pulse to limit measurement-induced dephasing was already suggested in~\cite{Motzoi15}.
Here, we expand on the conceptual solution we presented to the compensation pulse in \cref{eq:conceptual_output_solution} in the main text.
By sending a drive through the weak input port simultaneously with the existing pulse in the strong port, we can make nearly any pair of the four classical output states equal by solving $y^{kl}(\omega) = y^{mn}(\omega)$, where y is the qubit state dependent output field of~\cref{eq:theory_cascaded_system_of_equations_solution_fourier_domain}, $k,m$ denote the state of the first qubit and $l,n$ of the second qubit.
Solving this, we can obtain a general expression for the weak compensation measurement field $\epsilon^{\mathrm{w}}_{kl,mn}(\omega)$ as a function of the strong port measurement field $\epsilon^{\mathrm{s}}(\omega)$ and the desired pair of matching output states $\ket{kl}$ and $\ket{mn}$
\begin{equation}
\begin{aligned}
\label{eq:theory_compensation_pulse_solution}
\epsilon^{\mathrm{w}}\left(\omega\right) &= H^\mathrm{comp}(\omega) \epsilon^{\mathrm{s}}\left(\omega\right) \\
H^\mathrm{comp}(\omega) &= \frac{\sqrt{\eta_\mathrm{l}}e^{i\phi}\left(H^k_{a^R}\left(\omega\right)H^l_{b^R}\left(\omega\right) - H^m_{a^R}\left(\omega\right)H^n_{b^R}\left(\omega\right)\right)}{\sqrt{\kappa_{\mathrm{2}}^w}\left(H_{\mathrm{2}}^n\left(\omega\right) - H_{\mathrm{2}}^l\left(\omega\right)\right)}.
\end{aligned}
\end{equation}
Using this simple relation allows us to reduce dephasing of the two-qubit density matrix element $\rho_{kl,mn}$ and therefore create an odd ($y^{01}(\omega)=y^{10}(\omega)$) or even ($y^{00}(\omega)=y^{11}(\omega)$) parity state robust to fabrication variations in the chips.
The full parity measurement is a special case in this context, where two pairs of states are always matched, leading to an entangled state independent of the measurement result \cite{Riste13c}.
This is only possible with another symmetry in the system, namely $2 \chi = \kappa$ for both chips (leading to a steady-state phase shift of $180^{\circ}$), a condition we could not reach with these devices.
The number of free parameters available with two driving fields does not allow us to satisfy $y^{00}(\omega) = y^{11}(\omega)$ and $y^{01}(\omega) = y^{10}(\omega)$ simultaneously.

\subsection{Master equation model}\label{sec:master_equation}

\begin{figure*}
  \centering
  \includegraphics[width=\textwidth]{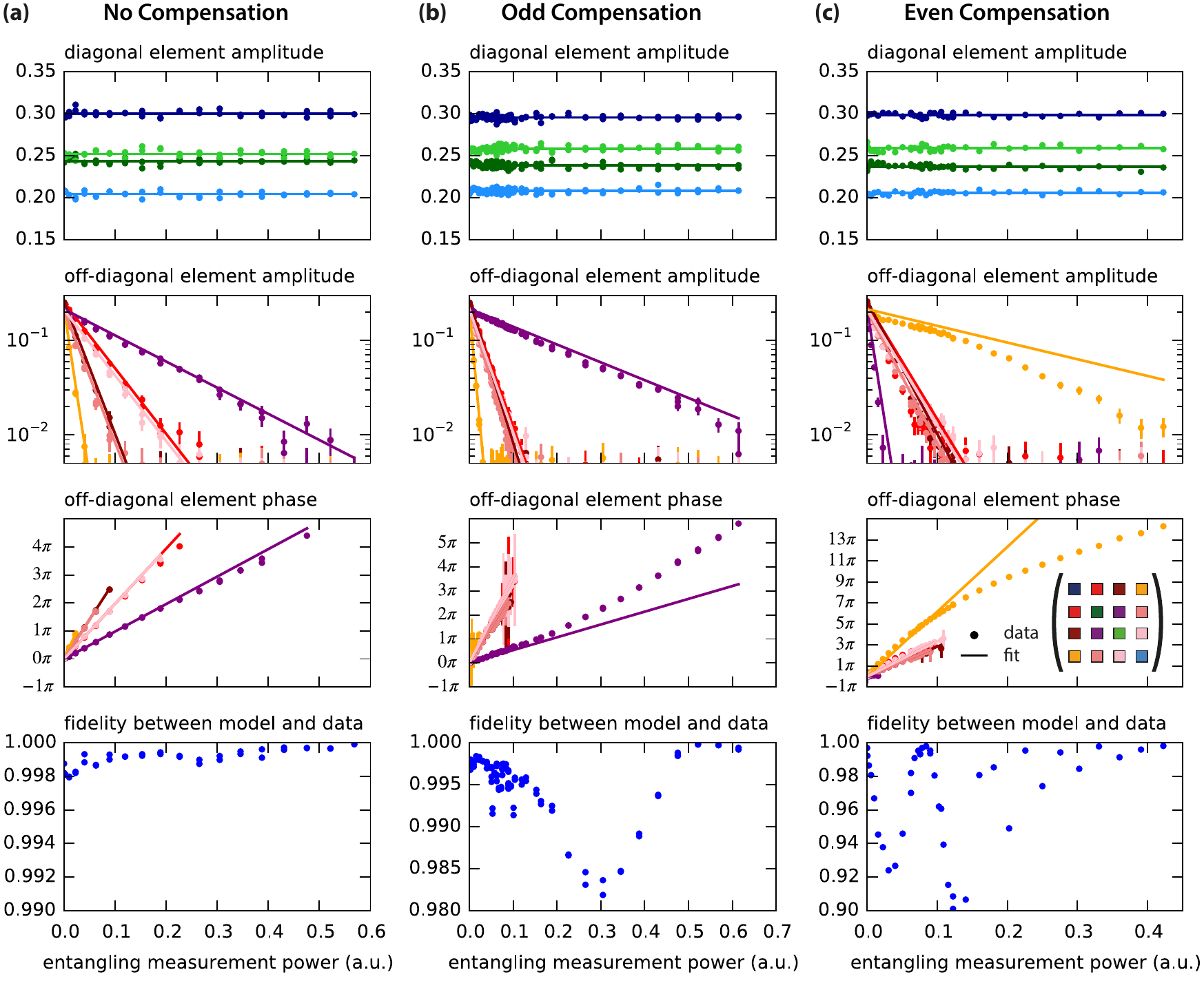}
  \caption{
    Master equation fits for the three cases: no-compensation \textbf{(a)}, the odd compensation \textbf{(b)} and the even compensation \textbf{(c)}.
    Populations are used to fit the $\Tone$ Lindblad operators and do not show a dependence on the measurement power.
    For the off-diagonal density matrix, amplitude and phase are plotted independently.
    The phase is only plotted if the amplitude of the off-diagonal density matrix elements is above 0.01.
    This no compensation case shows the best agreement between theory and experiment, while the odd and even compensation pulse cases deviate at larger amplitudes.
  }
  \label{fig:SOM_full_density_matrix_fits}
\end{figure*}

As derived in~\cite{Motzoi15}, in the dispersive regime we can fully model the average evolution of the qubit states using a qubit-only ME:
\begin{equation}\label{eq:SOM_master_equation}
\begin{aligned}
\dot\rho &= \sum_{ijkl}a_{ijkl}\left(t\right)P_{ij}\rho\left(t\right)P_{kl}+ \mathcal{L}_d\rho\left(t\right)\\
a_{ijkl}\left(t\right) &= 2i\chi_{\mathrm{1}}\left(1-\delta_{ik}\right)\left(\left(-1\right)^i\alpha^k\alpha^{*i}\right) \\
& + 2i\chi_{\mathrm{2}}\left(1-\delta_{jl}\right)\left(\left(-1\right)^i\beta^{kl}\beta^{*ij}\right),
\end{aligned}
\end{equation}
where $P_{ij} = \ket{ij}\bra{ij}$ are the two qubit projection operators, $\delta_{ij}$ the Kronecker delta function, $\mathcal{L}_d$ is given by standard Lindblad type $\mathcal{D}\left[A\right]\rho = A\rho A^\dagger - \frac{1}{2}\left(A^\dagger A \rho + \rho A^\dagger A\right)$  phenomenological qubit dissipation (with rate $\gamma^i$) and dephasing (with rate $\gamma^i_\phi$) operators $\mathcal{L}_d = \sum_{i=1}^{2} \gamma^i_\phi \mathcal{D}\left[\sigma_z^i\right] + \gamma^i\mathcal{D}\left[\sigma_-\right]$.
Adding the qubit relaxation operators makes this equation no longer exact but is still reasonably valid in the limit $\kappa^{\mathrm{s}}_{\mathrm{1}}, \kappa^{\mathrm{s}}_{\mathrm{2}} \gg \chi_{\mathrm{1}}, \chi_{\mathrm{2}}$~~\cite{Motzoi15}.
Since the resonators are traced out in this equation during the measurement process, we can see a trajectory with possible non-Markovian revival of coherence due to entangled photons leaking out of the resonators.

In order to fit the master equation model to the experimental data we take several steps.
As predicted by the model, the qubit populations are constant as a function of measurement power.
They only depend on the qubit $\Tone$ dissipation operators, which are fitted to the diagonal terms, since they fluctuate between datasets.
The density matrix at zero measurement power can be used to estimate $\gamma^i_\phi$.
We used a fit to extract the scaling factor between the AWG voltage at room temperature and the power that arrives at the experiment, as well as the inter-chip loss, fixing the other system parameters.
The results can be found in \cref{fig:SOM_full_density_matrix_fits}.
We find excellent agreement with theory for the no-compensation case, for the other two cases we simply applied the compensation in the model without re-fitting.
Similar to the integrated output power, agreement for the two compensation cases is considerably worse, which we attribute mostly to mixer imperfections that were not properly accounted for.
While we believe this is the main source of mismatch, we also reach the limits of the dispersive approximation due to the tuning qubits (see~\cref{sec:tuning_qubits}).
These effects can be included in a full two-qubit/two-cavity master equation including higher-order terms, but this would be computationally much more involved.

\subsection{Stochastic Master equation simulation}\label{sec:stochastic_master_equation}

\begin{figure}
  \centering
  \includegraphics[width=\columnwidth]{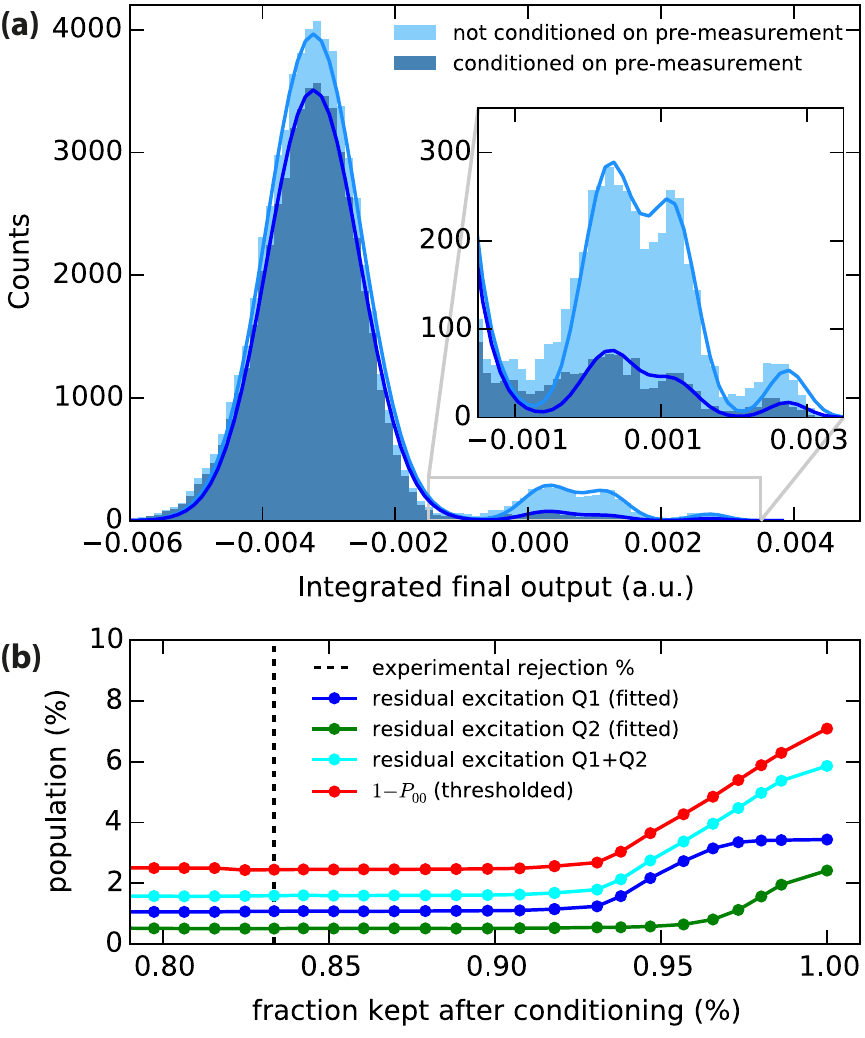}
  \caption{
    \textbf{(a)} Histogram of single-shot measurements of the system in the nominal ground state reveals additional peaks that coincide with the $\ket{01}$, $\ket{10}$ and $\ket{11}$ calibration points.
    \textbf{(b)} A multi-Gaussian fit can be used to estimate the residual populations of the two qubits for different conditioning on the pre-measurement.
    The conditioning on the ground state can bring down the residual populations to about $1\%$ ($0.5\%$) for qubit 1 (2) by rejecting $10\%$ of the experimental runs.
    In practice the rejection rate was closer $15-17\%$ indicated by the vertical line.
    A simple threshold estimate of the excited state population gives a slightly higher estimate.
    The constant offset could be due to tuning qubit excitations.
  }
  \label{fig:SOM_residual_excitations}
\end{figure}

\begin{figure}
  \centering
  \includegraphics[width=\columnwidth]{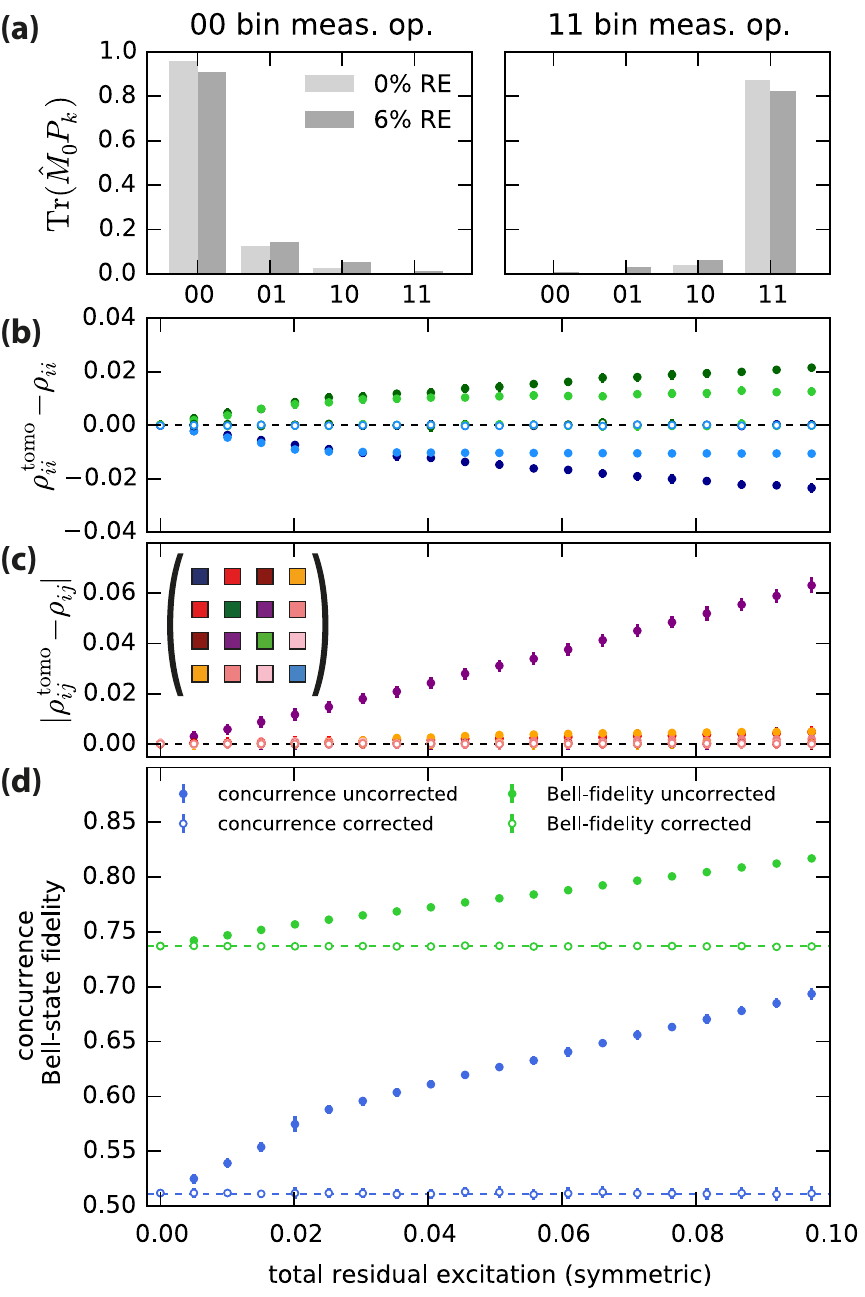}
  \caption{
    Simulation of the effects of residual excitations on quantum state tomography.
    Tomography runs are simulated giving the resulting $\rho^\mathrm{tomo}$ for an underlying density matrix $\rho$ corresponding to the highest concurrence state.
    \textbf{(a)} Change in measurement expectation values for the computational states without residual excitation (RE) and with 6\% total RE symmetrical on both qubits.
    \textbf{(b)} and \textbf{(c)} Change in populations and absolute coherence elements of the reconstructed density matrix as a function of total RE (filled markers).
    Taking into account known RE and fixing the measurement operators leads to the correct reconstruction of the density matrix (open markers).
    \textbf{(d)} Change entanglement measures as a function of total RE (filled markers), reconstructions taking the RE into account (open markers) lead to the correct values (lines).
    Statistical error bars are on the order of the marker size.
  }
  \label{fig:SOM_Tomo_thermal_excitation}
\end{figure}

The JPA in phase-insensitive mode, due to the squeezing, can be modeled as reading out a single quadrature of the output field.
Thus, we can define the angle $\theta$ along which we read out.
We approximate the imperfect readout due to photon loss up to the JPA and finite gain and bandwidth with a single quantity, the quantum efficiency of the measurement $\eta_\mathrm{l}$.
Modeling single runs of a homodyne measurement with angle $\theta$ and quantum efficiency $\eta_{\mathrm{m}}$ requires us to add another superoperator~$\mathcal{L}_\mathrm{m}$ to the right-hand side of \cref{eq:SOM_master_equation} adding the stochastic measurement dynamics to the ME.
This allows us to calculate the density matrix conditioned on the measurement result.
This gives a stochastic differential equation in It\^{o} form~\cite{Gardiner04} and~$\mathcal{L}_\mathrm{m}$ is given by~~\cite{Motzoi15}
\begin{equation}\label{eq:stochastic_measurement_operator}
\begin{aligned}
\mathcal{L}_\mathrm{m}\rho &= \sqrt{\eta_{\mathrm{m}}}\xi\left(t\right)\left[M\rho + \rho M^\dagger - \text{Tr}\left(M\rho + \rho M^\dagger\right)\rho\right],
\end{aligned}
\end{equation}
where $M=e^{i\theta}\left(-\sqrt{\kappa^{\mathrm{s}}_{\mathrm{1}}\eta_\mathrm{l}}\Pi_{\mathrm{1}} +\sqrt{\kappa^{\mathrm{s}}_{\mathrm{2}}}\Pi_{\mathrm{2}}\right)$, $\Pi_{\mathrm{1}}\left(t\right) = \sum_{i,j} P_{ij}\alpha^{ij}\left(t\right)$, $\Pi_{\mathrm{2}}\left(t\right) = \sum_{i,j} P_{ij}\beta^{ij}\left(t\right)$ are resonator-state-dependent qubit projection operators, $\xi\left(t\right)dt = dW$ is a white noise process satisfying $E\left[dW\right] = 0$ and $E\left[dW\left(t\right)dW\left(s\right)\right] = \delta\left(t-s\right)dt$ and $dW$ is a Wiener increment.
The measured output voltage corresponding to such a trajectory is given by~~\cite{Motzoi15}
\begin{equation}
V\left(t\right) = \sqrt{\eta_{\mathrm{m}}}\text{Re}\left (\langle {M} \rangle \right ) + \xi\left(t\right).
\end{equation}
This was used to simulate the measurement including post-selection and to generate the theory curves for \cref{fig:fig3}.

Although simulating the SME allows comparing individual trajectories at each point in time, we only looked at the qubit-only density matrix at a time where the resonators were back to the vacuum state.
Only at this point in time are the qubits a useful resource for remote information processing schemes.
Studying the trajectories themselves, on the other hand, is performed in more detail in~\cite{Roch14,Chantasri16}, which will be relevant for real-time feedback schemes.

\section{Quantum state tomography and SPAM errors}
\label{sec:tomography}

In this experiment, we diagnose the entanglement, the key figure of merit, via QST.
QST allows us to reconstruct the density matrix from which the entanglement measures are computed.
Our QST procedure consists of two steps.
First we do a set of calibration measurements with known input states to determine the observable $\hat{M}_0$.
For a joint dispersive readout, the measurement operator for a $d$-dimensional Hilbert space is of the simple form $\hat{M}_0 = \sum_{k=0}^{d-1} a_k P_k$ with $P_k = \ket{k}\bra{k}$~\cite{DiCarlo09}, so the coefficients $a_k$ can be directly read out from computational basis state inputs, e.g $a_k = \Tr{\hat{M}_0 P_k}$.
The second step is the reconstruction of an unknown $\rho$ using the now known measurement operator $\hat{M}_0$.
We can reduce this to a simple linear algebra problem where we need to estimate the $d^2-1$ independent basis coefficients of $\rho$ by measuring the expectation values  $\langle \hat{M}_i \rangle = \Tr{\hat{M}_i \rho} $ of at least $d^2-1$ orthogonal measurement operators $\hat{M}_i$ and solving the resulting system of equations.
The measurement operators $M_i$ can be effectively obtained by rotating $\rho$ before measurement using $\Tr{\hat{R_i}\hat{M}\hat{R_i^\dagger\rho}}= \Tr{\hat{M}\hat{R_i}^\dagger\rho\hat{R_i}}$.
In this experiment, we used the cardinal set (an overcomplete set of 36 single qubit rotations: $\left \{ I,\, X,\, X_{\nicefrac{\pi}{2}},\, X_{\nicefrac{-\pi}{2}},\, Y_{\nicefrac{\pi}{2}},\, Y_{\nicefrac{-\pi}{2}} \right\}^{\times 2}$) on both qubits. The 36 rotations together with 4 calibration points (each repeated 5 times) were measured sequentially and the whole sequence was repeated 12800 times.
We binned the measurement outcomes based on the calibration points, where one bin was mostly comprised of outcomes corresponding to $\ket{00}$ and another to those of $\ket{11}$.
Using proper normalization, the counts in bin $n$ for rotation $i$ corresponded to the expectation value $ \langle \hat{M}^n_i \rangle$ of the bin operator $\hat{M}^n_i$.
This resulted in an overcomplete set of $36 \times 2 = 72$ equations with 15 unknowns and was solved by performing standard maximum likelihood techniques with physicality constraints \cite{James01,Boyd04,Langford13}.

While QST is a widely used way to confirm entanglement, its accuracy is limited by state-preparation and measurement (SPAM) errors.
SPAM errors mainly impact the measurement operator, and thus arise in step 1, the calibration process.
They likely exceed the errors due to imperfect qubit gates.

Assuming the initial state to be perfectly $\ket{00}$ is an approximation.
The histograms of the measurement outcomes projected on one quadrature given in \cref{fig:SOM_residual_excitations}\textbf{(a)} clearly show multiple peaks which coincide with the average outcomes for the other computational states.
We conditioned on an additional initial measurement to reduce the residual excitation.
The results after post-selection are shown in \cref{fig:SOM_residual_excitations}\textbf{(b)} giving an estimated decrease of $6\%$ to less than $2\%$ total excitation in both qubits.
Re-excitation times calculated from $\Tone$ and the measured excitation fraction suggest that the conditioning should be limited to reducing the residual excitation to $\sim 0.5\%$.

The conditioning on the ground state paradoxically decreases the amount of entanglement in the tomography outcome.
As an example, for the run giving the highest entanglement keeping $25\%$ of the data, the extracted density matrix without conditioning resulted in $\mathcal{C}=0.58\pm0.01$ and $\mathcal{F}_\mathrm{B}=0.761\pm0.004$.
After conditioning , the same dataset resulted in $\mathcal{C}=0.57\pm0.01$ and $\mathcal{F}_\mathrm{B}=0.755\pm0.004$.
Reducing the amount of residual excitation should have increased $\mathcal{C}$ if the QST was accurate, because the conditioning should increase the purity of the initial state, which in turn would reduce the mixture in the final state.
This points to SPAM errors related to the residual excitation skewing the QST result.

Monte Carlo simulations of QST reproduce the effect, pointing to the flawed assumption of pure calibration points which are in reality mixed by residual excitation as seen in \cref{fig:SOM_Tomo_thermal_excitation}.
This skews the measurement operators obtained from calibration and artificially boosts the purity of the estimated density matrix.
For the optimum entangled state this results in a significant increase in $\mathcal{C}$.
Simulations also showed that beyond the limit of reducing residual excitation by conditioning, tomography can be further improved by taking the known mixture of the calibration points into account.
We can then correct the calibration of the measurement operators by assuming mixed input states $\tilde{P}_{ij}$ instead of pure projectors $P_{ij}$, which for $\ket{00}$ becomes
\begin{equation}\label{eq:som_te_correction}
\begin{aligned}
\tilde{P}_{00} =& \left(1-p_{e_{01}}\right)\left(1-p_{\mathrm{e}_{10}}\right)P_{00}  + p_{\mathrm{e}_{01}}\left(1-p_{\mathrm{e}_{10}}\right)P_{01} \\ + & p_{\mathrm{e}_{10}}\left(1-p_{\mathrm{e}_{01}}\right)P_{10} + p_{\mathrm{e}_{01}} p_{\mathrm{e}_{10}}P_{11},
 \end{aligned}
\end{equation}
where $p_{e_{01}}$ is the excitation fraction in qubit 2, $p_{e_{10}}$ the excitation fraction of qubit 1, and $P_{ij}$ the projector onto state $\ket{ij}$.
In simulation [\cref{fig:SOM_Tomo_thermal_excitation}] the correction leads to a more precise estimate of the density matrix given an accurate estimate of the residual excitation.
Systematic errors in tomography due to residual excitation likely exceed the statistical counting errors.

Correcting for the estimated residual excitations, the conditioning on the initial measurement now increases the entanglement as expected from $\mathcal{C}=0.46\pm0.01$ and $\mathcal{F}_\mathrm{B}=0.71\pm0.003$ to $\mathcal{C}=0.51\pm0.01$ and $\mathcal{F}_\mathrm{B}=0.734\pm0.005$.
The main text data was corrected for the estimated residual excitation in each experimental run, which remained $\sim1\%$ on both qubits after conditioning.

\section{Role of the tuning qubits}
\label{sec:tuning_qubits}

\begin{figure}
  \centering
  \includegraphics[width=\columnwidth]{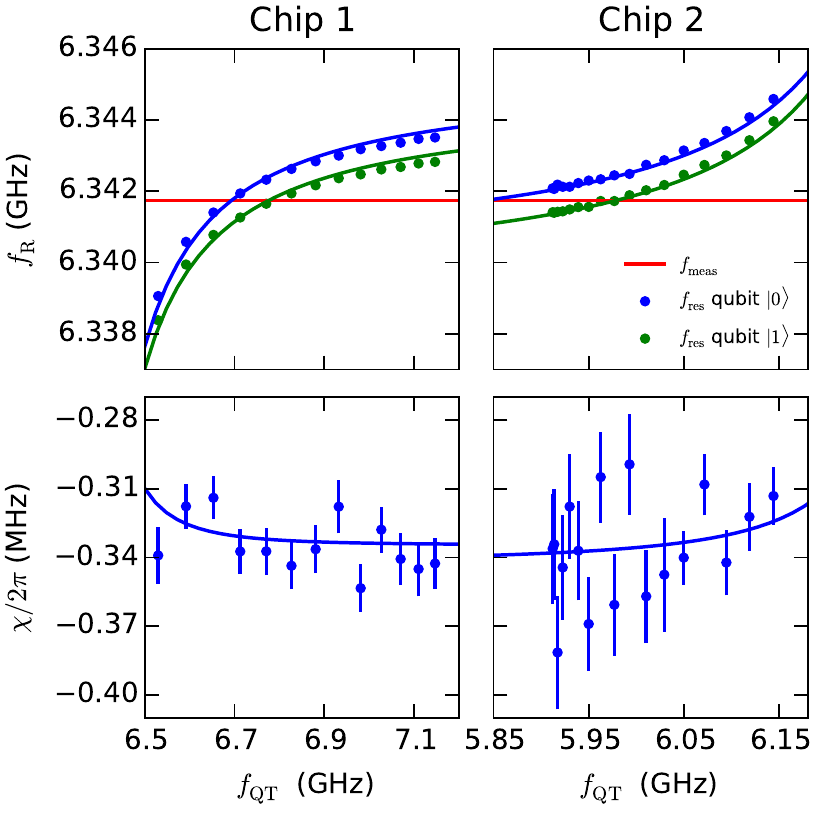}
  \caption{
    Matching resonator frequencies via the dispersive shifts of the tuning qubits.
    Resonator transmission is measured for the respective entanglement qubit in $\vert 0 \rangle$ and $\vert 1 \rangle$ and frequencies $f_\mathrm{R}$ are extracted from Lorentzian fits for both cases.
    They are plotted against the tuning qubit frequencies $f_\mathrm{QT}$.
    Tuning qubit frequency can be varied by changing the DC flux through its bias line.
    The two frequencies can be used to extract $\chi$.
    Tuning qubits have no measurable effect on the $\chi$ of the entangling qubits in this tuning range, as expected from the qutrit Tavis-Cummings model.
  }
  \label{fig:SOM_tuning_qubits}
\end{figure}

While the mismatch in $\kappa_{c}$ for the two chips was significant, this is an effect that can be fully corrected with the compensation pulse without affecting the performance of the protocol.
Differences in $\chi$ between the two chips could also be compensated without sacrificing performance.
However, resonator frequency mismatch has different implications and would have a strong impact on achievable entanglement.
The measurement-induced dephasing due to frequency mismatch could be eliminated using the compensation field, but the quantum efficiency would suffer.
This is due to the measurement with a phase-insensitive, low-bandwidth JPA, which is optimal in the symmetric readout condition.
For either qubit-resonator system which is not symmetrically driven, meaning that the measurement tone is halfway between the resonator frequencies for $\ket{0}$ and $\ket{1}$, the output information is not confined to one quadrature.
Any information in the de-amplified quadrature is lost, therefore realizing the symmetric driving condition simultaneously for both resonators is essential for maximizing the quantum efficiency.

Making identical microwave resonators to MHz precision is technically conceivable but challenging, partially due to the choice of niobium titanium nitride (NbTiN) film as the base superconductor.
NbTiN is a high-kinetic-inductance superconductor due to the low charge carrier density, which leads to a strong dependence on the film thickness.
Therefore, the two bare resonator frequencies are not identical within the linewidth $\kappa_i$.
The additional qubits are used to shift their respective resonators via their Lamb shift, allowing us to match the resonator frequencies at the cost of introducing additional Kerr-nonlinearity.

Tunable low-loss CPW-resonators have also been demonstrated using kinetic inductance~\cite{Vissers10} or via SQUID loops~\cite{Sandberg08,Yamamoto08}.
Each of the tuning methods leads to Kerr-nonlinearity in the resonator.
The effect of the tuning qubit on the ideal qubit-resonator system can be modeled via the Tavis-Cummings Hamiltonian~\cite{Tavis68}.
Tuning qubits were designed with a top and bottom sweet-spot sitting above and below the resonator.
As seen in \cref{fig:SOM_tuning_qubits}, the resonator frequency as a function of tuning-qubit frequency is well described by the model.
Our measurements were not accurate enough to resolve the small change in the $\chi$ of the entanglement-qubit in the range we measured.
At the operating points, $\Tone$ was found to be 3.9~$\mu$s and 4.1~$\mu$s for chip 1 and 2 respectively.
The critical photon numbers $n_{\mathrm{crit}} = \nicefrac{\Delta^2}{4g^2}$ calculated from the entanglement-qubit frequency and coupling are $193$ and $188$ for chip 1 and chip 2 respectively, a factor $\sim 10$ above the maximum photon numbers we reach in the protocol at the optimum entanglement amplitude.
The non-linearity from the tuning qubits can be inferred, using the qubit and bare-resonator frequencies and the coupling constants obtained from the fit above.
While photon numbers never reach $n_{\mathrm{crit}}$ of the entanglement qubits, they do for the tuning qubits.
We calculate $n_{\mathrm{crit}} = 50$ and $37$ for chips 1 and 2, respectively.
An additional unwanted effect is that residual excitations of the tuning-qubits would lead to additional noise in the readout signal, but it should not be correlated with the state of the entanglement-qubits, such that it would not skew the tomography result.


\begin{thebibliography}{72}%
\makeatletter
\providecommand \@ifxundefined [1]{%
 \@ifx{#1\undefined}
}%
\providecommand \@ifnum [1]{%
 \ifnum #1\expandafter \@firstoftwo
 \else \expandafter \@secondoftwo
 \fi
}%
\providecommand \@ifx [1]{%
 \ifx #1\expandafter \@firstoftwo
 \else \expandafter \@secondoftwo
 \fi
}%
\providecommand \natexlab [1]{#1}%
\providecommand \enquote  [1]{``#1''}%
\providecommand \bibnamefont  [1]{#1}%
\providecommand \bibfnamefont [1]{#1}%
\providecommand \citenamefont [1]{#1}%
\providecommand \href@noop [0]{\@secondoftwo}%
\providecommand \href [0]{\begingroup \@sanitize@url \@href}%
\providecommand \@href[1]{\@@startlink{#1}\@@href}%
\providecommand \@@href[1]{\endgroup#1\@@endlink}%
\providecommand \@sanitize@url [0]{\catcode `\\12\catcode `\$12\catcode
  `\&12\catcode `\#12\catcode `\^12\catcode `\_12\catcode `\%12\relax}%
\providecommand \@@startlink[1]{}%
\providecommand \@@endlink[0]{}%
\providecommand \url  [0]{\begingroup\@sanitize@url \@url }%
\providecommand \@url [1]{\endgroup\@href {#1}{\urlprefix }}%
\providecommand \urlprefix  [0]{URL }%
\providecommand \Eprint [0]{\href }%
\providecommand \doibase [0]{http://dx.doi.org/}%
\providecommand \selectlanguage [0]{\@gobble}%
\providecommand \bibinfo  [0]{\@secondoftwo}%
\providecommand \bibfield  [0]{\@secondoftwo}%
\providecommand \translation [1]{[#1]}%
\providecommand \BibitemOpen [0]{}%
\providecommand \bibitemStop [0]{}%
\providecommand \bibitemNoStop [0]{.\EOS\space}%
\providecommand \EOS [0]{\spacefactor3000\relax}%
\providecommand \BibitemShut  [1]{\csname bibitem#1\endcsname}%
\let\auto@bib@innerbib\@empty
\bibitem [{\citenamefont {DiVincenzo}(2009)}]{Divincenzo09}%
  \BibitemOpen
  \bibfield  {author} {\bibinfo {author} {\bibfnamefont {D.~P.}\ \bibnamefont
  {DiVincenzo}},\ }\href
  {https://arxiv.org/ct?url=http%3A%2F%2Fdx.doi.org%2F10%252E1088%2F0031-8949%2F2009%2FT137%2F014020&v=5eb633b4}
  {\bibfield  {journal} {\bibinfo  {journal} {Physica Scripta}\ }\textbf
  {\bibinfo {volume} {2009}},\ \bibinfo {pages} {014020} (\bibinfo {year}
  {2009})}\BibitemShut {NoStop}%
\bibitem [{\citenamefont {Helmer}\ \emph {et~al.}(2009)\citenamefont {Helmer},
  \citenamefont {Mariantoni}, \citenamefont {Fowler}, \citenamefont
  {Von~Delft}, \citenamefont {Solano},\ and\ \citenamefont
  {Marquardt}}]{Helmer09cavity}%
  \BibitemOpen
  \bibfield  {author} {\bibinfo {author} {\bibfnamefont {F.}~\bibnamefont
  {Helmer}}, \bibinfo {author} {\bibfnamefont {M.}~\bibnamefont {Mariantoni}},
  \bibinfo {author} {\bibfnamefont {A.}~\bibnamefont {Fowler}}, \bibinfo
  {author} {\bibfnamefont {J.}~\bibnamefont {Von~Delft}}, \bibinfo {author}
  {\bibfnamefont {E.}~\bibnamefont {Solano}}, \ and\ \bibinfo {author}
  {\bibfnamefont {F.}~\bibnamefont {Marquardt}},\ }\href
  {http://iopscience.iop.org/article/10.1209/0295-5075/85/50007/meta}
  {\bibfield  {journal} {\bibinfo  {journal} {Europhys. Lett.}\ }\textbf
  {\bibinfo {volume} {85}},\ \bibinfo {pages} {50007} (\bibinfo {year}
  {2009})}\BibitemShut {NoStop}%
\bibitem [{\citenamefont {Versluis}\ \emph {et~al.}(2017)\citenamefont
  {Versluis}, \citenamefont {Poletto}, \citenamefont {Khammassi}, \citenamefont
  {Tarasinski}, \citenamefont {Haider}, \citenamefont {Michalak}, \citenamefont
  {Bruno}, \citenamefont {Bertels},\ and\ \citenamefont
  {DiCarlo}}]{Versluis17}%
  \BibitemOpen
  \bibfield  {author} {\bibinfo {author} {\bibfnamefont {R.}~\bibnamefont
  {Versluis}}, \bibinfo {author} {\bibfnamefont {S.}~\bibnamefont {Poletto}},
  \bibinfo {author} {\bibfnamefont {N.}~\bibnamefont {Khammassi}}, \bibinfo
  {author} {\bibfnamefont {B.}~\bibnamefont {Tarasinski}}, \bibinfo {author}
  {\bibfnamefont {N.}~\bibnamefont {Haider}}, \bibinfo {author} {\bibfnamefont
  {D.~J.}\ \bibnamefont {Michalak}}, \bibinfo {author} {\bibfnamefont
  {A.}~\bibnamefont {Bruno}}, \bibinfo {author} {\bibfnamefont
  {K.}~\bibnamefont {Bertels}}, \ and\ \bibinfo {author} {\bibfnamefont
  {L.}~\bibnamefont {DiCarlo}},\ }\href {\doibase
  10.1103/PhysRevApplied.8.034021} {\bibfield  {journal} {\bibinfo  {journal}
  {Phys. Rev. Appl.}\ }\textbf {\bibinfo {volume} {8}},\ \bibinfo {pages}
  {034021} (\bibinfo {year} {2017})}\BibitemShut {NoStop}%
\bibitem [{\citenamefont {Hill}\ \emph {et~al.}(2015)\citenamefont {Hill},
  \citenamefont {Peretz}, \citenamefont {Hile}, \citenamefont {House},
  \citenamefont {Fuechsle}, \citenamefont {Rogge}, \citenamefont {Simmons},\
  and\ \citenamefont {Hollenberg}}]{Hill15}%
  \BibitemOpen
  \bibfield  {author} {\bibinfo {author} {\bibfnamefont {C.~D.}\ \bibnamefont
  {Hill}}, \bibinfo {author} {\bibfnamefont {E.}~\bibnamefont {Peretz}},
  \bibinfo {author} {\bibfnamefont {S.~J.}\ \bibnamefont {Hile}}, \bibinfo
  {author} {\bibfnamefont {M.~G.}\ \bibnamefont {House}}, \bibinfo {author}
  {\bibfnamefont {M.}~\bibnamefont {Fuechsle}}, \bibinfo {author}
  {\bibfnamefont {S.}~\bibnamefont {Rogge}}, \bibinfo {author} {\bibfnamefont
  {M.~Y.}\ \bibnamefont {Simmons}}, \ and\ \bibinfo {author} {\bibfnamefont
  {L.~C.~L.}\ \bibnamefont {Hollenberg}},\ }\href {\doibase
  10.1126/sciadv.1500707} {\bibfield  {journal} {\bibinfo  {journal} {Science
  Advances}\ }\textbf {\bibinfo {volume} {1}} (\bibinfo {year} {2015}),\
  10.1126/sciadv.1500707}\BibitemShut {NoStop}%
\bibitem [{\citenamefont {Li}\ \emph {et~al.}(2017)\citenamefont {Li},
  \citenamefont {Petit}, \citenamefont {Franke}, \citenamefont {Dehollain},
  \citenamefont {Helsen}, \citenamefont {Steudtner}, \citenamefont {Thomas},
  \citenamefont {Yoscovits}, \citenamefont {Singh}, \citenamefont {Wehner}
  \emph {et~al.}}]{Li17}%
  \BibitemOpen
  \bibfield  {author} {\bibinfo {author} {\bibfnamefont {R.}~\bibnamefont
  {Li}}, \bibinfo {author} {\bibfnamefont {L.}~\bibnamefont {Petit}}, \bibinfo
  {author} {\bibfnamefont {D.}~\bibnamefont {Franke}}, \bibinfo {author}
  {\bibfnamefont {J.}~\bibnamefont {Dehollain}}, \bibinfo {author}
  {\bibfnamefont {J.}~\bibnamefont {Helsen}}, \bibinfo {author} {\bibfnamefont
  {M.}~\bibnamefont {Steudtner}}, \bibinfo {author} {\bibfnamefont
  {N.}~\bibnamefont {Thomas}}, \bibinfo {author} {\bibfnamefont
  {Z.}~\bibnamefont {Yoscovits}}, \bibinfo {author} {\bibfnamefont
  {K.}~\bibnamefont {Singh}}, \bibinfo {author} {\bibfnamefont
  {S.}~\bibnamefont {Wehner}},  \emph {et~al.},\ }\href
  {https://arxiv.org/abs/1711.03807} {\bibfield  {journal} {\bibinfo  {journal}
  {arXiv:1711.03807}\ } (\bibinfo {year} {2017})}\BibitemShut {NoStop}%
\bibitem [{\citenamefont {Monroe}\ and\ \citenamefont {Kim}(2013)}]{Monroe13}%
  \BibitemOpen
  \bibfield  {author} {\bibinfo {author} {\bibfnamefont {C.}~\bibnamefont
  {Monroe}}\ and\ \bibinfo {author} {\bibfnamefont {J.}~\bibnamefont {Kim}},\
  }\href@noop {} {\bibfield  {journal} {\bibinfo  {journal} {Science}\ }\textbf
  {\bibinfo {volume} {339}},\ \bibinfo {pages} {1164} (\bibinfo {year}
  {2013})}\BibitemShut {NoStop}%
\bibitem [{\citenamefont {Nickerson}\ \emph {et~al.}(2013)\citenamefont
  {Nickerson}, \citenamefont {Li},\ and\ \citenamefont
  {Benjamin}}]{Nickerson13}%
  \BibitemOpen
  \bibfield  {author} {\bibinfo {author} {\bibfnamefont {N.~H.}\ \bibnamefont
  {Nickerson}}, \bibinfo {author} {\bibfnamefont {Y.}~\bibnamefont {Li}}, \
  and\ \bibinfo {author} {\bibfnamefont {S.~C.}\ \bibnamefont {Benjamin}},\
  }\href {http://dx.doi.org/10.1038/ncomms2773} {\bibfield  {journal} {\bibinfo
   {journal} {Nat.\ Commun.}\ }\textbf {\bibinfo {volume} {4}},\ \bibinfo
  {pages} {1756} (\bibinfo {year} {2013})}\BibitemShut {NoStop}%
\bibitem [{\citenamefont {Nemoto}\ \emph {et~al.}(2014)\citenamefont {Nemoto},
  \citenamefont {Trupke}, \citenamefont {Devitt}, \citenamefont {Stephens},
  \citenamefont {Scharfenberger}, \citenamefont {Buczak}, \citenamefont
  {N\"obauer}, \citenamefont {Everitt}, \citenamefont {Schmiedmayer},\ and\
  \citenamefont {Munro}}]{Nemoto14}%
  \BibitemOpen
  \bibfield  {author} {\bibinfo {author} {\bibfnamefont {K.}~\bibnamefont
  {Nemoto}}, \bibinfo {author} {\bibfnamefont {M.}~\bibnamefont {Trupke}},
  \bibinfo {author} {\bibfnamefont {S.~J.}\ \bibnamefont {Devitt}}, \bibinfo
  {author} {\bibfnamefont {A.~M.}\ \bibnamefont {Stephens}}, \bibinfo {author}
  {\bibfnamefont {B.}~\bibnamefont {Scharfenberger}}, \bibinfo {author}
  {\bibfnamefont {K.}~\bibnamefont {Buczak}}, \bibinfo {author} {\bibfnamefont
  {T.}~\bibnamefont {N\"obauer}}, \bibinfo {author} {\bibfnamefont {M.~S.}\
  \bibnamefont {Everitt}}, \bibinfo {author} {\bibfnamefont {J.}~\bibnamefont
  {Schmiedmayer}}, \ and\ \bibinfo {author} {\bibfnamefont {W.~J.}\
  \bibnamefont {Munro}},\ }\href {\doibase 10.1103/PhysRevX.4.031022}
  {\bibfield  {journal} {\bibinfo  {journal} {Phys. Rev. X}\ }\textbf {\bibinfo
  {volume} {4}},\ \bibinfo {pages} {031022} (\bibinfo {year}
  {2014})}\BibitemShut {NoStop}%
\bibitem [{\citenamefont {Brecht}\ \emph {et~al.}(2016)\citenamefont {Brecht},
  \citenamefont {Pfaff}, \citenamefont {Wang}, \citenamefont {Chu},
  \citenamefont {Frunzio}, \citenamefont {Devoret},\ and\ \citenamefont
  {Schoelkopf}}]{Brecht16}%
  \BibitemOpen
  \bibfield  {author} {\bibinfo {author} {\bibfnamefont {T.}~\bibnamefont
  {Brecht}}, \bibinfo {author} {\bibfnamefont {W.}~\bibnamefont {Pfaff}},
  \bibinfo {author} {\bibfnamefont {C.}~\bibnamefont {Wang}}, \bibinfo {author}
  {\bibfnamefont {Y.}~\bibnamefont {Chu}}, \bibinfo {author} {\bibfnamefont
  {L.}~\bibnamefont {Frunzio}}, \bibinfo {author} {\bibfnamefont {M.~H.}\
  \bibnamefont {Devoret}}, \ and\ \bibinfo {author} {\bibfnamefont {R.~J.}\
  \bibnamefont {Schoelkopf}},\ }\href
  {https://www.nature.com/articles/npjqi20162} {\bibfield  {journal} {\bibinfo
  {journal} {npj Quantum Information}\ }\textbf {\bibinfo {volume} {2}},\
  \bibinfo {pages} {16002} (\bibinfo {year} {2016})}\BibitemShut {NoStop}%
\bibitem [{\citenamefont {Bravyi}\ and\ \citenamefont
  {Kitaev}(1998)}]{Bravyi98}%
  \BibitemOpen
  \bibfield  {author} {\bibinfo {author} {\bibfnamefont {S.~B.}\ \bibnamefont
  {Bravyi}}\ and\ \bibinfo {author} {\bibfnamefont {A.~Y.}\ \bibnamefont
  {Kitaev}},\ }\href {https://arxiv.org/abs/quant-ph/9811052} {\bibfield
  {journal} {\bibinfo  {journal} {arXiv:quant-ph/9811052}\ } (\bibinfo {year}
  {1998})}\BibitemShut {NoStop}%
\bibitem [{\citenamefont {Fowler}\ \emph {et~al.}(2012)\citenamefont {Fowler},
  \citenamefont {Mariantoni}, \citenamefont {Martinis},\ and\ \citenamefont
  {Cleland}}]{Fowler12}%
  \BibitemOpen
  \bibfield  {author} {\bibinfo {author} {\bibfnamefont {A.~G.}\ \bibnamefont
  {Fowler}}, \bibinfo {author} {\bibfnamefont {M.}~\bibnamefont {Mariantoni}},
  \bibinfo {author} {\bibfnamefont {J.~M.}\ \bibnamefont {Martinis}}, \ and\
  \bibinfo {author} {\bibfnamefont {A.~N.}\ \bibnamefont {Cleland}},\ }\href
  {https://link.aps.org/doi/10.1103/PhysRevA.86.032324} {\bibfield  {journal}
  {\bibinfo  {journal} {Phys. Rev. A}\ }\textbf {\bibinfo {volume} {86}},\
  \bibinfo {pages} {032324} (\bibinfo {year} {2012})}\BibitemShut {NoStop}%
\bibitem [{\citenamefont {Blais}\ \emph {et~al.}(2004)\citenamefont {Blais},
  \citenamefont {Huang}, \citenamefont {Wallraff}, \citenamefont {Girvin},\
  and\ \citenamefont {Schoelkopf}}]{Blais04}%
  \BibitemOpen
  \bibfield  {author} {\bibinfo {author} {\bibfnamefont {A.}~\bibnamefont
  {Blais}}, \bibinfo {author} {\bibfnamefont {R.-S.}\ \bibnamefont {Huang}},
  \bibinfo {author} {\bibfnamefont {A.}~\bibnamefont {Wallraff}}, \bibinfo
  {author} {\bibfnamefont {S.~M.}\ \bibnamefont {Girvin}}, \ and\ \bibinfo
  {author} {\bibfnamefont {R.~J.}\ \bibnamefont {Schoelkopf}},\ }\href
  {https://link.aps.org/doi/10.1103/PhysRevA.69.062320} {\bibfield  {journal}
  {\bibinfo  {journal} {Phys. Rev. A}\ }\textbf {\bibinfo {volume} {69}},\
  \bibinfo {pages} {062320} (\bibinfo {year} {2004})}\BibitemShut {NoStop}%
\bibitem [{\citenamefont {Kerckhoff}\ \emph {et~al.}(2009)\citenamefont
  {Kerckhoff}, \citenamefont {Bouten}, \citenamefont {Silberfarb},\ and\
  \citenamefont {Mabuchi}}]{Kerckhoff09}%
  \BibitemOpen
  \bibfield  {author} {\bibinfo {author} {\bibfnamefont {J.}~\bibnamefont
  {Kerckhoff}}, \bibinfo {author} {\bibfnamefont {L.}~\bibnamefont {Bouten}},
  \bibinfo {author} {\bibfnamefont {A.}~\bibnamefont {Silberfarb}}, \ and\
  \bibinfo {author} {\bibfnamefont {H.}~\bibnamefont {Mabuchi}},\ }\href@noop
  {} {\bibfield  {journal} {\bibinfo  {journal} {Phys. Rev. A}\ }\textbf
  {\bibinfo {volume} {79}},\ \bibinfo {pages} {024305} (\bibinfo {year}
  {2009})}\BibitemShut {NoStop}%
\bibitem [{\citenamefont {Barrett}\ and\ \citenamefont
  {Kok}(2005)}]{Barrett05}%
  \BibitemOpen
  \bibfield  {author} {\bibinfo {author} {\bibfnamefont {S.~D.}\ \bibnamefont
  {Barrett}}\ and\ \bibinfo {author} {\bibfnamefont {P.}~\bibnamefont {Kok}},\
  }\href {\doibase 10.1103/PhysRevA.71.060310} {\bibfield  {journal} {\bibinfo
  {journal} {Phys. Rev. A}\ }\textbf {\bibinfo {volume} {71}},\ \bibinfo
  {pages} {060310} (\bibinfo {year} {2005})}\BibitemShut {NoStop}%
\bibitem [{\citenamefont {Roch}\ \emph {et~al.}(2014)\citenamefont {Roch},
  \citenamefont {Schwartz}, \citenamefont {Motzoi}, \citenamefont {Macklin},
  \citenamefont {Vijay}, \citenamefont {Eddins}, \citenamefont {Korotkov},
  \citenamefont {Whaley}, \citenamefont {Sarovar},\ and\ \citenamefont
  {Siddiqi}}]{Roch14}%
  \BibitemOpen
  \bibfield  {author} {\bibinfo {author} {\bibfnamefont {N.}~\bibnamefont
  {Roch}}, \bibinfo {author} {\bibfnamefont {M.~E.}\ \bibnamefont {Schwartz}},
  \bibinfo {author} {\bibfnamefont {F.}~\bibnamefont {Motzoi}}, \bibinfo
  {author} {\bibfnamefont {C.}~\bibnamefont {Macklin}}, \bibinfo {author}
  {\bibfnamefont {R.}~\bibnamefont {Vijay}}, \bibinfo {author} {\bibfnamefont
  {A.~W.}\ \bibnamefont {Eddins}}, \bibinfo {author} {\bibfnamefont {A.~N.}\
  \bibnamefont {Korotkov}}, \bibinfo {author} {\bibfnamefont {K.~B.}\
  \bibnamefont {Whaley}}, \bibinfo {author} {\bibfnamefont {M.}~\bibnamefont
  {Sarovar}}, \ and\ \bibinfo {author} {\bibfnamefont {I.}~\bibnamefont
  {Siddiqi}},\ }\href {\doibase 10.1103/PhysRevLett.112.170501} {\bibfield
  {journal} {\bibinfo  {journal} {Phys. Rev. Lett.}\ }\textbf {\bibinfo
  {volume} {112}},\ \bibinfo {pages} {170501} (\bibinfo {year}
  {2014})}\BibitemShut {NoStop}%
\bibitem [{\citenamefont {Chantasri}\ \emph {et~al.}(2016)\citenamefont
  {Chantasri}, \citenamefont {Kimchi-Schwartz}, \citenamefont {Roch},
  \citenamefont {Siddiqi},\ and\ \citenamefont {Jordan}}]{Chantasri16}%
  \BibitemOpen
  \bibfield  {author} {\bibinfo {author} {\bibfnamefont {A.}~\bibnamefont
  {Chantasri}}, \bibinfo {author} {\bibfnamefont {M.~E.}\ \bibnamefont
  {Kimchi-Schwartz}}, \bibinfo {author} {\bibfnamefont {N.}~\bibnamefont
  {Roch}}, \bibinfo {author} {\bibfnamefont {I.}~\bibnamefont {Siddiqi}}, \
  and\ \bibinfo {author} {\bibfnamefont {A.~N.}\ \bibnamefont {Jordan}},\
  }\href {\doibase 10.1103/PhysRevX.6.041052} {\bibfield  {journal} {\bibinfo
  {journal} {Phys. Rev. X}\ }\textbf {\bibinfo {volume} {6}},\ \bibinfo {pages}
  {041052} (\bibinfo {year} {2016})}\BibitemShut {NoStop}%
\bibitem [{\citenamefont {Roy}\ \emph {et~al.}(2015)\citenamefont {Roy},
  \citenamefont {Jiang}, \citenamefont {Stone},\ and\ \citenamefont
  {Devoret}}]{Roy15}%
  \BibitemOpen
  \bibfield  {author} {\bibinfo {author} {\bibfnamefont {A.}~\bibnamefont
  {Roy}}, \bibinfo {author} {\bibfnamefont {L.}~\bibnamefont {Jiang}}, \bibinfo
  {author} {\bibfnamefont {A.~D.}\ \bibnamefont {Stone}}, \ and\ \bibinfo
  {author} {\bibfnamefont {M.}~\bibnamefont {Devoret}},\ }\href {\doibase
  10.1103/PhysRevLett.115.150503} {\bibfield  {journal} {\bibinfo  {journal}
  {Phys. Rev. Lett.}\ }\textbf {\bibinfo {volume} {115}},\ \bibinfo {pages}
  {150503} (\bibinfo {year} {2015})}\BibitemShut {NoStop}%
\bibitem [{\citenamefont {Narla}\ \emph {et~al.}(2016)\citenamefont {Narla},
  \citenamefont {Shankar}, \citenamefont {Hatridge}, \citenamefont {Leghtas},
  \citenamefont {Sliwa}, \citenamefont {Zalys-Geller}, \citenamefont
  {Mundhada}, \citenamefont {Pfaff}, \citenamefont {Frunzio}, \citenamefont
  {Schoelkopf},\ and\ \citenamefont {Devoret}}]{Narla16}%
  \BibitemOpen
  \bibfield  {author} {\bibinfo {author} {\bibfnamefont {A.}~\bibnamefont
  {Narla}}, \bibinfo {author} {\bibfnamefont {S.}~\bibnamefont {Shankar}},
  \bibinfo {author} {\bibfnamefont {M.}~\bibnamefont {Hatridge}}, \bibinfo
  {author} {\bibfnamefont {Z.}~\bibnamefont {Leghtas}}, \bibinfo {author}
  {\bibfnamefont {K.~M.}\ \bibnamefont {Sliwa}}, \bibinfo {author}
  {\bibfnamefont {E.}~\bibnamefont {Zalys-Geller}}, \bibinfo {author}
  {\bibfnamefont {S.~O.}\ \bibnamefont {Mundhada}}, \bibinfo {author}
  {\bibfnamefont {W.}~\bibnamefont {Pfaff}}, \bibinfo {author} {\bibfnamefont
  {L.}~\bibnamefont {Frunzio}}, \bibinfo {author} {\bibfnamefont {R.~J.}\
  \bibnamefont {Schoelkopf}}, \ and\ \bibinfo {author} {\bibfnamefont {M.~H.}\
  \bibnamefont {Devoret}},\ }\href {\doibase 10.1103/PhysRevX.6.031036}
  {\bibfield  {journal} {\bibinfo  {journal} {Phys. Rev. X}\ }\textbf {\bibinfo
  {volume} {6}},\ \bibinfo {pages} {031036} (\bibinfo {year}
  {2016})}\BibitemShut {NoStop}%
\bibitem [{\citenamefont {Ohm}\ and\ \citenamefont {Hassler}(2017)}]{Ohm17}%
  \BibitemOpen
  \bibfield  {author} {\bibinfo {author} {\bibfnamefont {C.}~\bibnamefont
  {Ohm}}\ and\ \bibinfo {author} {\bibfnamefont {F.}~\bibnamefont {Hassler}},\
  }\href {http://stacks.iop.org/1367-2630/19/i=5/a=053018} {\bibfield
  {journal} {\bibinfo  {journal} {New Journal of Physics}\ }\textbf {\bibinfo
  {volume} {19}},\ \bibinfo {pages} {053018} (\bibinfo {year}
  {2017})}\BibitemShut {NoStop}%
\bibitem [{\citenamefont {Wenner}\ \emph {et~al.}(2014)\citenamefont {Wenner},
  \citenamefont {Yin}, \citenamefont {Chen}, \citenamefont {Barends},
  \citenamefont {Chiaro}, \citenamefont {Jeffrey}, \citenamefont {Kelly},
  \citenamefont {Megrant}, \citenamefont {Mutus}, \citenamefont {Neill},
  \citenamefont {O'Malley}, \citenamefont {Roushan}, \citenamefont {Sank},
  \citenamefont {Vainsencher}, \citenamefont {White}, \citenamefont {Korotkov},
  \citenamefont {Cleland},\ and\ \citenamefont {Martinis}}]{Wenner14}%
  \BibitemOpen
  \bibfield  {author} {\bibinfo {author} {\bibfnamefont {J.}~\bibnamefont
  {Wenner}}, \bibinfo {author} {\bibfnamefont {Y.}~\bibnamefont {Yin}},
  \bibinfo {author} {\bibfnamefont {Y.}~\bibnamefont {Chen}}, \bibinfo {author}
  {\bibfnamefont {R.}~\bibnamefont {Barends}}, \bibinfo {author} {\bibfnamefont
  {B.}~\bibnamefont {Chiaro}}, \bibinfo {author} {\bibfnamefont
  {E.}~\bibnamefont {Jeffrey}}, \bibinfo {author} {\bibfnamefont
  {J.}~\bibnamefont {Kelly}}, \bibinfo {author} {\bibfnamefont
  {A.}~\bibnamefont {Megrant}}, \bibinfo {author} {\bibfnamefont {J.~Y.}\
  \bibnamefont {Mutus}}, \bibinfo {author} {\bibfnamefont {C.}~\bibnamefont
  {Neill}}, \bibinfo {author} {\bibfnamefont {P.~J.~J.}\ \bibnamefont
  {O'Malley}}, \bibinfo {author} {\bibfnamefont {P.}~\bibnamefont {Roushan}},
  \bibinfo {author} {\bibfnamefont {D.}~\bibnamefont {Sank}}, \bibinfo {author}
  {\bibfnamefont {A.}~\bibnamefont {Vainsencher}}, \bibinfo {author}
  {\bibfnamefont {T.~C.}\ \bibnamefont {White}}, \bibinfo {author}
  {\bibfnamefont {A.~N.}\ \bibnamefont {Korotkov}}, \bibinfo {author}
  {\bibfnamefont {A.~N.}\ \bibnamefont {Cleland}}, \ and\ \bibinfo {author}
  {\bibfnamefont {J.~M.}\ \bibnamefont {Martinis}},\ }\href {\doibase
  10.1103/PhysRevLett.112.210501} {\bibfield  {journal} {\bibinfo  {journal}
  {Phys. Rev. Lett.}\ }\textbf {\bibinfo {volume} {112}},\ \bibinfo {pages}
  {210501} (\bibinfo {year} {2014})}\BibitemShut {NoStop}%
\bibitem [{\citenamefont {Pfaff}\ \emph {et~al.}(2017)\citenamefont {Pfaff},
  \citenamefont {J.~Axline}, \citenamefont {D.~Burkhart}, \citenamefont {Vool},
  \citenamefont {Reinhold}, \citenamefont {Frunzio}, \citenamefont {Jiang},
  \citenamefont {H.~Devoret},\ and\ \citenamefont {J.~Schoelkopf}}]{Pfaff17}%
  \BibitemOpen
  \bibfield  {author} {\bibinfo {author} {\bibfnamefont {W.}~\bibnamefont
  {Pfaff}}, \bibinfo {author} {\bibfnamefont {C.}~\bibnamefont {J.~Axline}},
  \bibinfo {author} {\bibfnamefont {L.}~\bibnamefont {D.~Burkhart}}, \bibinfo
  {author} {\bibfnamefont {U.}~\bibnamefont {Vool}}, \bibinfo {author}
  {\bibfnamefont {P.}~\bibnamefont {Reinhold}}, \bibinfo {author}
  {\bibfnamefont {L.}~\bibnamefont {Frunzio}}, \bibinfo {author} {\bibfnamefont
  {L.}~\bibnamefont {Jiang}}, \bibinfo {author} {\bibfnamefont
  {M.}~\bibnamefont {H.~Devoret}}, \ and\ \bibinfo {author} {\bibfnamefont
  {R.}~\bibnamefont {J.~Schoelkopf}},\ }\href
  {https://www.nature.com/articles/nphys4143} {\bibfield  {journal} {\bibinfo
  {journal} {Nat.\ Phys.}\ }\textbf {\bibinfo {volume} {13}},\ \bibinfo {pages}
  {882} (\bibinfo {year} {2017})}\BibitemShut {NoStop}%
\bibitem [{\citenamefont {Kelly}\ \emph {et~al.}(2015)\citenamefont {Kelly},
  \citenamefont {Barends}, \citenamefont {Fowler}, \citenamefont {Megrant},
  \citenamefont {Jeffrey}, \citenamefont {White}, \citenamefont {Sank},
  \citenamefont {Mutus}, \citenamefont {Campbell}, \citenamefont {Chen} \emph
  {et~al.}}]{Kelly15}%
  \BibitemOpen
  \bibfield  {author} {\bibinfo {author} {\bibfnamefont {J.}~\bibnamefont
  {Kelly}}, \bibinfo {author} {\bibfnamefont {R.}~\bibnamefont {Barends}},
  \bibinfo {author} {\bibfnamefont {A.}~\bibnamefont {Fowler}}, \bibinfo
  {author} {\bibfnamefont {A.}~\bibnamefont {Megrant}}, \bibinfo {author}
  {\bibfnamefont {E.}~\bibnamefont {Jeffrey}}, \bibinfo {author} {\bibfnamefont
  {T.}~\bibnamefont {White}}, \bibinfo {author} {\bibfnamefont
  {D.}~\bibnamefont {Sank}}, \bibinfo {author} {\bibfnamefont {J.}~\bibnamefont
  {Mutus}}, \bibinfo {author} {\bibfnamefont {B.}~\bibnamefont {Campbell}},
  \bibinfo {author} {\bibfnamefont {Y.}~\bibnamefont {Chen}},  \emph {et~al.},\
  }\href
  {https://www.nature.com/nature/journal/v519/n7541/full/nature14270.html}
  {\bibfield  {journal} {\bibinfo  {journal} {Nature}\ }\textbf {\bibinfo
  {volume} {519}},\ \bibinfo {pages} {66} (\bibinfo {year} {2015})}\BibitemShut
  {NoStop}%
\bibitem [{\citenamefont {Rist\`{e}}\ \emph {et~al.}(2015)\citenamefont
  {Rist\`{e}}, \citenamefont {Poletto}, \citenamefont {Huang}, \citenamefont
  {Bruno}, \citenamefont {Vesterinen}, \citenamefont {Saira},\ and\
  \citenamefont {DiCarlo}}]{Riste15}%
  \BibitemOpen
  \bibfield  {author} {\bibinfo {author} {\bibfnamefont {D.}~\bibnamefont
  {Rist\`{e}}}, \bibinfo {author} {\bibfnamefont {S.}~\bibnamefont {Poletto}},
  \bibinfo {author} {\bibfnamefont {M.~Z.}\ \bibnamefont {Huang}}, \bibinfo
  {author} {\bibfnamefont {A.}~\bibnamefont {Bruno}}, \bibinfo {author}
  {\bibfnamefont {V.}~\bibnamefont {Vesterinen}}, \bibinfo {author}
  {\bibfnamefont {O.~P.}\ \bibnamefont {Saira}}, \ and\ \bibinfo {author}
  {\bibfnamefont {L.}~\bibnamefont {DiCarlo}},\ }\href
  {https://www.nature.com/articles/ncomms7983} {\bibfield  {journal} {\bibinfo
  {journal} {Nat.\ Commun.}\ }\textbf {\bibinfo {volume} {{6}}},\ \bibinfo
  {pages} {6983} (\bibinfo {year} {{2015}})}\BibitemShut {NoStop}%
\bibitem [{\citenamefont {Song}\ \emph {et~al.}(2017)\citenamefont {Song},
  \citenamefont {Xu}, \citenamefont {Liu}, \citenamefont {Yang}, \citenamefont
  {Zheng}, \citenamefont {Deng}, \citenamefont {Xie}, \citenamefont {Huang},
  \citenamefont {Guo}, \citenamefont {Zhang}, \citenamefont {Zhang},
  \citenamefont {Xu}, \citenamefont {Zheng}, \citenamefont {Zhu}, \citenamefont
  {Wang}, \citenamefont {Chen}, \citenamefont {Lu}, \citenamefont {Han},\ and\
  \citenamefont {Pan}}]{Song17}%
  \BibitemOpen
  \bibfield  {author} {\bibinfo {author} {\bibfnamefont {C.}~\bibnamefont
  {Song}}, \bibinfo {author} {\bibfnamefont {K.}~\bibnamefont {Xu}}, \bibinfo
  {author} {\bibfnamefont {W.}~\bibnamefont {Liu}}, \bibinfo {author}
  {\bibfnamefont {C.-p.}\ \bibnamefont {Yang}}, \bibinfo {author}
  {\bibfnamefont {S.-B.}\ \bibnamefont {Zheng}}, \bibinfo {author}
  {\bibfnamefont {H.}~\bibnamefont {Deng}}, \bibinfo {author} {\bibfnamefont
  {Q.}~\bibnamefont {Xie}}, \bibinfo {author} {\bibfnamefont {K.}~\bibnamefont
  {Huang}}, \bibinfo {author} {\bibfnamefont {Q.}~\bibnamefont {Guo}}, \bibinfo
  {author} {\bibfnamefont {L.}~\bibnamefont {Zhang}}, \bibinfo {author}
  {\bibfnamefont {P.}~\bibnamefont {Zhang}}, \bibinfo {author} {\bibfnamefont
  {D.}~\bibnamefont {Xu}}, \bibinfo {author} {\bibfnamefont {D.}~\bibnamefont
  {Zheng}}, \bibinfo {author} {\bibfnamefont {X.}~\bibnamefont {Zhu}}, \bibinfo
  {author} {\bibfnamefont {H.}~\bibnamefont {Wang}}, \bibinfo {author}
  {\bibfnamefont {Y.-A.}\ \bibnamefont {Chen}}, \bibinfo {author}
  {\bibfnamefont {C.-Y.}\ \bibnamefont {Lu}}, \bibinfo {author} {\bibfnamefont
  {S.}~\bibnamefont {Han}}, \ and\ \bibinfo {author} {\bibfnamefont {J.-W.}\
  \bibnamefont {Pan}},\ }\href {\doibase 10.1103/PhysRevLett.119.180511}
  {\bibfield  {journal} {\bibinfo  {journal} {Phys. Rev. Lett.}\ }\textbf
  {\bibinfo {volume} {119}},\ \bibinfo {pages} {180511} (\bibinfo {year}
  {2017})}\BibitemShut {NoStop}%
\bibitem [{\citenamefont {Takita}\ \emph {et~al.}(2017)\citenamefont {Takita},
  \citenamefont {Cross}, \citenamefont {C\'orcoles}, \citenamefont {Chow},\
  and\ \citenamefont {Gambetta}}]{Takita17}%
  \BibitemOpen
  \bibfield  {author} {\bibinfo {author} {\bibfnamefont {M.}~\bibnamefont
  {Takita}}, \bibinfo {author} {\bibfnamefont {A.~W.}\ \bibnamefont {Cross}},
  \bibinfo {author} {\bibfnamefont {A.~D.}\ \bibnamefont {C\'orcoles}},
  \bibinfo {author} {\bibfnamefont {J.~M.}\ \bibnamefont {Chow}}, \ and\
  \bibinfo {author} {\bibfnamefont {J.~M.}\ \bibnamefont {Gambetta}},\ }\href
  {https://journals.aps.org/prl/abstract/10.1103/PhysRevLett.119.180501}
  {\bibfield  {journal} {\bibinfo  {journal} {Phys. Rev. Lett.}\ }\textbf
  {\bibinfo {volume} {119}},\ \bibinfo {pages} {180501} (\bibinfo {year}
  {2017})}\BibitemShut {NoStop}%
\bibitem [{\citenamefont {Barends}\ \emph {et~al.}(2014)\citenamefont
  {Barends}, \citenamefont {Kelly}, \citenamefont {Megrant}, \citenamefont
  {Veitia}, \citenamefont {Sank}, \citenamefont {Jeffrey}, \citenamefont
  {White}, \citenamefont {Mutus}, \citenamefont {Fowler}, \citenamefont
  {Campbell}, \citenamefont {Chen}, \citenamefont {Chen}, \citenamefont
  {Chiaro}, \citenamefont {Dunsworth}, \citenamefont {Neill}, \citenamefont
  {O'Malley}, \citenamefont {Roushan}, \citenamefont {Vainsencher},
  \citenamefont {Wenner}, \citenamefont {Korotkov}, \citenamefont {Cleland},\
  and\ \citenamefont {Martinis}}]{Barends14}%
  \BibitemOpen
  \bibfield  {author} {\bibinfo {author} {\bibfnamefont {R.}~\bibnamefont
  {Barends}}, \bibinfo {author} {\bibfnamefont {J.}~\bibnamefont {Kelly}},
  \bibinfo {author} {\bibfnamefont {A.}~\bibnamefont {Megrant}}, \bibinfo
  {author} {\bibfnamefont {A.}~\bibnamefont {Veitia}}, \bibinfo {author}
  {\bibfnamefont {D.}~\bibnamefont {Sank}}, \bibinfo {author} {\bibfnamefont
  {E.}~\bibnamefont {Jeffrey}}, \bibinfo {author} {\bibfnamefont {T.~C.}\
  \bibnamefont {White}}, \bibinfo {author} {\bibfnamefont {J.}~\bibnamefont
  {Mutus}}, \bibinfo {author} {\bibfnamefont {A.~G.}\ \bibnamefont {Fowler}},
  \bibinfo {author} {\bibfnamefont {B.}~\bibnamefont {Campbell}}, \bibinfo
  {author} {\bibfnamefont {Y.}~\bibnamefont {Chen}}, \bibinfo {author}
  {\bibfnamefont {Z.}~\bibnamefont {Chen}}, \bibinfo {author} {\bibfnamefont
  {B.}~\bibnamefont {Chiaro}}, \bibinfo {author} {\bibfnamefont
  {A.}~\bibnamefont {Dunsworth}}, \bibinfo {author} {\bibfnamefont
  {C.}~\bibnamefont {Neill}}, \bibinfo {author} {\bibfnamefont
  {P.}~\bibnamefont {O'Malley}}, \bibinfo {author} {\bibfnamefont
  {P.}~\bibnamefont {Roushan}}, \bibinfo {author} {\bibfnamefont
  {A.}~\bibnamefont {Vainsencher}}, \bibinfo {author} {\bibfnamefont
  {J.}~\bibnamefont {Wenner}}, \bibinfo {author} {\bibfnamefont {A.~N.}\
  \bibnamefont {Korotkov}}, \bibinfo {author} {\bibfnamefont {A.~N.}\
  \bibnamefont {Cleland}}, \ and\ \bibinfo {author} {\bibfnamefont {J.~M.}\
  \bibnamefont {Martinis}},\ }\href
  {http://www.nature.com/nature/journal/v508/n7497/abs/nature13171.html}
  {\bibfield  {journal} {\bibinfo  {journal} {Nature}\ }\textbf {\bibinfo
  {volume} {508}},\ \bibinfo {pages} {500} (\bibinfo {year}
  {2014})}\BibitemShut {NoStop}%
\bibitem [{\citenamefont {Castellanos-Beltran}\ \emph
  {et~al.}(2008)\citenamefont {Castellanos-Beltran}, \citenamefont {Irwin},
  \citenamefont {Hilton}, \citenamefont {Vale},\ and\ \citenamefont
  {Lehnert}}]{Castellanos-Beltran08}%
  \BibitemOpen
  \bibfield  {author} {\bibinfo {author} {\bibfnamefont {M.~A.}\ \bibnamefont
  {Castellanos-Beltran}}, \bibinfo {author} {\bibfnamefont {K.~D.}\
  \bibnamefont {Irwin}}, \bibinfo {author} {\bibfnamefont {G.~C.}\ \bibnamefont
  {Hilton}}, \bibinfo {author} {\bibfnamefont {L.~R.}\ \bibnamefont {Vale}}, \
  and\ \bibinfo {author} {\bibfnamefont {K.~W.}\ \bibnamefont {Lehnert}},\
  }\href {\doibase 10.1038/nphys1090} {\bibfield  {journal} {\bibinfo
  {journal} {Nat.\ Phys.}\ }\textbf {\bibinfo {volume} {4}},\ \bibinfo {pages}
  {929} (\bibinfo {year} {2008})}\BibitemShut {NoStop}%
\bibitem [{\citenamefont {Gambetta}\ \emph {et~al.}(2008)\citenamefont
  {Gambetta}, \citenamefont {Blais}, \citenamefont {Boissonneault},
  \citenamefont {Houck}, \citenamefont {Schuster},\ and\ \citenamefont
  {Girvin}}]{Gambetta08}%
  \BibitemOpen
  \bibfield  {author} {\bibinfo {author} {\bibfnamefont {J.}~\bibnamefont
  {Gambetta}}, \bibinfo {author} {\bibfnamefont {A.}~\bibnamefont {Blais}},
  \bibinfo {author} {\bibfnamefont {M.}~\bibnamefont {Boissonneault}}, \bibinfo
  {author} {\bibfnamefont {A.~A.}\ \bibnamefont {Houck}}, \bibinfo {author}
  {\bibfnamefont {D.~I.}\ \bibnamefont {Schuster}}, \ and\ \bibinfo {author}
  {\bibfnamefont {S.~M.}\ \bibnamefont {Girvin}},\ }\href
  {https://journals.aps.org/pra/abstract/10.1103/PhysRevA.77.012112} {\bibfield
   {journal} {\bibinfo  {journal} {Phys. Rev. A}\ }\textbf {\bibinfo {volume}
  {77}},\ \bibinfo {pages} {012112} (\bibinfo {year} {2008})}\BibitemShut
  {NoStop}%
\bibitem [{\citenamefont {Bultink}\ \emph {et~al.}(2017)\citenamefont
  {Bultink}, \citenamefont {Tarasinski}, \citenamefont {Haandbaek},
  \citenamefont {Poletto}, \citenamefont {Haider}, \citenamefont {Michalak},
  \citenamefont {Bruno},\ and\ \citenamefont {DiCarlo}}]{Bultink17}%
  \BibitemOpen
  \bibfield  {author} {\bibinfo {author} {\bibfnamefont {C.~C.}\ \bibnamefont
  {Bultink}}, \bibinfo {author} {\bibfnamefont {B.}~\bibnamefont {Tarasinski}},
  \bibinfo {author} {\bibfnamefont {N.}~\bibnamefont {Haandbaek}}, \bibinfo
  {author} {\bibfnamefont {S.}~\bibnamefont {Poletto}}, \bibinfo {author}
  {\bibfnamefont {N.}~\bibnamefont {Haider}}, \bibinfo {author} {\bibfnamefont
  {D.~J.}\ \bibnamefont {Michalak}}, \bibinfo {author} {\bibfnamefont
  {A.}~\bibnamefont {Bruno}}, \ and\ \bibinfo {author} {\bibfnamefont
  {L.}~\bibnamefont {DiCarlo}},\ }\href {https://arxiv.org/abs/1711.05336}
  {\bibfield  {journal} {\bibinfo  {journal} {arXiv:1711.05336}\ } (\bibinfo
  {year} {2017})}\BibitemShut {NoStop}%
\bibitem [{\citenamefont {Tornberg}\ and\ \citenamefont
  {Johansson}(2010)}]{Tornberg10}%
  \BibitemOpen
  \bibfield  {author} {\bibinfo {author} {\bibfnamefont {L.}~\bibnamefont
  {Tornberg}}\ and\ \bibinfo {author} {\bibfnamefont {G.}~\bibnamefont
  {Johansson}},\ }\href {\doibase 10.1103/PhysRevA.82.012329} {\bibfield
  {journal} {\bibinfo  {journal} {Phys. Rev. A}\ }\textbf {\bibinfo {volume}
  {82}},\ \bibinfo {pages} {012329} (\bibinfo {year} {2010})}\BibitemShut
  {NoStop}%
\bibitem [{\citenamefont {Motzoi}\ \emph {et~al.}(2015)\citenamefont {Motzoi},
  \citenamefont {Whaley},\ and\ \citenamefont {Sarovar}}]{Motzoi15}%
  \BibitemOpen
  \bibfield  {author} {\bibinfo {author} {\bibfnamefont {F.}~\bibnamefont
  {Motzoi}}, \bibinfo {author} {\bibfnamefont {K.~B.}\ \bibnamefont {Whaley}},
  \ and\ \bibinfo {author} {\bibfnamefont {M.}~\bibnamefont {Sarovar}},\ }\href
  {\doibase 10.1103/PhysRevA.92.032308} {\bibfield  {journal} {\bibinfo
  {journal} {Phys. Rev. A}\ }\textbf {\bibinfo {volume} {92}},\ \bibinfo
  {pages} {032308} (\bibinfo {year} {2015})}\BibitemShut {NoStop}%
\bibitem [{\citenamefont {Chow}\ \emph
  {et~al.}(2010{\natexlab{a}})\citenamefont {Chow}, \citenamefont {DiCarlo},
  \citenamefont {Gambetta}, \citenamefont {Nunnenkamp}, \citenamefont {Bishop},
  \citenamefont {Frunzio}, \citenamefont {Devoret}, \citenamefont {Girvin},\
  and\ \citenamefont {Schoelkopf}}]{Chow10}%
  \BibitemOpen
  \bibfield  {author} {\bibinfo {author} {\bibfnamefont {J.~M.}\ \bibnamefont
  {Chow}}, \bibinfo {author} {\bibfnamefont {L.}~\bibnamefont {DiCarlo}},
  \bibinfo {author} {\bibfnamefont {J.~M.}\ \bibnamefont {Gambetta}}, \bibinfo
  {author} {\bibfnamefont {A.}~\bibnamefont {Nunnenkamp}}, \bibinfo {author}
  {\bibfnamefont {L.~S.}\ \bibnamefont {Bishop}}, \bibinfo {author}
  {\bibfnamefont {L.}~\bibnamefont {Frunzio}}, \bibinfo {author} {\bibfnamefont
  {M.~H.}\ \bibnamefont {Devoret}}, \bibinfo {author} {\bibfnamefont {S.~M.}\
  \bibnamefont {Girvin}}, \ and\ \bibinfo {author} {\bibfnamefont {R.~J.}\
  \bibnamefont {Schoelkopf}},\ }\href {\doibase 10.1103/PhysRevA.81.062325}
  {\bibfield  {journal} {\bibinfo  {journal} {Phys. Rev. A}\ }\textbf {\bibinfo
  {volume} {81}},\ \bibinfo {pages} {062325} (\bibinfo {year}
  {2010}{\natexlab{a}})}\BibitemShut {NoStop}%
\bibitem [{\citenamefont {Filipp}\ \emph {et~al.}(2009)\citenamefont {Filipp},
  \citenamefont {Maurer}, \citenamefont {Leek}, \citenamefont {Baur},
  \citenamefont {Bianchetti}, \citenamefont {Fink}, \citenamefont {G\"oppl},
  \citenamefont {Steffen}, \citenamefont {Gambetta}, \citenamefont {Blais},\
  and\ \citenamefont {Wallraff}}]{Filipp09}%
  \BibitemOpen
  \bibfield  {author} {\bibinfo {author} {\bibfnamefont {S.}~\bibnamefont
  {Filipp}}, \bibinfo {author} {\bibfnamefont {P.}~\bibnamefont {Maurer}},
  \bibinfo {author} {\bibfnamefont {P.~J.}\ \bibnamefont {Leek}}, \bibinfo
  {author} {\bibfnamefont {M.}~\bibnamefont {Baur}}, \bibinfo {author}
  {\bibfnamefont {R.}~\bibnamefont {Bianchetti}}, \bibinfo {author}
  {\bibfnamefont {J.~M.}\ \bibnamefont {Fink}}, \bibinfo {author}
  {\bibfnamefont {M.}~\bibnamefont {G\"oppl}}, \bibinfo {author} {\bibfnamefont
  {L.}~\bibnamefont {Steffen}}, \bibinfo {author} {\bibfnamefont {J.~M.}\
  \bibnamefont {Gambetta}}, \bibinfo {author} {\bibfnamefont {A.}~\bibnamefont
  {Blais}}, \ and\ \bibinfo {author} {\bibfnamefont {A.}~\bibnamefont
  {Wallraff}},\ }\href {\doibase 10.1103/PhysRevLett.102.200402} {\bibfield
  {journal} {\bibinfo  {journal} {Phys. Rev. Lett.}\ }\textbf {\bibinfo
  {volume} {102}},\ \bibinfo {pages} {200402} (\bibinfo {year}
  {2009})}\BibitemShut {NoStop}%
\bibitem [{\citenamefont {Nelder}\ and\ \citenamefont {Mead}(1965)}]{Nelder65}%
  \BibitemOpen
  \bibfield  {author} {\bibinfo {author} {\bibfnamefont {J.~A.}\ \bibnamefont
  {Nelder}}\ and\ \bibinfo {author} {\bibfnamefont {R.}~\bibnamefont {Mead}},\
  }\href@noop {} {\bibfield  {journal} {\bibinfo  {journal} {The Computer
  Journal}\ }\textbf {\bibinfo {volume} {7}},\ \bibinfo {pages} {308} (\bibinfo
  {year} {1965})}\BibitemShut {NoStop}%
\bibitem [{\citenamefont {Wootters}(1998)}]{Wootters98}%
  \BibitemOpen
  \bibfield  {author} {\bibinfo {author} {\bibfnamefont {W.~K.}\ \bibnamefont
  {Wootters}},\ }\href@noop {} {\bibfield  {journal} {\bibinfo  {journal}
  {Phys. Rev. Lett.}\ }\textbf {\bibinfo {volume} {80}},\ \bibinfo {pages}
  {2245} (\bibinfo {year} {1998})}\BibitemShut {NoStop}%
\bibitem [{\citenamefont {Ryan}\ \emph {et~al.}(2015)\citenamefont {Ryan},
  \citenamefont {Johnson}, \citenamefont {Gambetta}, \citenamefont {Chow},
  \citenamefont {da~Silva}, \citenamefont {Dial},\ and\ \citenamefont
  {Ohki}}]{Ryan15}%
  \BibitemOpen
  \bibfield  {author} {\bibinfo {author} {\bibfnamefont {C.~A.}\ \bibnamefont
  {Ryan}}, \bibinfo {author} {\bibfnamefont {B.~R.}\ \bibnamefont {Johnson}},
  \bibinfo {author} {\bibfnamefont {J.~M.}\ \bibnamefont {Gambetta}}, \bibinfo
  {author} {\bibfnamefont {J.~M.}\ \bibnamefont {Chow}}, \bibinfo {author}
  {\bibfnamefont {M.~P.}\ \bibnamefont {da~Silva}}, \bibinfo {author}
  {\bibfnamefont {O.~E.}\ \bibnamefont {Dial}}, \ and\ \bibinfo {author}
  {\bibfnamefont {T.~A.}\ \bibnamefont {Ohki}},\ }\href
  {https://journals.aps.org/pra/pdf/10.1103/PhysRevA.91.022118} {\bibfield
  {journal} {\bibinfo  {journal} {Phys. Rev. A}\ }\textbf {\bibinfo {volume}
  {91}},\ \bibinfo {pages} {022118} (\bibinfo {year} {2015})}\BibitemShut
  {NoStop}%
\bibitem [{\citenamefont {Magesan}\ \emph {et~al.}(2015)\citenamefont
  {Magesan}, \citenamefont {Gambetta}, \citenamefont {C\'orcoles},\ and\
  \citenamefont {Chow}}]{Magesan15}%
  \BibitemOpen
  \bibfield  {author} {\bibinfo {author} {\bibfnamefont {E.}~\bibnamefont
  {Magesan}}, \bibinfo {author} {\bibfnamefont {J.~M.}\ \bibnamefont
  {Gambetta}}, \bibinfo {author} {\bibfnamefont {A.~D.}\ \bibnamefont
  {C\'orcoles}}, \ and\ \bibinfo {author} {\bibfnamefont {J.~M.}\ \bibnamefont
  {Chow}},\ }\href
  {https://journals.aps.org/prl/pdf/10.1103/PhysRevLett.114.200501} {\bibfield
  {journal} {\bibinfo  {journal} {Phys. Rev. Lett.}\ }\textbf {\bibinfo
  {volume} {114}},\ \bibinfo {pages} {200501} (\bibinfo {year}
  {2015})}\BibitemShut {NoStop}%
\bibitem [{\citenamefont {Pedregosa}\ \emph {et~al.}(2011)\citenamefont
  {Pedregosa}, \citenamefont {Varoquaux}, \citenamefont {Gramfort},
  \citenamefont {Michel}, \citenamefont {Thirion}, \citenamefont {Grisel},
  \citenamefont {Blondel}, \citenamefont {Prettenhofer}, \citenamefont {Weiss},
  \citenamefont {Dubourg}, \citenamefont {Vanderplas}, \citenamefont {Passos},
  \citenamefont {Cournapeau}, \citenamefont {Brucher}, \citenamefont {Perrot},\
  and\ \citenamefont {Duchesnay}}]{scikit-learn}%
  \BibitemOpen
  \bibfield  {author} {\bibinfo {author} {\bibfnamefont {F.}~\bibnamefont
  {Pedregosa}}, \bibinfo {author} {\bibfnamefont {G.}~\bibnamefont
  {Varoquaux}}, \bibinfo {author} {\bibfnamefont {A.}~\bibnamefont {Gramfort}},
  \bibinfo {author} {\bibfnamefont {V.}~\bibnamefont {Michel}}, \bibinfo
  {author} {\bibfnamefont {B.}~\bibnamefont {Thirion}}, \bibinfo {author}
  {\bibfnamefont {O.}~\bibnamefont {Grisel}}, \bibinfo {author} {\bibfnamefont
  {M.}~\bibnamefont {Blondel}}, \bibinfo {author} {\bibfnamefont
  {P.}~\bibnamefont {Prettenhofer}}, \bibinfo {author} {\bibfnamefont
  {R.}~\bibnamefont {Weiss}}, \bibinfo {author} {\bibfnamefont
  {V.}~\bibnamefont {Dubourg}}, \bibinfo {author} {\bibfnamefont
  {J.}~\bibnamefont {Vanderplas}}, \bibinfo {author} {\bibfnamefont
  {A.}~\bibnamefont {Passos}}, \bibinfo {author} {\bibfnamefont
  {D.}~\bibnamefont {Cournapeau}}, \bibinfo {author} {\bibfnamefont
  {M.}~\bibnamefont {Brucher}}, \bibinfo {author} {\bibfnamefont
  {M.}~\bibnamefont {Perrot}}, \ and\ \bibinfo {author} {\bibfnamefont
  {E.}~\bibnamefont {Duchesnay}},\ }\href@noop {} {\bibfield  {journal}
  {\bibinfo  {journal} {Journal of Machine Learning Research}\ }\textbf
  {\bibinfo {volume} {12}},\ \bibinfo {pages} {2825} (\bibinfo {year}
  {2011})}\BibitemShut {NoStop}%
\bibitem [{\citenamefont {Horodecki}\ \emph {et~al.}(2009)\citenamefont
  {Horodecki}, \citenamefont {Horodecki}, \citenamefont {Horodecki},\ and\
  \citenamefont {Horodecki}}]{Horodecki09}%
  \BibitemOpen
  \bibfield  {author} {\bibinfo {author} {\bibfnamefont {R.}~\bibnamefont
  {Horodecki}}, \bibinfo {author} {\bibfnamefont {P.}~\bibnamefont
  {Horodecki}}, \bibinfo {author} {\bibfnamefont {M.}~\bibnamefont
  {Horodecki}}, \ and\ \bibinfo {author} {\bibfnamefont {K.}~\bibnamefont
  {Horodecki}},\ }\href@noop {} {\bibfield  {journal} {\bibinfo  {journal}
  {Rev. Mod. Phys.}\ }\textbf {\bibinfo {volume} {81}},\ \bibinfo {pages} {865}
  (\bibinfo {year} {2009})}\BibitemShut {NoStop}%
\bibitem [{\citenamefont {Vidal}\ and\ \citenamefont {Werner}(2002)}]{Vidal02}%
  \BibitemOpen
  \bibfield  {author} {\bibinfo {author} {\bibfnamefont {G.}~\bibnamefont
  {Vidal}}\ and\ \bibinfo {author} {\bibfnamefont {R.~F.}\ \bibnamefont
  {Werner}},\ }\href {\doibase 10.1103/PhysRevA.65.032314} {\bibfield
  {journal} {\bibinfo  {journal} {Phys. Rev. A}\ }\textbf {\bibinfo {volume}
  {65}},\ \bibinfo {pages} {032314} (\bibinfo {year} {2002})}\BibitemShut
  {NoStop}%
\bibitem [{\citenamefont {Rist\`e}\ \emph {et~al.}(2012)\citenamefont
  {Rist\`e}, \citenamefont {van Leeuwen}, \citenamefont {Ku}, \citenamefont
  {Lehnert},\ and\ \citenamefont {DiCarlo}}]{Riste12}%
  \BibitemOpen
  \bibfield  {author} {\bibinfo {author} {\bibfnamefont {D.}~\bibnamefont
  {Rist\`e}}, \bibinfo {author} {\bibfnamefont {J.~G.}\ \bibnamefont {van
  Leeuwen}}, \bibinfo {author} {\bibfnamefont {H.-S.}\ \bibnamefont {Ku}},
  \bibinfo {author} {\bibfnamefont {K.~W.}\ \bibnamefont {Lehnert}}, \ and\
  \bibinfo {author} {\bibfnamefont {L.}~\bibnamefont {DiCarlo}},\ }\href
  {\doibase 10.1103/PhysRevLett.109.050507} {\bibfield  {journal} {\bibinfo
  {journal} {Phys. Rev. Lett.}\ }\textbf {\bibinfo {volume} {109}},\ \bibinfo
  {pages} {050507} (\bibinfo {year} {2012})}\BibitemShut {NoStop}%
\bibitem [{\citenamefont {Chapman}\ \emph {et~al.}(2017)\citenamefont
  {Chapman}, \citenamefont {Rosenthal}, \citenamefont {Kerckhoff},
  \citenamefont {Moores}, \citenamefont {Vale}, \citenamefont {Mates},
  \citenamefont {Hilton}, \citenamefont {Lalumi\`ere}, \citenamefont {Blais},\
  and\ \citenamefont {Lehnert}}]{Chapman17}%
  \BibitemOpen
  \bibfield  {author} {\bibinfo {author} {\bibfnamefont {B.~J.}\ \bibnamefont
  {Chapman}}, \bibinfo {author} {\bibfnamefont {E.~I.}\ \bibnamefont
  {Rosenthal}}, \bibinfo {author} {\bibfnamefont {J.}~\bibnamefont
  {Kerckhoff}}, \bibinfo {author} {\bibfnamefont {B.~A.}\ \bibnamefont
  {Moores}}, \bibinfo {author} {\bibfnamefont {L.~R.}\ \bibnamefont {Vale}},
  \bibinfo {author} {\bibfnamefont {J.~A.~B.}\ \bibnamefont {Mates}}, \bibinfo
  {author} {\bibfnamefont {G.~C.}\ \bibnamefont {Hilton}}, \bibinfo {author}
  {\bibfnamefont {K.}~\bibnamefont {Lalumi\`ere}}, \bibinfo {author}
  {\bibfnamefont {A.}~\bibnamefont {Blais}}, \ and\ \bibinfo {author}
  {\bibfnamefont {K.~W.}\ \bibnamefont {Lehnert}},\ }\href {\doibase
  10.1103/PhysRevX.7.041043} {\bibfield  {journal} {\bibinfo  {journal} {Phys.
  Rev. X}\ }\textbf {\bibinfo {volume} {7}},\ \bibinfo {pages} {041043}
  (\bibinfo {year} {2017})}\BibitemShut {NoStop}%
\bibitem [{\citenamefont {Roy}\ \emph {et~al.}(2016)\citenamefont {Roy},
  \citenamefont {Stone},\ and\ \citenamefont {Jiang}}]{Roy16}%
  \BibitemOpen
  \bibfield  {author} {\bibinfo {author} {\bibfnamefont {A.}~\bibnamefont
  {Roy}}, \bibinfo {author} {\bibfnamefont {A.~D.}\ \bibnamefont {Stone}}, \
  and\ \bibinfo {author} {\bibfnamefont {L.}~\bibnamefont {Jiang}},\ }\href
  {\doibase 10.1103/PhysRevA.94.032333} {\bibfield  {journal} {\bibinfo
  {journal} {Phys. Rev. A}\ }\textbf {\bibinfo {volume} {94}},\ \bibinfo
  {pages} {032333} (\bibinfo {year} {2016})}\BibitemShut {NoStop}%
\bibitem [{\citenamefont {Deutsch}\ \emph {et~al.}(1996)\citenamefont
  {Deutsch}, \citenamefont {Ekert}, \citenamefont {Jozsa}, \citenamefont
  {Macchiavello}, \citenamefont {Popescu},\ and\ \citenamefont
  {Sanpera}}]{Deutsch96}%
  \BibitemOpen
  \bibfield  {author} {\bibinfo {author} {\bibfnamefont {D.}~\bibnamefont
  {Deutsch}}, \bibinfo {author} {\bibfnamefont {A.}~\bibnamefont {Ekert}},
  \bibinfo {author} {\bibfnamefont {R.}~\bibnamefont {Jozsa}}, \bibinfo
  {author} {\bibfnamefont {C.}~\bibnamefont {Macchiavello}}, \bibinfo {author}
  {\bibfnamefont {S.}~\bibnamefont {Popescu}}, \ and\ \bibinfo {author}
  {\bibfnamefont {A.}~\bibnamefont {Sanpera}},\ }\href {\doibase
  10.1103/PhysRevLett.77.2818} {\bibfield  {journal} {\bibinfo  {journal}
  {Phys. Rev. Lett.}\ }\textbf {\bibinfo {volume} {77}},\ \bibinfo {pages}
  {2818} (\bibinfo {year} {1996})}\BibitemShut {NoStop}%
\bibitem [{\citenamefont {Bennett}\ \emph {et~al.}(1996)\citenamefont
  {Bennett}, \citenamefont {Brassard}, \citenamefont {Popescu}, \citenamefont
  {Schumacher}, \citenamefont {Smolin},\ and\ \citenamefont
  {Wootters}}]{Bennett96b}%
  \BibitemOpen
  \bibfield  {author} {\bibinfo {author} {\bibfnamefont {C.~H.}\ \bibnamefont
  {Bennett}}, \bibinfo {author} {\bibfnamefont {G.}~\bibnamefont {Brassard}},
  \bibinfo {author} {\bibfnamefont {S.}~\bibnamefont {Popescu}}, \bibinfo
  {author} {\bibfnamefont {B.}~\bibnamefont {Schumacher}}, \bibinfo {author}
  {\bibfnamefont {J.~A.}\ \bibnamefont {Smolin}}, \ and\ \bibinfo {author}
  {\bibfnamefont {W.~K.}\ \bibnamefont {Wootters}},\ }\href {\doibase
  10.1103/PhysRevLett.76.722} {\bibfield  {journal} {\bibinfo  {journal} {Phys.
  Rev. Lett.}\ }\textbf {\bibinfo {volume} {76}},\ \bibinfo {pages} {722}
  (\bibinfo {year} {1996})}\BibitemShut {NoStop}%
\bibitem [{\citenamefont {Shankar}\ \emph {et~al.}(2013)\citenamefont
  {Shankar}, \citenamefont {Hatridge}, \citenamefont {Leghtas}, \citenamefont
  {Sliwa}, \citenamefont {Narla}, \citenamefont {Vool}, \citenamefont {Girvin},
  \citenamefont {Frunzio}, \citenamefont {Mirrahimi},\ and\ \citenamefont
  {Devoret}}]{Shankar13}%
  \BibitemOpen
  \bibfield  {author} {\bibinfo {author} {\bibfnamefont {S.}~\bibnamefont
  {Shankar}}, \bibinfo {author} {\bibfnamefont {M.}~\bibnamefont {Hatridge}},
  \bibinfo {author} {\bibfnamefont {Z.}~\bibnamefont {Leghtas}}, \bibinfo
  {author} {\bibfnamefont {K.~M.}\ \bibnamefont {Sliwa}}, \bibinfo {author}
  {\bibfnamefont {A.}~\bibnamefont {Narla}}, \bibinfo {author} {\bibfnamefont
  {U.}~\bibnamefont {Vool}}, \bibinfo {author} {\bibfnamefont {S.~M.}\
  \bibnamefont {Girvin}}, \bibinfo {author} {\bibfnamefont {L.}~\bibnamefont
  {Frunzio}}, \bibinfo {author} {\bibfnamefont {M.}~\bibnamefont {Mirrahimi}},
  \ and\ \bibinfo {author} {\bibfnamefont {M.~H.}\ \bibnamefont {Devoret}},\
  }\href {https://www.nature.com/articles/nature12802} {\bibfield  {journal}
  {\bibinfo  {journal} {Nature}\ }\textbf {\bibinfo {volume} {504}},\ \bibinfo
  {pages} {419} (\bibinfo {year} {2013})}\BibitemShut {NoStop}%
\bibitem [{\citenamefont {Motzoi}\ \emph {et~al.}(2016)\citenamefont {Motzoi},
  \citenamefont {Halperin}, \citenamefont {Wang}, \citenamefont {Whaley},\ and\
  \citenamefont {Schirmer}}]{Motzoi16}%
  \BibitemOpen
  \bibfield  {author} {\bibinfo {author} {\bibfnamefont {F.}~\bibnamefont
  {Motzoi}}, \bibinfo {author} {\bibfnamefont {E.}~\bibnamefont {Halperin}},
  \bibinfo {author} {\bibfnamefont {X.}~\bibnamefont {Wang}}, \bibinfo {author}
  {\bibfnamefont {K.~B.}\ \bibnamefont {Whaley}}, \ and\ \bibinfo {author}
  {\bibfnamefont {S.}~\bibnamefont {Schirmer}},\ }\href {\doibase
  10.1103/PhysRevA.94.032313} {\bibfield  {journal} {\bibinfo  {journal} {Phys.
  Rev. A}\ }\textbf {\bibinfo {volume} {94}},\ \bibinfo {pages} {032313}
  (\bibinfo {year} {2016})}\BibitemShut {NoStop}%
\bibitem [{\citenamefont {Kimchi-Schwartz}\ \emph {et~al.}(2016)\citenamefont
  {Kimchi-Schwartz}, \citenamefont {Martin}, \citenamefont {Flurin},
  \citenamefont {Aron}, \citenamefont {Kulkarni}, \citenamefont {Tureci},\ and\
  \citenamefont {Siddiqi}}]{Kimchi-Schwartz16}%
  \BibitemOpen
  \bibfield  {author} {\bibinfo {author} {\bibfnamefont {M.~E.}\ \bibnamefont
  {Kimchi-Schwartz}}, \bibinfo {author} {\bibfnamefont {L.}~\bibnamefont
  {Martin}}, \bibinfo {author} {\bibfnamefont {E.}~\bibnamefont {Flurin}},
  \bibinfo {author} {\bibfnamefont {C.}~\bibnamefont {Aron}}, \bibinfo {author}
  {\bibfnamefont {M.}~\bibnamefont {Kulkarni}}, \bibinfo {author}
  {\bibfnamefont {H.~E.}\ \bibnamefont {Tureci}}, \ and\ \bibinfo {author}
  {\bibfnamefont {I.}~\bibnamefont {Siddiqi}},\ }\href {\doibase
  10.1103/PhysRevLett.116.240503} {\bibfield  {journal} {\bibinfo  {journal}
  {Phys. Rev. Lett.}\ }\textbf {\bibinfo {volume} {116}},\ \bibinfo {pages}
  {240503} (\bibinfo {year} {2016})}\BibitemShut {NoStop}%
\bibitem [{\citenamefont {Martin}\ \emph {et~al.}(2015)\citenamefont {Martin},
  \citenamefont {Motzoi}, \citenamefont {Li}, \citenamefont {Sarovar},\ and\
  \citenamefont {Whaley}}]{Martin15}%
  \BibitemOpen
  \bibfield  {author} {\bibinfo {author} {\bibfnamefont {L.}~\bibnamefont
  {Martin}}, \bibinfo {author} {\bibfnamefont {F.}~\bibnamefont {Motzoi}},
  \bibinfo {author} {\bibfnamefont {H.}~\bibnamefont {Li}}, \bibinfo {author}
  {\bibfnamefont {M.}~\bibnamefont {Sarovar}}, \ and\ \bibinfo {author}
  {\bibfnamefont {K.~B.}\ \bibnamefont {Whaley}},\ }\href {\doibase
  10.1103/PhysRevA.92.062321} {\bibfield  {journal} {\bibinfo  {journal} {Phys.
  Rev. A}\ }\textbf {\bibinfo {volume} {92}},\ \bibinfo {pages} {062321}
  (\bibinfo {year} {2015})}\BibitemShut {NoStop}%
\bibitem [{\citenamefont {Barends}\ \emph {et~al.}(2011)\citenamefont
  {Barends}, \citenamefont {Wenner}, \citenamefont {Lenander}, \citenamefont
  {Chen}, \citenamefont {Bialczak}, \citenamefont {Kelly}, \citenamefont
  {Lucero}, \citenamefont {O'Malley}, \citenamefont {Mariantoni}, \citenamefont
  {Sank}, \citenamefont {Wang}, \citenamefont {White}, \citenamefont {Yin},
  \citenamefont {Zhao}, \citenamefont {Cleland}, \citenamefont {Martinis},\
  and\ \citenamefont {Baselmans}}]{Barends11}%
  \BibitemOpen
  \bibfield  {author} {\bibinfo {author} {\bibfnamefont {R.}~\bibnamefont
  {Barends}}, \bibinfo {author} {\bibfnamefont {J.}~\bibnamefont {Wenner}},
  \bibinfo {author} {\bibfnamefont {M.}~\bibnamefont {Lenander}}, \bibinfo
  {author} {\bibfnamefont {Y.}~\bibnamefont {Chen}}, \bibinfo {author}
  {\bibfnamefont {R.~C.}\ \bibnamefont {Bialczak}}, \bibinfo {author}
  {\bibfnamefont {J.}~\bibnamefont {Kelly}}, \bibinfo {author} {\bibfnamefont
  {E.}~\bibnamefont {Lucero}}, \bibinfo {author} {\bibfnamefont
  {P.}~\bibnamefont {O'Malley}}, \bibinfo {author} {\bibfnamefont
  {M.}~\bibnamefont {Mariantoni}}, \bibinfo {author} {\bibfnamefont
  {D.}~\bibnamefont {Sank}}, \bibinfo {author} {\bibfnamefont {H.}~\bibnamefont
  {Wang}}, \bibinfo {author} {\bibfnamefont {T.~C.}\ \bibnamefont {White}},
  \bibinfo {author} {\bibfnamefont {Y.}~\bibnamefont {Yin}}, \bibinfo {author}
  {\bibfnamefont {J.}~\bibnamefont {Zhao}}, \bibinfo {author} {\bibfnamefont
  {A.~N.}\ \bibnamefont {Cleland}}, \bibinfo {author} {\bibfnamefont {J.~M.}\
  \bibnamefont {Martinis}}, \ and\ \bibinfo {author} {\bibfnamefont {J.~J.~A.}\
  \bibnamefont {Baselmans}},\ }\href
  {http://scitation.aip.org/content/aip/journal/apl/99/11/10.1063/1.3638063}
  {\bibfield  {journal} {\bibinfo  {journal} {Appl. Phys. Lett.}\ }\textbf
  {\bibinfo {volume} {99}},\ \bibinfo {pages} {113507} (\bibinfo {year}
  {2011})}\BibitemShut {NoStop}%
\bibitem [{\citenamefont {Langford}\ \emph {et~al.}(2017)\citenamefont
  {Langford}, \citenamefont {Sagastizabal}, \citenamefont {Kounalakis},
  \citenamefont {Dickel}, \citenamefont {Bruno}, \citenamefont {Luthi},
  \citenamefont {Thoen}, \citenamefont {Endo},\ and\ \citenamefont
  {DiCarlo}}]{Langford17}%
  \BibitemOpen
  \bibfield  {author} {\bibinfo {author} {\bibfnamefont {N.~K.}\ \bibnamefont
  {Langford}}, \bibinfo {author} {\bibfnamefont {R.}~\bibnamefont
  {Sagastizabal}}, \bibinfo {author} {\bibfnamefont {M.}~\bibnamefont
  {Kounalakis}}, \bibinfo {author} {\bibfnamefont {C.}~\bibnamefont {Dickel}},
  \bibinfo {author} {\bibfnamefont {A.}~\bibnamefont {Bruno}}, \bibinfo
  {author} {\bibfnamefont {F.}~\bibnamefont {Luthi}}, \bibinfo {author}
  {\bibfnamefont {D.}~\bibnamefont {Thoen}}, \bibinfo {author} {\bibfnamefont
  {A.}~\bibnamefont {Endo}}, \ and\ \bibinfo {author} {\bibfnamefont
  {L.}~\bibnamefont {DiCarlo}},\ }\href
  {https://www.nature.com/articles/s41467-017-01061-x?WT.feed_name=subjects_physics}
  {\bibfield  {journal} {\bibinfo  {journal} {Nature Communications}\ }\textbf
  {\bibinfo {volume} {8}},\ \bibinfo {pages} {1715} (\bibinfo {year}
  {2017})}\BibitemShut {NoStop}%
\bibitem [{\citenamefont {Pop}\ \emph {et~al.}(2012)\citenamefont {Pop},
  \citenamefont {Fournier}, \citenamefont {Crozes}, \citenamefont {Lecocq},
  \citenamefont {Matei}, \citenamefont {Pannetier}, \citenamefont {Buisson},\
  and\ \citenamefont {Guichard}}]{Pop12}%
  \BibitemOpen
  \bibfield  {author} {\bibinfo {author} {\bibfnamefont {I.~M.}\ \bibnamefont
  {Pop}}, \bibinfo {author} {\bibfnamefont {T.}~\bibnamefont {Fournier}},
  \bibinfo {author} {\bibfnamefont {T.}~\bibnamefont {Crozes}}, \bibinfo
  {author} {\bibfnamefont {F.}~\bibnamefont {Lecocq}}, \bibinfo {author}
  {\bibfnamefont {I.}~\bibnamefont {Matei}}, \bibinfo {author} {\bibfnamefont
  {B.}~\bibnamefont {Pannetier}}, \bibinfo {author} {\bibfnamefont
  {O.}~\bibnamefont {Buisson}}, \ and\ \bibinfo {author} {\bibfnamefont
  {W.}~\bibnamefont {Guichard}},\ }\href {\doibase 10.1116/1.3673790}
  {\bibfield  {journal} {\bibinfo  {journal} {J. Vac. Sci. Technol. B}\
  }\textbf {\bibinfo {volume} {30}},\ \bibinfo {pages} {010607} (\bibinfo
  {year} {2012})}\BibitemShut {NoStop}%
\bibitem [{\citenamefont {Ambegaokar}\ and\ \citenamefont
  {Baratoff}(1963)}]{Ambegaokar63}%
  \BibitemOpen
  \bibfield  {author} {\bibinfo {author} {\bibfnamefont {V.}~\bibnamefont
  {Ambegaokar}}\ and\ \bibinfo {author} {\bibfnamefont {A.}~\bibnamefont
  {Baratoff}},\ }\href {\doibase 10.1103/PhysRevLett.10.486} {\bibfield
  {journal} {\bibinfo  {journal} {Phys. Rev. Lett.}\ }\textbf {\bibinfo
  {volume} {10}},\ \bibinfo {pages} {486} (\bibinfo {year} {1963})}\BibitemShut
  {NoStop}%
\bibitem [{\citenamefont {Motzoi}\ \emph {et~al.}(2009)\citenamefont {Motzoi},
  \citenamefont {Gambetta}, \citenamefont {Rebentrost},\ and\ \citenamefont
  {Wilhelm}}]{Motzoi09}%
  \BibitemOpen
  \bibfield  {author} {\bibinfo {author} {\bibfnamefont {F.}~\bibnamefont
  {Motzoi}}, \bibinfo {author} {\bibfnamefont {J.~M.}\ \bibnamefont
  {Gambetta}}, \bibinfo {author} {\bibfnamefont {P.}~\bibnamefont
  {Rebentrost}}, \ and\ \bibinfo {author} {\bibfnamefont {F.~K.}\ \bibnamefont
  {Wilhelm}},\ }\href
  {https://journals.aps.org/prl/abstract/10.1103/PhysRevLett.103.110501}
  {\bibfield  {journal} {\bibinfo  {journal} {Phys. Rev. Lett.}\ }\textbf
  {\bibinfo {volume} {103}},\ \bibinfo {pages} {110501} (\bibinfo {year}
  {2009})}\BibitemShut {NoStop}%
\bibitem [{\citenamefont {Chow}\ \emph
  {et~al.}(2010{\natexlab{b}})\citenamefont {Chow}, \citenamefont {DiCarlo},
  \citenamefont {Gambetta}, \citenamefont {Motzoi}, \citenamefont {Frunzio},
  \citenamefont {Girvin},\ and\ \citenamefont {Schoelkopf}}]{Chow10b}%
  \BibitemOpen
  \bibfield  {author} {\bibinfo {author} {\bibfnamefont {J.~M.}\ \bibnamefont
  {Chow}}, \bibinfo {author} {\bibfnamefont {L.}~\bibnamefont {DiCarlo}},
  \bibinfo {author} {\bibfnamefont {J.~M.}\ \bibnamefont {Gambetta}}, \bibinfo
  {author} {\bibfnamefont {F.}~\bibnamefont {Motzoi}}, \bibinfo {author}
  {\bibfnamefont {L.}~\bibnamefont {Frunzio}}, \bibinfo {author} {\bibfnamefont
  {S.~M.}\ \bibnamefont {Girvin}}, \ and\ \bibinfo {author} {\bibfnamefont
  {R.~J.}\ \bibnamefont {Schoelkopf}},\ }\href
  {https://journals.aps.org/pra/abstract/10.1103/PhysRevA.82.040305} {\bibfield
   {journal} {\bibinfo  {journal} {Phys. Rev. A}\ }\textbf {\bibinfo {volume}
  {82}},\ \bibinfo {pages} {040305} (\bibinfo {year}
  {2010}{\natexlab{b}})}\BibitemShut {NoStop}%
\bibitem [{\citenamefont {Reed}(2013)}]{thesisReed13}%
  \BibitemOpen
  \bibfield  {author} {\bibinfo {author} {\bibfnamefont {M.~D.}\ \bibnamefont
  {Reed}},\ }\emph {\bibinfo {title} {Entanglement and Quantum Error Correction
  with Superconducting Qubits}},\ \href@noop {} {\bibinfo {type} {Ph{D}
  {D}issertation}},\ \bibinfo  {school} {Yale University} (\bibinfo {year}
  {2013})\BibitemShut {NoStop}%
\bibitem [{\citenamefont {Bhattacharyya}(1946)}]{Bhattacharyya46}%
  \BibitemOpen
  \bibfield  {author} {\bibinfo {author} {\bibfnamefont {A.}~\bibnamefont
  {Bhattacharyya}},\ }\href@noop {} {\bibfield  {journal} {\bibinfo  {journal}
  {Sankhy{\=a}: the indian journal of statistics}\ }\textbf {\bibinfo {volume}
  {7}},\ \bibinfo {pages} {401} (\bibinfo {year} {1946})}\BibitemShut {NoStop}%
\bibitem [{\citenamefont {de~Lange}\ \emph {et~al.}(2014)\citenamefont
  {de~Lange}, \citenamefont {Rist\`e}, \citenamefont {Tiggelman}, \citenamefont
  {Eichler}, \citenamefont {Tornberg}, \citenamefont {Johansson}, \citenamefont
  {Wallraff}, \citenamefont {Schouten},\ and\ \citenamefont
  {DiCarlo}}]{deLange14}%
  \BibitemOpen
  \bibfield  {author} {\bibinfo {author} {\bibfnamefont {G.}~\bibnamefont
  {de~Lange}}, \bibinfo {author} {\bibfnamefont {D.}~\bibnamefont {Rist\`e}},
  \bibinfo {author} {\bibfnamefont {M.~J.}\ \bibnamefont {Tiggelman}}, \bibinfo
  {author} {\bibfnamefont {C.}~\bibnamefont {Eichler}}, \bibinfo {author}
  {\bibfnamefont {L.}~\bibnamefont {Tornberg}}, \bibinfo {author}
  {\bibfnamefont {G.}~\bibnamefont {Johansson}}, \bibinfo {author}
  {\bibfnamefont {A.}~\bibnamefont {Wallraff}}, \bibinfo {author}
  {\bibfnamefont {R.~N.}\ \bibnamefont {Schouten}}, \ and\ \bibinfo {author}
  {\bibfnamefont {L.}~\bibnamefont {DiCarlo}},\ }\href@noop {} {\bibfield
  {journal} {\bibinfo  {journal} {Phys. Rev. Lett.}\ }\textbf {\bibinfo
  {volume} {112}},\ \bibinfo {pages} {080501} (\bibinfo {year}
  {2014})}\BibitemShut {NoStop}%
\bibitem [{\citenamefont {Macklin}\ \emph {et~al.}(2015)\citenamefont
  {Macklin}, \citenamefont {O{\textquoteright}Brien}, \citenamefont {Hover},
  \citenamefont {Schwartz}, \citenamefont {Bolkhovsky}, \citenamefont {Zhang},
  \citenamefont {Oliver},\ and\ \citenamefont {Siddiqi}}]{Macklin15}%
  \BibitemOpen
  \bibfield  {author} {\bibinfo {author} {\bibfnamefont {C.}~\bibnamefont
  {Macklin}}, \bibinfo {author} {\bibfnamefont {K.}~\bibnamefont
  {O{\textquoteright}Brien}}, \bibinfo {author} {\bibfnamefont
  {D.}~\bibnamefont {Hover}}, \bibinfo {author} {\bibfnamefont {M.~E.}\
  \bibnamefont {Schwartz}}, \bibinfo {author} {\bibfnamefont {V.}~\bibnamefont
  {Bolkhovsky}}, \bibinfo {author} {\bibfnamefont {X.}~\bibnamefont {Zhang}},
  \bibinfo {author} {\bibfnamefont {W.~D.}\ \bibnamefont {Oliver}}, \ and\
  \bibinfo {author} {\bibfnamefont {I.}~\bibnamefont {Siddiqi}},\ }\href
  {http://www.sciencemag.org/cgi/doi/10.1126/science.aaa8525} {\bibfield
  {journal} {\bibinfo  {journal} {Science}\ }\textbf {\bibinfo {volume}
  {350}},\ \bibinfo {pages} {307} (\bibinfo {year} {2015})}\BibitemShut
  {NoStop}%
\bibitem [{\citenamefont {Walls}\ and\ \citenamefont
  {Milburn}(2007)}]{walls2007quantum}%
  \BibitemOpen
  \bibfield  {author} {\bibinfo {author} {\bibfnamefont {D.~F.}\ \bibnamefont
  {Walls}}\ and\ \bibinfo {author} {\bibfnamefont {G.~J.}\ \bibnamefont
  {Milburn}},\ }\href@noop {} {\emph {\bibinfo {title} {Quantum optics}}}\
  (\bibinfo  {publisher} {Springer Science \& Business Media},\ \bibinfo {year}
  {2007})\BibitemShut {NoStop}%
\bibitem [{\citenamefont {McClure}\ \emph {et~al.}(2016)\citenamefont
  {McClure}, \citenamefont {Paik}, \citenamefont {Bishop}, \citenamefont
  {Steffen}, \citenamefont {Chow},\ and\ \citenamefont {Gambetta}}]{McClure16}%
  \BibitemOpen
  \bibfield  {author} {\bibinfo {author} {\bibfnamefont {D.~T.}\ \bibnamefont
  {McClure}}, \bibinfo {author} {\bibfnamefont {H.}~\bibnamefont {Paik}},
  \bibinfo {author} {\bibfnamefont {L.~S.}\ \bibnamefont {Bishop}}, \bibinfo
  {author} {\bibfnamefont {M.}~\bibnamefont {Steffen}}, \bibinfo {author}
  {\bibfnamefont {J.~M.}\ \bibnamefont {Chow}}, \ and\ \bibinfo {author}
  {\bibfnamefont {J.~M.}\ \bibnamefont {Gambetta}},\ }\href
  {https://journals.aps.org/prapplied/abstract/10.1103/PhysRevApplied.5.011001}
  {\bibfield  {journal} {\bibinfo  {journal} {Phys. Rev. Appl.}\ }\textbf
  {\bibinfo {volume} {5}},\ \bibinfo {pages} {011001} (\bibinfo {year}
  {2016})}\BibitemShut {NoStop}%
\bibitem [{\citenamefont {Bultink}\ \emph {et~al.}(2016)\citenamefont
  {Bultink}, \citenamefont {Rol}, \citenamefont {O'Brien}, \citenamefont {Fu},
  \citenamefont {Dikken}, \citenamefont {Dickel}, \citenamefont {Vermeulen},
  \citenamefont {de~Sterke}, \citenamefont {Bruno}, \citenamefont {Schouten},\
  and\ \citenamefont {DiCarlo}}]{Bultink16}%
  \BibitemOpen
  \bibfield  {author} {\bibinfo {author} {\bibfnamefont {C.~C.}\ \bibnamefont
  {Bultink}}, \bibinfo {author} {\bibfnamefont {M.~A.}\ \bibnamefont {Rol}},
  \bibinfo {author} {\bibfnamefont {T.~E.}\ \bibnamefont {O'Brien}}, \bibinfo
  {author} {\bibfnamefont {X.}~\bibnamefont {Fu}}, \bibinfo {author}
  {\bibfnamefont {B.~C.~S.}\ \bibnamefont {Dikken}}, \bibinfo {author}
  {\bibfnamefont {C.}~\bibnamefont {Dickel}}, \bibinfo {author} {\bibfnamefont
  {R.~F.~L.}\ \bibnamefont {Vermeulen}}, \bibinfo {author} {\bibfnamefont
  {J.~C.}\ \bibnamefont {de~Sterke}}, \bibinfo {author} {\bibfnamefont
  {A.}~\bibnamefont {Bruno}}, \bibinfo {author} {\bibfnamefont {R.~N.}\
  \bibnamefont {Schouten}}, \ and\ \bibinfo {author} {\bibfnamefont
  {L.}~\bibnamefont {DiCarlo}},\ }\href
  {https://link.aps.org/doi/10.1103/PhysRevApplied.6.034008} {\bibfield
  {journal} {\bibinfo  {journal} {Phys. Rev. Appl.}\ }\textbf {\bibinfo
  {volume} {6}},\ \bibinfo {pages} {034008} (\bibinfo {year}
  {2016})}\BibitemShut {NoStop}%
\bibitem [{\citenamefont {Rist\`{e}}\ \emph {et~al.}(2013)\citenamefont
  {Rist\`{e}}, \citenamefont {Dukalski}, \citenamefont {Watson}, \citenamefont
  {de~Lange}, \citenamefont {Tiggelman}, \citenamefont {Blanter}, \citenamefont
  {Lehnert}, \citenamefont {Schouten},\ and\ \citenamefont
  {DiCarlo}}]{Riste13c}%
  \BibitemOpen
  \bibfield  {author} {\bibinfo {author} {\bibfnamefont {D.}~\bibnamefont
  {Rist\`{e}}}, \bibinfo {author} {\bibfnamefont {M.}~\bibnamefont {Dukalski}},
  \bibinfo {author} {\bibfnamefont {C.~A.}\ \bibnamefont {Watson}}, \bibinfo
  {author} {\bibfnamefont {G.}~\bibnamefont {de~Lange}}, \bibinfo {author}
  {\bibfnamefont {M.~J.}\ \bibnamefont {Tiggelman}}, \bibinfo {author}
  {\bibfnamefont {Y.~M.}\ \bibnamefont {Blanter}}, \bibinfo {author}
  {\bibfnamefont {K.~W.}\ \bibnamefont {Lehnert}}, \bibinfo {author}
  {\bibfnamefont {R.~N.}\ \bibnamefont {Schouten}}, \ and\ \bibinfo {author}
  {\bibfnamefont {L.}~\bibnamefont {DiCarlo}},\ }\href {\doibase
  10.1038/nature12513} {\bibfield  {journal} {\bibinfo  {journal} {Nature}\
  }\textbf {\bibinfo {volume} {502}},\ \bibinfo {pages} {350} (\bibinfo {year}
  {2013})}\BibitemShut {NoStop}%
\bibitem [{\citenamefont {Gardiner}\ and\ \citenamefont
  {Zoller}(2004)}]{Gardiner04}%
  \BibitemOpen
  \bibfield  {author} {\bibinfo {author} {\bibfnamefont {C.}~\bibnamefont
  {Gardiner}}\ and\ \bibinfo {author} {\bibfnamefont {P.}~\bibnamefont
  {Zoller}},\ }\href@noop {} {\emph {\bibinfo {title} {{Quantum Noise}}}},\
  \bibinfo {edition} {3rd}\ ed.\ (\bibinfo  {publisher} {Springer, Berlin,
  Germany},\ \bibinfo {year} {2004})\BibitemShut {NoStop}%
\bibitem [{\citenamefont {Di{C}arlo}\ \emph {et~al.}(2009)\citenamefont
  {Di{C}arlo}, \citenamefont {Chow}, \citenamefont {Gambetta}, \citenamefont
  {Bishop}, \citenamefont {Johnson}, \citenamefont {Schuster}, \citenamefont
  {Majer}, \citenamefont {Blais}, \citenamefont {Frunzio}, \citenamefont
  {Girvin},\ and\ \citenamefont {Schoelkopf}}]{DiCarlo09}%
  \BibitemOpen
  \bibfield  {author} {\bibinfo {author} {\bibfnamefont {L.}~\bibnamefont
  {Di{C}arlo}}, \bibinfo {author} {\bibfnamefont {J.~M.}\ \bibnamefont {Chow}},
  \bibinfo {author} {\bibfnamefont {J.~M.}\ \bibnamefont {Gambetta}}, \bibinfo
  {author} {\bibfnamefont {L.~S.}\ \bibnamefont {Bishop}}, \bibinfo {author}
  {\bibfnamefont {B.~R.}\ \bibnamefont {Johnson}}, \bibinfo {author}
  {\bibfnamefont {D.~I.}\ \bibnamefont {Schuster}}, \bibinfo {author}
  {\bibfnamefont {J.}~\bibnamefont {Majer}}, \bibinfo {author} {\bibfnamefont
  {A.}~\bibnamefont {Blais}}, \bibinfo {author} {\bibfnamefont
  {L.}~\bibnamefont {Frunzio}}, \bibinfo {author} {\bibfnamefont {S.~M.}\
  \bibnamefont {Girvin}}, \ and\ \bibinfo {author} {\bibfnamefont {R.~J.}\
  \bibnamefont {Schoelkopf}},\ }\href
  {http://www.nature.com/nature/journal/v460/n7252/abs/nature08121.html}
  {\bibfield  {journal} {\bibinfo  {journal} {Nature}\ }\textbf {\bibinfo
  {volume} {460}},\ \bibinfo {pages} {240} (\bibinfo {year}
  {2009})}\BibitemShut {NoStop}%
\bibitem [{\citenamefont {James}\ \emph {et~al.}(2001)\citenamefont {James},
  \citenamefont {Kwiat}, \citenamefont {Munro},\ and\ \citenamefont
  {White}}]{James01}%
  \BibitemOpen
  \bibfield  {author} {\bibinfo {author} {\bibfnamefont {D.~F.~V.}\
  \bibnamefont {James}}, \bibinfo {author} {\bibfnamefont {P.~G.}\ \bibnamefont
  {Kwiat}}, \bibinfo {author} {\bibfnamefont {W.~J.}\ \bibnamefont {Munro}}, \
  and\ \bibinfo {author} {\bibfnamefont {A.~G.}\ \bibnamefont {White}},\ }\href
  {http://journals.aps.org/pra/abstract/10.1103/PhysRevA.64.052312} {\bibfield
  {journal} {\bibinfo  {journal} {Phys. Rev. A}\ }\textbf {\bibinfo {volume}
  {64}},\ \bibinfo {pages} {052312} (\bibinfo {year} {2001})}\BibitemShut
  {NoStop}%
\bibitem [{\citenamefont {Boyd}(2004)}]{Boyd04}%
  \BibitemOpen
  \bibfield  {author} {\bibinfo {author} {\bibfnamefont {S.}~\bibnamefont
  {Boyd}},\ }\href@noop {} {\emph {\bibinfo {title} {{Convex Optimization}}}},\
  \bibinfo {edition} {1st}\ ed.\ (\bibinfo  {publisher} {Cambridge University
  Press},\ \bibinfo {year} {2004})\BibitemShut {NoStop}%
\bibitem [{\citenamefont {Langford}(2013)}]{Langford13}%
  \BibitemOpen
  \bibfield  {author} {\bibinfo {author} {\bibfnamefont {N.~K.}\ \bibnamefont
  {Langford}},\ }\href
  {http://iopscience.iop.org/article/10.1088/1367-2630/15/3/035003} {\bibfield
  {journal} {\bibinfo  {journal} {New J.\ Phys.}\ }\textbf {\bibinfo {volume}
  {15}},\ \bibinfo {pages} {035003} (\bibinfo {year} {2013})}\BibitemShut
  {NoStop}%
\bibitem [{\citenamefont {Vissers}\ \emph {et~al.}(2010)\citenamefont
  {Vissers}, \citenamefont {Gao}, \citenamefont {Wisbey}, \citenamefont {Hite},
  \citenamefont {Tsuei}, \citenamefont {Corcoles}, \citenamefont {Steffen},\
  and\ \citenamefont {Pappas}}]{Vissers10}%
  \BibitemOpen
  \bibfield  {author} {\bibinfo {author} {\bibfnamefont {M.~R.}\ \bibnamefont
  {Vissers}}, \bibinfo {author} {\bibfnamefont {J.}~\bibnamefont {Gao}},
  \bibinfo {author} {\bibfnamefont {D.~S.}\ \bibnamefont {Wisbey}}, \bibinfo
  {author} {\bibfnamefont {D.~A.}\ \bibnamefont {Hite}}, \bibinfo {author}
  {\bibfnamefont {C.~C.}\ \bibnamefont {Tsuei}}, \bibinfo {author}
  {\bibfnamefont {A.~D.}\ \bibnamefont {Corcoles}}, \bibinfo {author}
  {\bibfnamefont {M.}~\bibnamefont {Steffen}}, \ and\ \bibinfo {author}
  {\bibfnamefont {D.~P.}\ \bibnamefont {Pappas}},\ }\href {\doibase
  doi:10.1063/1.3517252} {\bibfield  {journal} {\bibinfo  {journal} {Appl.
  Phys. Lett.}\ }\textbf {\bibinfo {volume} {97}},\ \bibinfo {pages} {232509}
  (\bibinfo {year} {2010})}\BibitemShut {NoStop}%
\bibitem [{\citenamefont {Sandberg}\ \emph {et~al.}(2008)\citenamefont
  {Sandberg}, \citenamefont {Wilson}, \citenamefont {Persson}, \citenamefont
  {Bauch}, \citenamefont {Johansson}, \citenamefont {Shumeiko}, \citenamefont
  {Duty},\ and\ \citenamefont {Delsing}}]{Sandberg08}%
  \BibitemOpen
  \bibfield  {author} {\bibinfo {author} {\bibfnamefont {M.}~\bibnamefont
  {Sandberg}}, \bibinfo {author} {\bibfnamefont {C.~M.}\ \bibnamefont
  {Wilson}}, \bibinfo {author} {\bibfnamefont {F.}~\bibnamefont {Persson}},
  \bibinfo {author} {\bibfnamefont {T.}~\bibnamefont {Bauch}}, \bibinfo
  {author} {\bibfnamefont {G.}~\bibnamefont {Johansson}}, \bibinfo {author}
  {\bibfnamefont {V.}~\bibnamefont {Shumeiko}}, \bibinfo {author}
  {\bibfnamefont {T.}~\bibnamefont {Duty}}, \ and\ \bibinfo {author}
  {\bibfnamefont {P.}~\bibnamefont {Delsing}},\ }\href
  {http://dx.doi.org/10.1063/1.2929367} {\bibfield  {journal} {\bibinfo
  {journal} {Appl. Phys. Lett.}\ }\textbf {\bibinfo {volume} {92}},\ \bibinfo
  {pages} {203501} (\bibinfo {year} {2008})}\BibitemShut {NoStop}%
\bibitem [{\citenamefont {Yamamoto}\ \emph {et~al.}(2008)\citenamefont
  {Yamamoto}, \citenamefont {Inomata}, \citenamefont {Watanabe}, \citenamefont
  {Matsuba}, \citenamefont {Miyazaki}, \citenamefont {Oliver}, \citenamefont
  {Nakamura},\ and\ \citenamefont {Tsai}}]{Yamamoto08}%
  \BibitemOpen
  \bibfield  {author} {\bibinfo {author} {\bibfnamefont {T.}~\bibnamefont
  {Yamamoto}}, \bibinfo {author} {\bibfnamefont {K.}~\bibnamefont {Inomata}},
  \bibinfo {author} {\bibfnamefont {M.}~\bibnamefont {Watanabe}}, \bibinfo
  {author} {\bibfnamefont {K.}~\bibnamefont {Matsuba}}, \bibinfo {author}
  {\bibfnamefont {T.}~\bibnamefont {Miyazaki}}, \bibinfo {author}
  {\bibfnamefont {W.~D.}\ \bibnamefont {Oliver}}, \bibinfo {author}
  {\bibfnamefont {Y.}~\bibnamefont {Nakamura}}, \ and\ \bibinfo {author}
  {\bibfnamefont {J.~S.}\ \bibnamefont {Tsai}},\ }\href
  {http://dx.doi.org/10.1063/1.2964182} {\bibfield  {journal} {\bibinfo
  {journal} {Appl. Phys. Lett.}\ }\textbf {\bibinfo {volume} {93}},\ \bibinfo
  {pages} {042510} (\bibinfo {year} {2008})}\BibitemShut {NoStop}%
\bibitem [{\citenamefont {Tavis}\ and\ \citenamefont
  {Cummings}(1968)}]{Tavis68}%
  \BibitemOpen
  \bibfield  {author} {\bibinfo {author} {\bibfnamefont {M.}~\bibnamefont
  {Tavis}}\ and\ \bibinfo {author} {\bibfnamefont {F.~W.}\ \bibnamefont
  {Cummings}},\ }\href
  {https://journals.aps.org/pr/abstract/10.1103/PhysRev.170.379} {\bibfield
  {journal} {\bibinfo  {journal} {Phys. Rev.}\ }\textbf {\bibinfo {volume}
  {170}},\ \bibinfo {pages} {379} (\bibinfo {year} {1968})}\BibitemShut
  {NoStop}%
\end{thebibliography}
\end{document}